%% file: main.tex
\documentclass[twocolumn]{aastex631}
\usepackage[utf8]{inputenc}
\usepackage{graphicx}
\usepackage{comment}
\usepackage{natbib}
\usepackage{hyperref}
\usepackage{ulem,color}
\usepackage{eqnarray,amsmath}

\usepackage{listings}
\usepackage{xcolor}
\usepackage{xspace}
\definecolor{codegreen}{rgb}{0,0.6,0}
\definecolor{codegray}{rgb}{0.5,0.5,0.5}
\definecolor{codepurple}{rgb}{0.58,0,0.82}
\definecolor{backcolour}{rgb}{1,1,1}

\lstdefinestyle{mystyle}{
    backgroundcolor=\color{backcolour},   
    commentstyle=\color{codegreen},
    keywordstyle=\color{magenta},
    numberstyle=\tiny\color{codegray},
    stringstyle=\color{codepurple},
    basicstyle=\ttfamily\footnotesize,
    breakatwhitespace=false,         
    breaklines=true,                 
    captionpos=b,                    
    keepspaces=true,                 
    numbers=none,                    
    numbersep=5pt,                  
    showspaces=false,                
    showstringspaces=false,
    showtabs=false,                  
    tabsize=2
}

\newcommand{\rsun}{\ensuremath{R_{\odot}}}
\newcommand{\kms}{km~s$^{-1}$\xspace}
\newcommand{\msun}{\ensuremath{M_{\odot}}}
\newcommand{\lsun}{\ensuremath{L_{\odot}}}
\newcommand{\bh}{\textit{Gaia}~DR3 4373465352415301632}

\begin{document}

\title{A non-interacting Galactic black hole candidate in a binary system with a main-sequence star}

%\author{Joe\altaffiliation{1}}
%\affil{$^{1}$ School of Physics and Astronomy, Rochester Institute of Technology, 1 Lomb Memorial Dr, Rochester, NY 14623}

\author{Sukanya Chakrabarti}
\affiliation{Department of Physics and Astronomy, University of Alabama, Huntsville, 301 North Sparkman Drive, Huntsville, USA}
\author{Joshua D. Simon} 
\affiliation{Observatories of the Carnegie Institution for Science, 813 Santa Barbara St., Pasadena, CA 91101, USA}
\author{Peter A. Craig}
\affiliation{School of Physics and Astronomy, Rochester Institute of Technology, 1 Lomb Memorial Dr, Rochester, NY 14623}
\author{Henrique Reggiani}
\affiliation{Observatories of the Carnegie Institution for Science, 813 Santa Barbara St., Pasadena, CA 91101, USA}
\author{Timothy D. Brandt}
\affiliation{Department of Physics, University of California, Santa Barbara, Santa Barbara, CA 93106, USA}
\author{Puragra Guhathakurta}
\affiliation{University of California Santa Cruz, UCO/Lick Observatory, 1156 High St., Santa Cruz, CA 95064, USA}
\author{Paul A.\ Dalba}
\affiliation{Heising-Simons 51 Pegasi b Postdoctoral Fellow}
\affiliation{Department of Astronomy and Astrophysics, University of California, Santa Cruz, CA 95064, USA}
\affiliation{SETI Institute, Carl Sagan Center, 339 Bernardo Ave, Suite 200, Mountain View, CA 94043, USA}
\author{Evan N. Kirby}
\affiliation{Department of Physics and Astronomy, University of Notre Dame, 225 Nieuwland Science Hall, Notre Dame, IN 46556, USA}
\author{Philip Chang}
\affiliation{Department of Physics, University of Wisconsin-Milwaukee, 3135 North Maryland Avenue, Milwaukee, WI 53211, USA}
\author{Daniel R. Hey}
\affiliation{Institute for Astronomy, University of Hawai'i, Honolulu, HI 96822, USA}
\author{Alessandro Savino}
\affiliation{Department of Astronomy, University of California, Berkeley, Berkeley, CA, 94720, USA}
\author{Marla Geha}
\affiliation{Department of Astronomy, Yale University, 52 Hillhouse Avenue, New Haven, CT 06520, USA}
\author{Ian B. Thompson}
\affiliation{Observatories of the Carnegie Institution for Science, 813 Santa Barbara St., Pasadena, CA 91101, USA}
%\affil{$^{1}$Department of Physics and Astronomy, University of Alabama, Huntsville, 301 North Sparkman Drive, Huntsville, USA}
%\affil{$^{2}$ Observatories of the Carnegie Institution for Science, 813 Santa Barbara St., Pasadena, CA 91101, USA}
%\affil{$^{3}$ School of Physics and Astronomy, Rochester Institute of Technology, 1 Lomb Memorial Dr, Rochester, NY 14623}
%\affil{$^{2}$ Observatories of the Carnegie Institution for Science, 813 Santa Barbara St., Pasadena, CA 91101, USA}
%\affil{$^{5}$ University of California Santa Cruz, UCO/Lick Observatory, 1156 High St., Santa Cruz, CA 95064, USA}
%\affil{$^{6}$ Heising-Simons 51 Pegasi b Postdoctoral Fellow}
%\affil{$^{7}$ Department of Astronomy and Astrophysics, University of California, Santa Cruz, CA 95064, USA}
%\affil{$^{8}$ SETI Institute, Carl Sagan Center, 339 Bernardo Ave, Suite 200, Mountain View, CA 94043, USA}
%\affil{$^{9}$ Department of Physics and Astronomy, University of Notre Dame, 225 Nieuwland Science Hall, Notre Dame, IN 46556, USA}
%\affil{$^{10}$ Department of Physics, University of Wisconsin-Milwaukee, 3135 North Maryland Avenue, Milwaukee, WI 53211, USA}
%\affil{$^{11}$ Institute for Astronomy, University of Hawai'i, Honolulu, HI 96822, USA}
%\affil{$^{12}$ Department of Astronomy, University of California, Berkeley, Berkeley, CA, 94720, USA}
%\affil{$^{13}$ Department of Astronomy, Yale University, 52 Hillhouse Avenue, New Haven, CT 06520, USA}

\begin{abstract}

We describe the discovery of a solar neighborhood (\textbf{$d=468$} pc) binary system with a main-sequence sunlike star and a massive non-interacting black hole candidate. The spectral energy distribution (SED) of the visible star is described by a single stellar model. We derive stellar parameters from a high signal-to-noise Magellan/MIKE spectrum, classifying the star as a main-sequence star with $T_{\rm eff} = 5972 \rm K$, $\log{g} = 4.54$, and $M = 0.91$ \msun. The spectrum shows no indication of a second luminous component. To determine the spectroscopic orbit of the binary, we measured radial velocities of this system with the Automated Planet Finder, Magellan, and Keck over four months. We show that the velocity data are consistent with the \textit{Gaia} astrometric orbit and provide independent evidence for a massive dark companion. From a combined fit of our spectroscopic data and the astrometry, we derive a companion mass of $11.39^{+1.51}_{-1.31}$\msun. We conclude that this binary system harbors a massive black hole on an eccentric $(e =0.46 \pm 0.02)$, $185.4 \pm 0.1$ d orbit. These conclusions are independent of \cite{ElBadry2022Disc}, who recently reported the discovery of the same system. A joint fit to all available data (including \cite{ElBadry2022Disc}'s) yields a comparable period solution, but a lower companion mass of $9.32^{+0.22}_{-0.21} M_{\odot}$. Radial velocity fits to all available data produce a unimodal solution for the period that is not possible with either data set alone. The combination of both data sets yields the most accurate orbit currently available. 
 
\end{abstract}

\section{Introduction}
\label{s:intro}

Simple stellar population calculations suggest that stellar mass black holes should be abundant \citep[e.g.,][]{Breivik2017,Wiktorowicz2019}, with $\sim10^{8}$ present in the Milky Way \citep[e.g.,][]{Olejak2020}.  
%The recent discovery of a large sample of merging black hole binary systems via gravitational waves has significantly increased interest in these objects.  Determining the mass function of black holes in the Galaxy will have implications for the origin of gravitational wave sources, the evolution of massive binary stars, and the physics of supernova explosions.
However, black holes in this mass range are difficult to identify observationally.  Isolated black holes can only be detected with gravitational microlensing \citep[e.g.,][]{Lam2022,Sahu2022},
while black holes in binary systems are easily detectable only when the companion star is close enough for an accretion disk to form.  
Although dozens of X-ray binaries with black hole candidates in short-period orbits have been identified \citep[e.g.,][]{Corral2013,Corral-Santana2016,Pan22,Russell2022}, wider binaries have proven challenging to find.

The detection of gravitational waves from black hole merger events with LIGO \citep[e.g.,][]{abbott2016,abbott2019,abbott2021} provided a new avenue for discovering binary black holes beyond the X-ray bright population \citep{Corral-Santana2016}.  However, like the X-ray binaries, LIGO black hole mergers are necessarily found in very tight orbits.  With the publication of the large catalog of binary systems included in the third \textit{Gaia} data release (DR3; \citealt{2016A&A...595A...1G,GaiaDR3,Halbwachs2022,Vallenari2022}) and the anticipated even larger data set to come in DR4, we can expect to significantly expand upon discoveries of black holes in binary systems, and particularly for black holes in long-period systems that can be characterized by measuring accelerations from astrometric and spectroscopic data \citep[e.g.,][]{Breivik2017,Mashian2017,Yamaguchi2018,Wiktorowicz2020,Chawla2022,Janssens2022}.  Since binaries are coeval, studying these systems may help in understanding the dependence of the formation rate of black holes in binaries on metallicity and age, and thereby indirectly on the formation channel.  
 
Several mechanisms for the formation of black holes in binary systems have been discussed in the literature.  Binary black holes may form via common envelope evolution \citep{Belczynski2016}, stable mass transfer \citep{VandenHeuvel2017,Neijssel2019,Bavera2020,Marchant2021}, chemically homogeneous evolution \citep{DeMink_Mandel2016}, and dynamical processes in dense stellar environments \citep{Antonini_Rasio2016}.  Early studies of massive LIGO black holes attributed the formation of these systems to pristine low-metallicity environments, which may be found in dwarf galaxies or massive galaxies at high redshift \citep{Lamberts2016}.  Another possibility is that these massive black holes may form in the outer \ion{H}{1} disks of spiral galaxies \citep{Chakrabarti2017,Chakrabarti2018}, where low metallicities are also prevalent \citep[e.g.,][]{Kennicutt2003,Bresolin2012,Berg2020}.  Studies of the black hole-main sequence (BH-MS) binary population discovered from \textit{Gaia} can now reveal potential differences with respect to the black hole-black hole (BH-BH) and black hole-neutron star (BH-NS) demographics.  Although it is now clear that there is a strong observed anticorrelation between metallicity and close (periods $ <10^{4}~\rm d$) binary fraction for solar type stars \citep{Gao2014,Moe2019}, the reverse trend is apparent for the high-mass X-ray binary population \citep{Lehmer2021}.

%Using a simple model, \cite{Fischer2004} showed that conservation of angular momentum leads to the binary period distribution being anticorrelated with the mass ratio of the binary in the context of isolated star formation.  

Searches for non-interacting black holes in binary systems with luminous companions have intensified in recent years, with a number of claimed detections \citep[e.g.,][]{Liu2019,Rivinius2020,Jayasinghe2021,Jayasinghe2022,Lennon2021,Saracino2022}.  Most of these systems have been rapidly disputed or ruled out \citep[e.g.,][]{Abdul-Masih2020,Bodensteiner2020,El-Badry_Quataert20202,Eldridge2020,Shenar2020,El-Badry2021,El-BadryBurdge2022,ElBadry2022,El-BadryBurdgeMroz2022,Frost2022}.
%Although non-interacting black holes are expected to be abundant in the Galaxy \citep{Breivik2017,Wiktorowicz2020}, 
To date only a few likely black holes\footnote{Here we are setting aside several black hole binary candidates in globular clusters \citep{Giesers2018,Giesers2019}, which may have different formation mechanisms.} have survived this community vetting: a $>3$~M$_{\odot}$ black hole with a red giant companion \citep[][although see \citealt{vandenHeuvel2020}]{Thompson2019}, a $>9$~M$_{\odot}$ black hole orbiting a massive O star in the Large Magellanic Cloud \citep{Shenar2022}, and a $>$  7 $M_{\odot}$ black hole candidate orbiting the Galactic O star HD~130298 \citep{Mahy2022}.  The object that we discuss in this paper has also been studied earlier by \cite{ElBadry2022Disc}. 

In work that is contemporaneous with ours (posted to the archive on September 14, 2022), \cite{ElBadry2022Disc} reported the independent discovery and analysis of the same source.  Here we present our independent analysis of a main-sequence G star on an eccentric 185~d orbit around a $\sim10$~\msun\ dark companion, which we selected from the \textit{Gaia} DR3 binary catalog.  Given the now available data published from \cite{ElBadry2022Disc}, we also present RV-only fits to the combined data set (ours and that of \citeauthor{ElBadry2022Disc}) as well as joint fits that incorporate both sets of RV data and the astrometry. 

Our search for black hole candidates in binary systems with luminous companions begins with the \textit{Gaia} DR3 catalog of binary masses. \textit{Gaia} DR3 provides the largest sample of binary stars produced by the astronomical community thus far, yielding orbital solutions derived from astrometry and/or spectroscopy for $\sim 3 \times 10^{5}$ stars \citep{GaiaDR3,Halbwachs2022,Holl2022}.   

The paper is organized as follows.  In \S \ref{s:methods}, we briefly review our methodology, including our selection from the \textit{Gaia} DR3 binary mass catalog, and additional tests.  We provide all available photometry for the visible star, and describe our follow-up spectroscopic observations over the last several months and our procedures for measuring radial velocities.  In \S \ref{s:results}, we describe our analysis of the observed spectral energy distribution (SED), fitting the photometric data with single source and multiple source models to find that the visible star is adequately described by a single stellar photosphere.  In \S \ref{s:orbits}, we discuss the \textit{Gaia} astrometric orbital solution, and compute orbits from our velocity data alone and from the combination of the velocities and the astrometry.  We discuss our results in \S \ref{s:discuss} and conclude in \S \ref{s:conclude}.

\section{Methods and Data Collection}
\label{s:methods}

The Gaia DR3 non\_single\_stars catalog contains 195,315 binary systems with lower limits on the mass of the secondary star.  From this set, we selected those for which the secondary was constrained to be more massive than the primary and at least 5~\msun, well above the mass range of white dwarfs and neutron stars.
We removed double-lined spectroscopic binaries, and focused on bright ($G < 14$) companions that would allow for follow-up observations on a wide range of telescopes.  For spectroscopic solutions, we retained sources that have at least ten RV points as reported by \textit{Gaia} and $\mathrm{significance} > 5$.  As an Astronomical Data Query Language search, this selection is given by

\begin{lstlisting}[language=SQL]
select * from 
    gaiadr3.gaia_source inner join
    gaiadr3.nss_two_body_orbit
    using (source_id) inner join 
    gaiadr3.binary_masses
    using (source_id)
where 
    m2_lower > 5 and
    nss_solution_type != 'SB2' and
    phot_g_mean_mag < 14 and
    rv_nb_transits > 10 and
    significance > 5
\end{lstlisting}

In addition, we required that sources lie within $\sim0.5$ magnitude of the \textit{Gaia} main sequence.  Sources with spectral energy distributions (SEDs) that are well-described by a single star and are located on the main sequence are strong candidates for hosting dark companions, as a luminous star that is more massive than the primary should be easily detectable \citep[e.g.,][]{ElBadryRix2022}.  %\citet{ElBadryRix2022} have discussed spectroscopic binaries that lie near the \textit{Gaia} main sequence whose UV-IR SEDs cannot be fit by a single source and shown that they are mass-transfer binaries.  As we show in \S \ref{s:sed}, \bh\ can be fit successfully by a single source, and there is no improvement in the fit when using two sources.  
Finally, we excluded sources that show variability in their TESS light curve that would indicate that they are eclipsing binaries.  These cuts left us with a set of sources on which we obtained follow-up RV measurements (to be discussed elsewhere). 

\begin{figure}[h]        
\begin{center}
\hspace*{-0.05in}\includegraphics[scale=0.6]{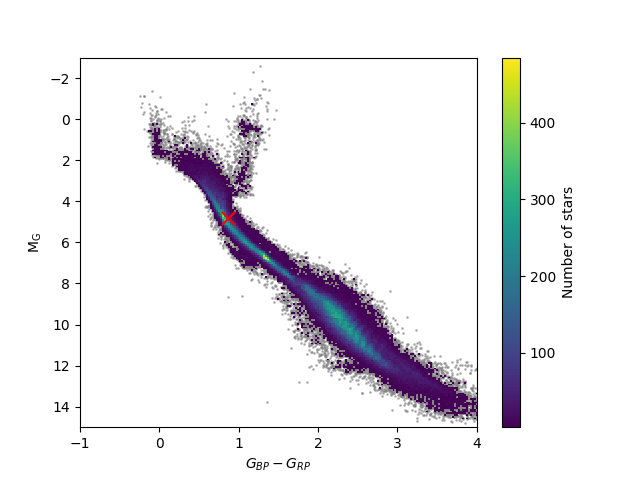}
\caption{\textit{Gaia} color-magnitude diagram.  The extinction-corrected absolute magnitude and color of \bh, based on the \textit{Gaia} parallax and G-band extinction, are marked with a red cross; the Gaia main-sequence is apparent.}
\label{f:cmd}
\end{center}
\end{figure}

In the remainder of this paper, we describe our results and analysis of one star, \bh, for which an astrometric orbit was reported in DR3.  We display the position of this object in the \textit{Gaia} color-magnitude diagram in Figure~\ref{f:cmd}. In the DR3 binary catalog, \bh\ has a primary\footnote{For this paper, we adopt the \textit{Gaia} convention of referring to the luminous star as the primary and the non-luminous companion as the secondary even though the latter may be more massive.} mass of 0.95~\msun, a secondary mass of $12.8^{+2.8}_{-2.3}$~\msun, a period of $185.8 \pm 0.3$~d, and an eccentricity of $0.49 \pm 0.07$.  However, the only radial velocity information provided by \textit{Gaia} is a single measurement of $v = 23.0 \pm 2.6$~\kms.  No spectroscopic orbital solution is included in DR3, indicating that a spectroscopic orbit could not be determined from the 13 independent radial velocity measurements obtained by \textit{Gaia}.  Because the astrometric orbit is at the extreme edge of the distribution of companion masses in the DR3 catalog, the lack of spectroscopic confirmation raises the concern that the astrometric solution, and the parameters derived from it, could be erroneous \citep{Halbwachs2022,GaiaDR3}.  We focus in this paper on this source because it is the most compelling black hole candidate (meeting our criteria above) for which we have obtained RV data thus far.  Spectroscopic observations for a limited number of other possible candidates are ongoing; analysis of those objects is beyond the scope of this paper. %Our analysis of \bh\ shows that it is located near or on the \textit{Gaia} main-sequence, with an SED that is successfully described by single stellar atmosphere models, with a spectroscopic orbit that is consistent with the \textit{Gaia} astrometric orbit, confirming that the star is on a long-period orbit around a massive dark companion. 

We note that, in addition to the comprehensive analysis by \citet{ElBadry2022Disc}, two other recent studies of the DR3 data set have also discussed this star as a candidate black hole binary.  \citet{Andrews2022} selected 24 candidate compact object binaries from the DR3 astrometric orbital solutions, and obtained follow-up spectroscopy of some of them, including \bh.  However, they discarded it because the orbital period is close to an integer multiple of the \textit{Gaia} scanning law, suggesting that the derived orbit could be an artifact.  \citet{Shahaf2022} used the astrometric mass ratio function to identify 177 candidate binaries containing compact objects.  In their sample, \bh\ has the highest secondary mass of the Class III sources, for which a compact object is the only viable type of companion, but no follow-up observations were obtained.

\subsection{Photometry}
\begin{figure}[h]
\begin{center}
\includegraphics[scale=0.3]{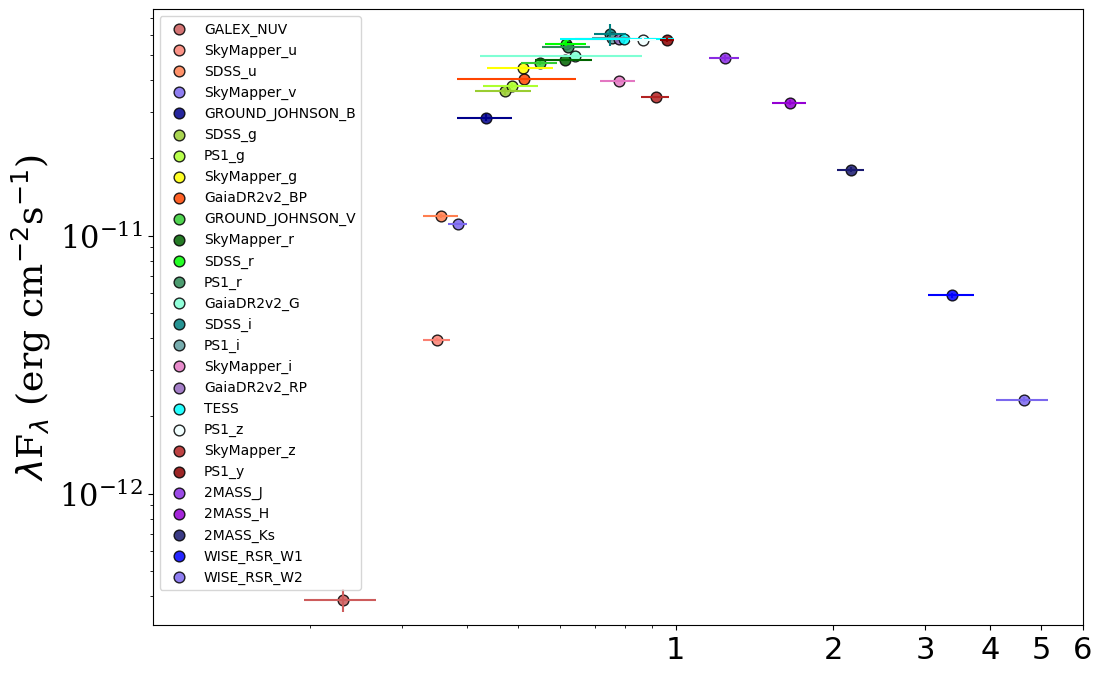}
\caption{Available photometry of \bh\ from the near-UV to the mid-infrared, with the data sources given in the legend.}
\label{f:allphotometry}
\end{center}
\end{figure}

We retrieved UV-IR photometry for this source from GALEX \citep{Martin2005}, SDSS \citep{Fukugita1996}, the APASS survey \citep{Henden2016}, \textit{Gaia} DR3 \citep{Riello2021}, PanSTARRS \citep{Tonry2012}, SkyMapper \citep{Wolf2018}, 2MASS \citep{Skrutskie2006}, and WISE \citep{Wright2010}.  Figure \ref{f:allphotometry} displays the available photometry for this source and Table~\ref{photom_table} lists the measurements and associated errors in each band. We analyze the spectral energy distribution of this source below and discuss our choice of photometry.

\input{4373_photom_table}

\subsection{Spectroscopy}

\subsubsection{Observations and data reduction}
We obtained follow-up high-resolution spectroscopy of \bh\ in order to determine its stellar parameters, search for signatures of a second luminous component in the binary, and measure the radial velocity as a function of time.  We observed the star for five epochs with the Automated Planet Finder (APF) telescope \citep{Vogt2014} at Lick Observatory, four epochs with the DEIMOS spectrograph \citep{Faber03} on the Keck~II telescope, and obtained a high signal-to-noise ratio spectrum with the MIKE spectrograph \citep{Bernstein2003} on the Magellan/Clay telescope (as well as a second, shorter MIKE spectrum) at Las Campanas Observatory.

The APF observations employed the Levy Spectrometer \citep{Radovan2010} with a $2\arcsec \times 8\arcsec$ slit, providing $R=80000$ spectra covering 3730--10200~\AA.  We observed \bh\ on 2022 June 28, 2022 August 28, 2022 September 2, 2022 September 12, and 2022 October 4, obtaining three 1000~s exposures per epoch.  The APF spectra were reduced with the California Planet Search pipeline \citep{Butler96,Howard10,Fulton15}.

\begin{figure*}       
\includegraphics[scale=0.47]{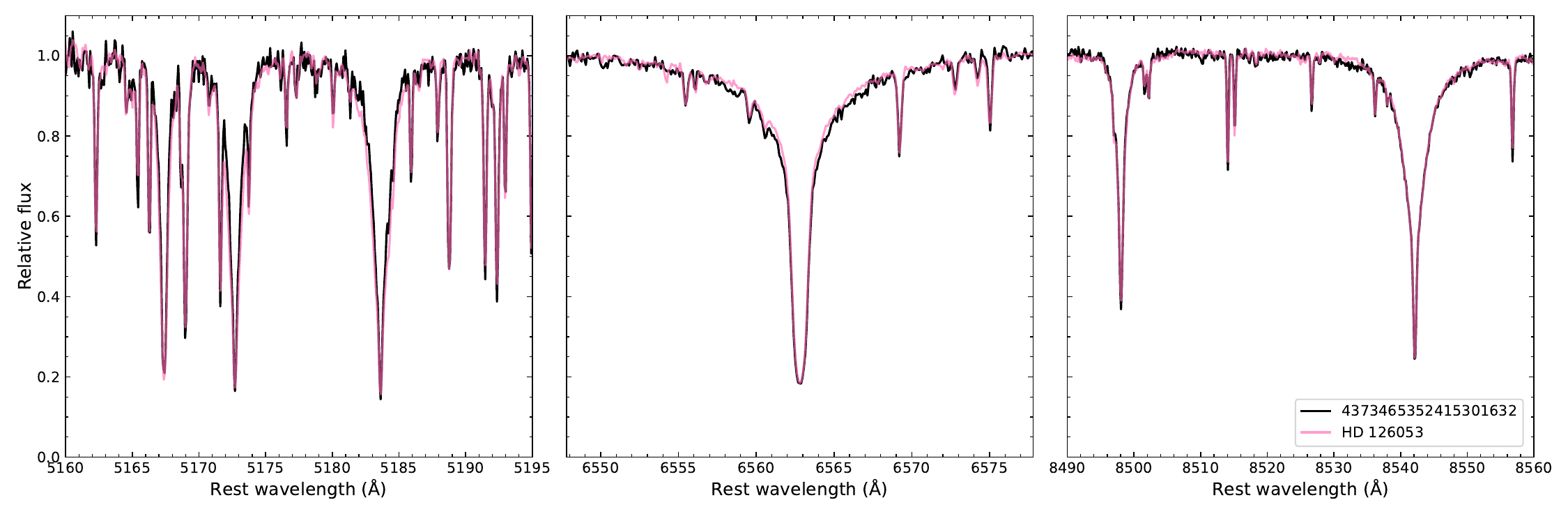}
%\plotone{4373_spectrum_cutouts.pdf}
\caption{Selected portions of the Magellan/MIKE spectrum of \bh\ (\textbf{from left to right:} the Mg triplet lines, H$\alpha$, and the Ca triplet).  The spectrum of HD~126053, which has a spectral type of G1.5V \citep{Gray03}, is overplotted in pink as a comparison. The close resemblance between the two stars in terms of line strengths and profiles is evident.  The stellar parameters of HD~126053 are $T_{\textrm{eff}} = 5700$~K, $\log{g} = 4.5$, $\textrm{[Fe/H]} = -0.35$ \citep[e.g.,][]{Brewer16,Casali20}.  The small deviations in the wings of some strong lines are likely caused by a combination of continuum normalization uncertainties. and the $\sim200$~K difference in temperature between the two stars.
\label{f:spectrum}}
\end{figure*}

\begin{figure*}       
\includegraphics[scale=0.47]{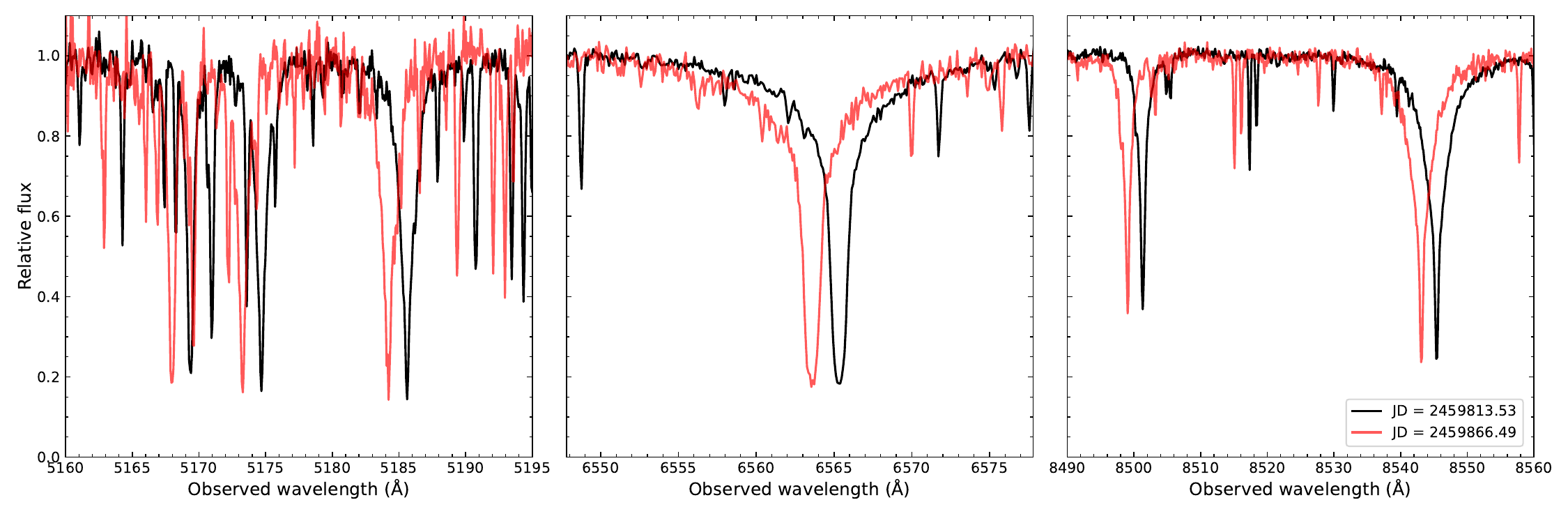}
%\plotone{4373_spectrum_cutouts.pdf}
\caption{Selected portions of the Magellan/MIKE spectrum of \bh\ (the Mg triplet lines, H$\alpha$, and the Ca triplet), shown here in the observed frame at the radial velocity extremes spanned by our data set.
\label{f:spectrum_refereerequest}}
\end{figure*}

We observed \bh\ with Magellan/MIKE for 2700~s on 2022 August 22, using a 0\farcs7 slit to obtain an $R\approx30000$ spectrum from 3300--9400~\AA.  We reduced the MIKE spectrum with the Carnegie Python data reduction package \citep{Kelson2003}.  To provide a comparison spectrum, we also obtained MIKE observations of the G1.5V radial velocity standard star HD~126053.  Portions of the spectra of these two stars are displayed in Fig.~\ref{f:spectrum}.  
 We also observed \bh\ for 745~s with MIKE during twilight on 2022 October 13 to obtain an additional velocity measurement.  Fig. \ref{f:spectrum_refereerequest} compares the two MIKE spectra of \bh, which represent the radial velocity extremes in our data set. 

We observed \bh\ with DEIMOS on 2022 September 19, 2022 September 24, 2022 September 27, and 2022 October 28 using the 1200G grating and a 0\farcs7 long slit to obtain $R\approx7000$ spectra from 6500--9000~\AA.  A single 300~s exposure was obtained in twilight on each night.  We reduced the data with a modified version of the DEEP2 reduction pipeline \citep{Cooper12,Newman13, Kirby15}.  The long-slit mask employed for these observations did not contain enough slits to adequately constrain the quadratic correction to the sky line wavelengths as a function of position on the mask described by \citet{Kirby15}.  Instead, we examined the night sky emission line wavelengths in the extracted spectrum, and used the measured wavelength shifts near the A-band and Ca triplet lines (which ranged from 0--4~\kms) to correct the wavelength scale.

\subsubsection{Velocity measurements}

We measured the velocity of \bh\ from the APF and MIKE spectra by performing $\chi^{2}$ fits to high S/N spectra of radial velocity standard stars shifted to the rest frame.  The velocity template for the MIKE spectra was HD~126053, for which we assumed a velocity of $v_{hel} = -19.21$~\kms\ from \textit{Gaia}~DR3, which is in excellent agreement with the measurement of $v_{hel} = -19.45$~\kms\ from \citet{Stefanik1999}.  For the APF velocity measurements, we used an APF observation of the G2V standard star HD~12846 \citep{Gray03,Soubiran2018}.  We assumed the \textit{Gaia}~DR3 velocity of $v_{hel} = -4.65$~\kms, which is within 0.2~\kms\ of the pre-\textit{Gaia} measurement of \citeauthor{Soubiran2018}.
The fit procedures were based on those described by \citet{Simon2017} and subsequent papers, adapted for echelle spectroscopy.  We fit each order of the target spectra independently with the matching order of the template spectrum, discarding orders or portions of orders affected by telluric or interstellar absorption.  For the MIKE observation we used only the data from the red spectrograph (covering $\sim$4900--8900~\AA) for velocities because the stellar continuum was more difficult to define at blue wavelengths and the 23 clean red orders already provided plenty of signal.  The S/N of the APF data was much lower, but the velocity could still be determined accurately over $\sim30$ spectral orders from $\sim$4500--6800~\AA.  For each observation, we averaged the velocities from all measured orders.  The resulting statistical uncertainty was $\sim0.1$~\kms\ for both the MIKE spectrum and the APF spectra.  However, we imposed a systematic error floor of 1.0~\kms\ for MIKE based on the results of \citet{Ji2020}.  Given the potential for velocity offsets between different instruments, we assumed the same minimum uncertainty of 1.0~\kms\ for the APF observations.

For the DEIMOS spectrum, we used the set of radial velocity templates described by \citet{Kirby15}, fitting only to the Ca triplet wavelength range.  We found that the metal-poor K1 dwarf HD~103095 produced the best match to the \bh\ spectrum, so we used that template to determine the velocity.  The velocity of the HD~103095 template spectrum was tied to that of the metal-poor giant HD~122563, for which we assumed a velocity of $-26.51$~\kms\ \citep{Chubak12}, which agrees with the \textit{Gaia} measurements within 0.4~\kms.  We corrected for slit centering errors by carrying out a separate fit to the telluric A-band absorption of the rapidly rotating star HR~7346 \citep{Sohn07,SG07}.  Based on the scatter of sky lines around their true wavelengths, we assume a velocity uncertainty of 2.0~\kms\ for each DEIMOS measurement \citep[cf.][]{SG07,Kirby15}.

All of our velocity measurements, along with the two archival data points from LAMOST DR7 \citep{Cui12,Luo22}, are listed in Table~\ref{vel_table}.

\input{4373_vel_table}

\section{Results \& Analysis}
\label{s:results}

\subsection{The Spectral Energy Distribution}\label{s:sed}

All of the archival photometry of \bh\ that we were able to locate, including measurements from GALEX \citep{Morrissey07}, SDSS DR16 \cite{Ahumada20}, APASS DR10 \citep{Henden19}, Pan-STARRS DR2 \citep{Flewelling20}, SkyMapper DR2 \citep{Onken19}, 2MASS \citep{Cutri03}, and WISE \citep{Cutri14}, is displayed in Figure \ref{f:allphotometry} and listed in Table~\ref{photom_table}.
%The spectral energy distribution (SED) of \bh\ with all available photometry is shown in Figure \ref{f:allphotometry}. 
As is clear, the SkyMapper $v, i, z$ bands are discrepant with other photometric measurements at comparable wavelengths, and we therefore exclude them from our SED fits.  Using the publicly available SED fitting code ARIADNE \citep{Vines_Jenkins2022}, we fit the observed photometry (excluding SkyMapper as well as the wide \textit{Gaia} bands) to derive stellar parameters.  Although \textit{Gaia} offers extremely precise optical photometry of 
\bh, we note that there may be contamination in the $G_{BP}$ and $G_{RP}$ bands as a result of the star's location in the direction of the Galactic bulge.  The phot\_bp\_rp\_excess\_factor for this source is 1.24 and the $G_{BP}-G_{RP}$ color is 1.17, which indicates that the blending probability is $\sim30$\%\ \citep{Riello2021}.  

\begin{figure}[h]        
\begin{center}
\includegraphics[scale=0.24]{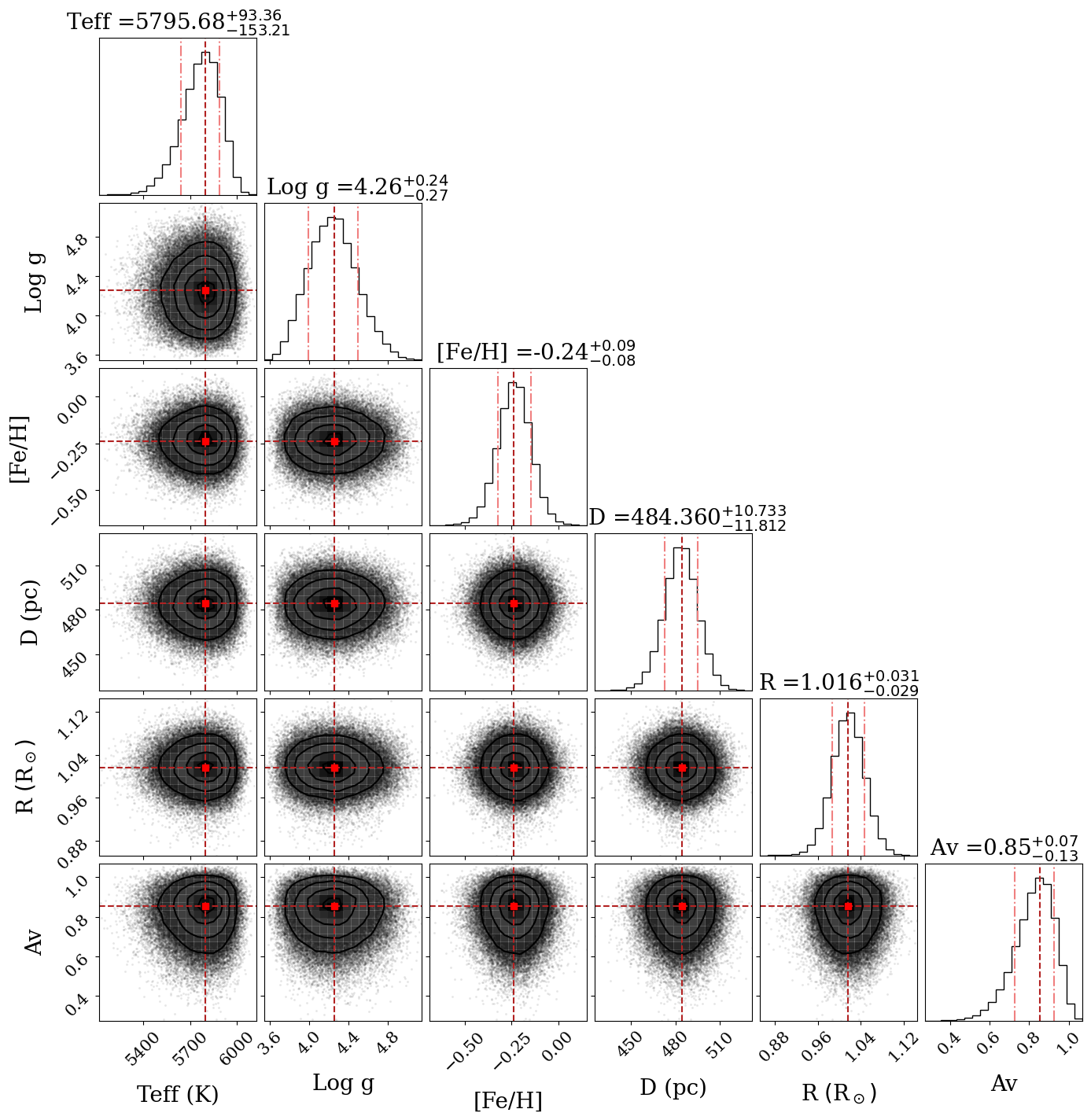}
\includegraphics[scale=0.31]{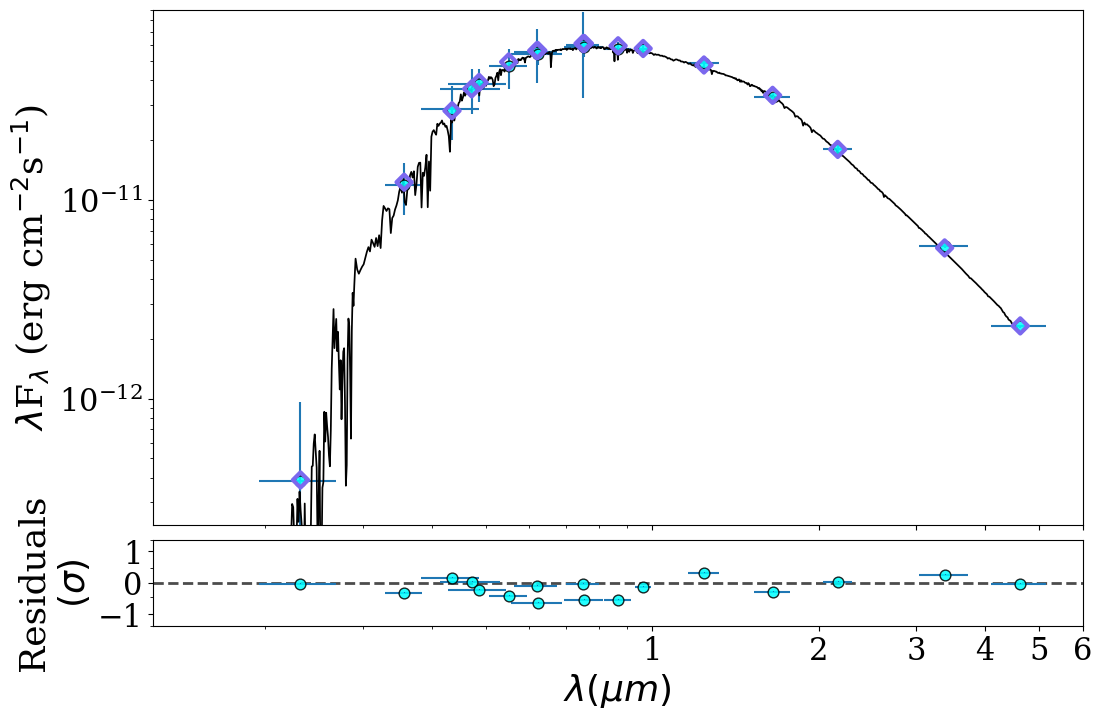}
\caption{(top) Posterior distributions of stellar parameters from ARIADNE using all available photometry except SkyMapper and \textit{Gaia}, with the distance prior set from the measured Gaia parallax. (bottom) Resultant best-fit SED.  There is no evidence for an excess at short or long wavelengths that could indicate a second source contributing to the observed fluxes.  The same points are shown in the residual plot below in circles (in the SED plot, they are overplotted with diamonds).}
\label{f:noskynoGaiaparallaxprior}
\end{center}
\end{figure}

The resultant posterior distributions are shown in the corner plot for the stellar parameters, along with the best-fit extinction in Figure \ref{f:noskynoGaiaparallaxprior}.  ARIADNE fits broadband photometry to a variety of stellar atmosphere models.  In this case, the best fit is obtained with the Phoenix models \citep{phoenix2013}.  Here, we have used a prior on the distance that is consistent with the measured \textit{Gaia} parallax for this source.  Including the \textit{Gaia} $G$-band magnitude in the fit does not significantly change any of the derived parameters.  We obtain similar values for the effective temperature and other derived parameters fitting only to GALEX photometry, along with $B$ and $V$ bands from APASS, SDSS $g,r,i,z$, 2MASS $J, H, K_{s}$, and WISE photometry.  Similar posterior distributions are also obtained if Pan-STARRS photometry is used instead of SDSS.  The best-fit ARIADNE parameters are:  $\rm T_{eff} = 5796^{+93}_{-153}$~K, $\log{g} = 4.26^{+0.24}_{-0.27}$, $R = 1.02^{+0.031}_{-0.029}$~\rsun, and $\textrm{[Fe/H]} = -0.24^{+0.09}_{-0.08}$.  As the corner plot shows, the posterior distributions are tightly constrained.  The SED fit, along with the residuals, is shown in the bottom panel of Figure \ref{f:noskynoGaiaparallaxprior}.

\begin{figure}[h]        
\begin{center}
\includegraphics[scale=0.57]{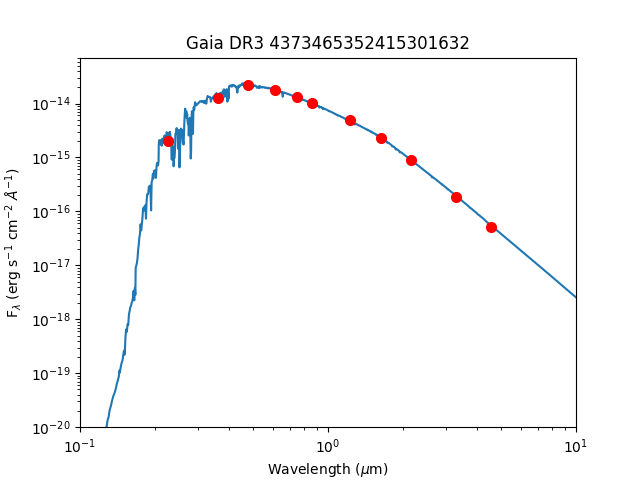}
\includegraphics[scale=0.57]{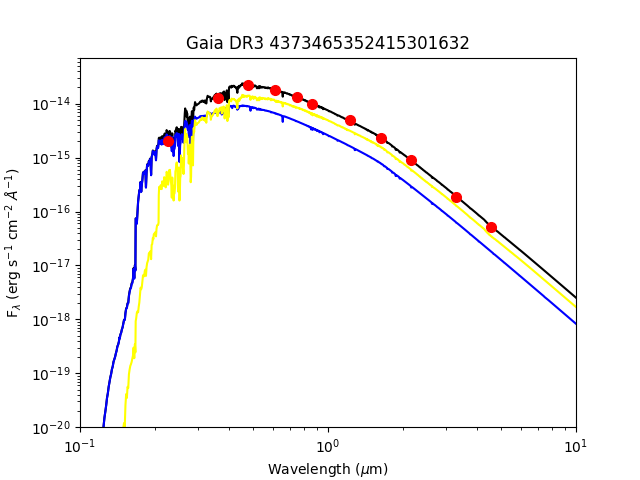}
\caption{(a) Single SED fit using the BaSeL model and the \citet{Fitzpatrick99} extinction law. (b) Composite SED fit, with the cooler component shown in yellow, hotter component shown in blue, and the composite spectrum shown in black.}
\label{f:sed}
\end{center}
\end{figure}

We have also used the  BaSeL model spectral library \citep{Lejeune1997,Lejeune1998} to fit the observed photometry with both a combined $\chi^{2}$ and genetic algorithm optimization and a Markov Chain Monte Carlo (MCMC) approach.  Here we assume a \citet{Fitzpatrick99} extinction law with $R_{V} = 3.1$ and reddening $E(B-V) = 0.3$~mag.  The best-fit SEDs are shown in Figure~\ref{f:sed}a for a single source, and the fit assuming two sources is shown in Figure \ref{f:sed}b.  In the single source case, we obtain a reduced $\chi^{2}$ of 3.7, best-fit parameters $T_{\rm eff} = 5927~\rm K$, $\log{g} = 4.0$,
$L= 1.08$~\lsun\ and metallicity of $Z = 0.009$.  For a two source fit, we obtain a reduced $\chi^{2}$ of 5.5, with $T_{\rm eff} = 6241$~K, $\log{g} = 4.99$, $L= 0.42$~\lsun, and $Z = 0.0006$ for the hotter component and $T_{\rm eff} = 5763$~K, $\log{g} = 4.01$,
$L = 0.67$~\lsun, and $Z = 0.01$ for the cooler component. Although the raw $\chi^{2}$ value does improve upon adding a second component (17 for the composite spectrum vs. 25 for the single SED fit), this is not enough to compensate for the increase in the number of model parameters in the reduced $\chi^{2}$.  We conclude that there is no significant evidence in the SED for the presence of more than one luminous component.
%it does not enable a substantial improvement in the SED fit, as is visually clear.

\subsection{Spectroscopic Analysis and Stellar Parameter Estimation}

\subsubsection{Full spectral fitting}

Visual examination of the MIKE spectrum of \bh\ shows that it closely resembles an early G star (see Fig.~\ref{f:spectrum}).  No sign of a secondary component is visible in either the metallic or hydrogen lines.  Given the spectral resolution, a companion star that makes a non-negligible contribution to the flux should be detectable unless its velocity is within $\sim10$~\kms\ of that of the primary.  As we will show below, the velocity of the primary at the time of observation was more than 40~\kms\ away from the center-of-mass velocity of the system (also see Fig.~\ref{f:spectrum_refereerequest}), which indicates that the secondary spectrum should be resolvable from the primary at this epoch if it is a luminous main-sequence star.

To provide an initial estimate of the stellar parameters from its spectrum rather than the broadband photometry, we fit the MIKE spectrum with the empirical spectral fitting code \texttt{SpecMatch-Emp} \citep{Yee17}.  Using the spectral range 5100--5700~\AA, the best-fit parameters determined from a linear combination of stars in the template library are: $T_{\rm eff} = 5917 \pm 110$~K, $R = 1.20 \pm 0.18$~\rsun, $\log{g} = 4.24 \pm 0.14$,\footnote{To determine $\log{g}$, we make use of the result from the detailed spectroscopic analysis below that the mass of the star is 0.91~\msun.} and $\textrm{[Fe/H]} = -0.22 \pm 0.09$, with HD~44985 as the single closest match to \bh.  These parameters are in excellent agreement with those from the SED fit.

\subsubsection{Fundamental and photospheric stellar parameters} 
\label{sec:stellarparams}

Next, we determined the stellar parameters of \bh\ more rigorously using a combination of the classical spectroscopic approach\footnote{Based on simultaneously minimizing the difference between the abundances determined from \ion{Fe}{1} and \ion{Fe}{2} lines, as well as their dependencies on transition excitation potential and reduced equivalent width.} and a comparison with theoretical isochrones to infer accurate, precise, and self-consistent parameters. 

The inputs to this parameter inference procedure include the equivalent widths of \ion{Fe}{1} and \ion{Fe}{2} atomic absorption lines, as well as multiwavelength photometry from \textit{Gaia} DR3 G-band \citep{Riello2021} and 2MASS (J, H, and K$_{\textrm{s}}$), the \textit{Gaia} DR3 parallax \citep{Lindegren21}, and we adopt the V-band extinction from the three-dimensional Stilism reddening map \citep{lallement2014,capitanio2017,lallement2018}.  The absorption line data are based on the line lists presented in \cite{melendez2014} and \citet{reggiani2018}. We measured the equivalent widths by fitting Gaussian profiles with the Spectroscopy Made Harder (\texttt{smhr}) software package to the continuum-normalized MIKE spectrum.  The continuum normalization was performed using \texttt{smhr}.  We assume solar abundances from \citet{asplund2021} and follow the steps described in \citet{reggiani2022a,reggiani2022b} to obtain the fundamental and photospheric stellar parameters from a combination of spectral information and a fit to the MESA Isochrones and Stellar Tracks \cite[MIST;][]{dotter2016,choi2016,paxton2011,paxton2019} stellar models. We do the fitting with the \texttt{isochrones} package\footnote{\url{https://github.com/timothydmorton/isochrones}} \citep{morton2015}, which uses \texttt{MultiNest}\footnote{\url{https://ccpforge.cse.rl.ac.uk/gf/project/multinest/}} \citep{feroz2008,feroz2009,feroz2019} via \texttt{PyMultinest} \citep{buchner2014}.

\begin{figure}
\epsscale{1.2}
\plotone{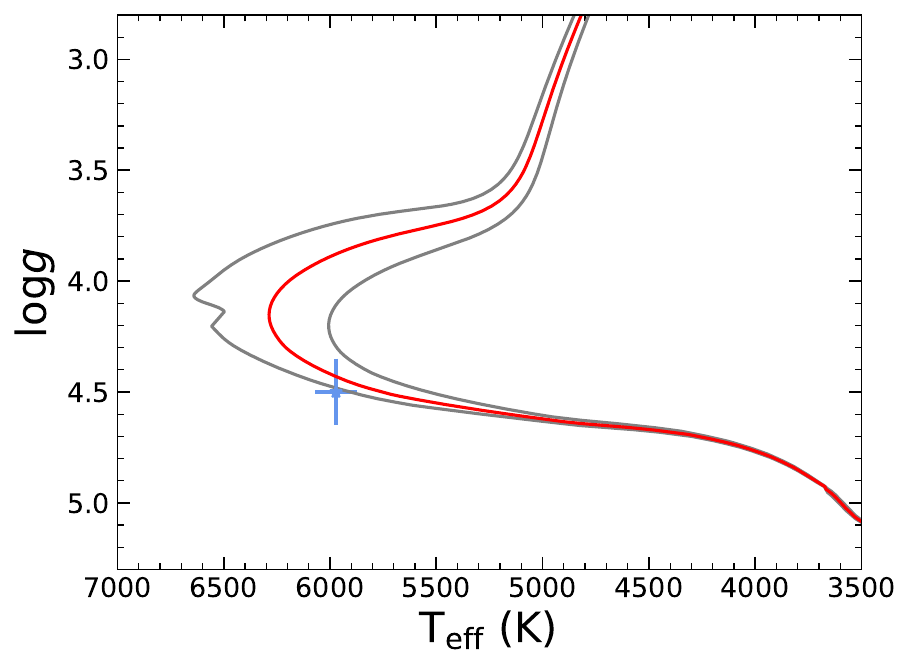}
\caption{Effective temperature and surface gravity of \bh\ compared to MIST isochrones.  The blue point shows the derived position of the star and the red curve is the 0.91~\msun\ isochrone that provides the best fit to the combined photometric and spectroscopic data.  The gray curves represent 0.89~\msun\ and 0.93~\msun\ isochrones for comparison.}
\label{f:isochrone}
\end{figure}

Our derived fundamental and photospheric parameters for \bh\ are listed in Table \ref{stellar_params_table}.  Its position relative to the MIST isochrones is illustrated in Figure~\ref{f:isochrone}.  The effective temperature is in excellent agreement with the values determined from the single-source SED fit as well as the full spectral fitting.  According to the effective temperature calibration given by \citet{Gray09}, the spectral classification would be G0.
%As a further check to the effective temperature of the star, we also used the \texttt{colte}\footnote{\url{https://github.com/casaluca/colte}} code \citep{casagrande2021} to estimate the temperature via the infrared flux method (IRFM). The IRFM effective temperature is $\mathrm{T_{eff}}=5817\pm70$~K, fully consistent with both the spectroscopic analysis and the SED fits in Section~\ref{s:sed}.  
The surface gravity required to enforce ionization balance on the Fe lines is somewhat higher than determined via other techniques, but the disagreement is only at the $1.6\sigma$ level.  The spectroscopic metallicity of \bh\ is also modestly lower than the photometric value.  The overall good agreement on the stellar parameters achieved through multiple independent methods indicates that systematic uncertainties, e.g., related to the choice of stellar evolution tracks, are smaller than the adopted uncertainties (Table~\ref{stellar_params_table}).  We conclude that \bh\ is an old, moderately metal-poor main sequence star, slightly less massive than the Sun. 

\begin{deluxetable}{lcc}
\tablecaption{Derived Stellar Parameters\label{stellar_params_table}}
\tablewidth{0pt}
\tablehead{
\colhead{Property} & \colhead{Value} & \colhead{Unit}}
\startdata
Effective temperature ($T_{\text{eff}}$) & $5972\pm100$ & K \\
Surface gravity ($\log{g}$) & $4.54 \pm 0.15$ & $\log{\textrm{cm s}^{-2}}$ \\
Metallicity ($[\text{Fe/H}]$) &  $-0.30\pm0.10$ & \\
Microturbulent velocity $(\xi$) & $1.10 \pm 0.10$ & km s$^{-1}$ \\
Stellar mass ($M_{\ast}$) & $0.91 \pm 0.10$ & $M_{\odot}$ \\
Stellar radius ($R_{\ast}$) & $1.003\pm0.075$ & $R_{\odot}$ \\
Luminosity ($L_{\ast}$) & $1.186 \pm 0.232$ & $L_{\odot}$ \\
Isochrone-based age ($\tau_{\text{iso}}$) & $7.1\pm3.8$ & Gyr 
\enddata
\tablecomments{The parameters listed here are determined via the combined analysis of the high-resolution stellar spectrum along with theoretical isochrones, as described in Section~\ref{sec:stellarparams}.}
\end{deluxetable}

\subsubsection{Chemical abundances}
We used the photospheric stellar parameters from Table \ref{stellar_params_table} to calculate the abundances of \ion{C}{1}, \ion{O}{1}, \ion{Na}{1}, \ion{Mg}{1}, \ion{Al}{1}, \ion{Si}{1}, \ion{S}{1}, \ion{K}{1}, \ion{Ca}{1}, \ion{Sc}{2}, \ion{Ti}{1}, \ion{Ti}{2}, \ion{V}{1}, \ion{Cr}{1}, \ion{Cr}{2}, \ion{Mn}{1}, \ion{Fe}{1}, \ion{Fe}{2}, \ion{Co}{1}, \ion{Ni}{1}, \ion{Cu}{1}, \ion{Zn}{1}, \ion{Sr}{2}, \ion{Y}{2},  \ion{Zr}{2}, \ion{Ba}{2}, and \ion{Ce}{2}, from the equivalent widths (EWs) of spectroscopic absorption lines.  We measured equivalent widths from the continuum-normalized MIKE spectrum by fitting Gaussian profiles with \texttt{smhr}. We use the 1D plane-parallel, $\alpha$-enhanced, ATLAS9 model atmospheres \citep{Castelli03} and the 2019 version of \texttt{MOOG} \citep{sneden1973} to infer elemental abundances based on each equivalent width measurement, initially assuming local thermodynamic equilibrium (LTE). Our calculations include hyperfine structure splitting for \ion{Sc}{2} (based on the Kurucz line list\footnote{http://kurucz.harvard.edu/linelists.html}), \ion{V}{1} \citep{prochaska2000b}, \ion{Mn}{1} \citep{prochaska2000a,Blackwell-Whitehead2005}, \ion{Cu}{1} \citep[][and Kurucz line lists]{prochaska2000b}, and \ion{Y}{2} (Kurucz linelists),  and we account for isotopic splitting for \ion{Ba}{2} \citep{mcwilliam1998}.  We report the adopted atomic data, equivalent width measurements, and individual line-based abundances in Table~\ref{measured_ews}.  Our final abundances are reported in Table \ref{chem_abundances}.

\begin{deluxetable*}{lccccc}
\tablecaption{Atomic data, Equivalent Widths and line Abundances. Full version online.\label{measured_ews}}
\tablewidth{0pt}
\tablehead{
\colhead{Wavelength} & \colhead{Species} &
\colhead{Excitation Potential} & \colhead{log($gf$)} &
\colhead{EW} & \colhead{$\log_\epsilon(\rm{X})$} \\ 
\colhead{(\AA)} &  & \colhead{(eV)} & & (m\AA) & }
\startdata
$6154.225$ & \ion{Na}{1} & $2.102$ & $-1.547$ & $20.80$ & $5.993$\\ 
$6160.747$ & \ion{Na}{1} & $2.104$ & $-1.246$ & $31.52$ & $5.927$\\ 
$4571.095$ & \ion{Mg}{1} & $0.000$ & $-5.623$ & $89.51$ & $7.395$\\ 
$4730.040$ & \ion{Mg}{1} & $4.340$ & $-2.389$ & $43.79$ & $7.381$\\ 
$5711.088$ & \ion{Mg}{1} & $4.345$ & $-1.729$ & $80.01$ & $7.244$\\ 
$6318.717$ & \ion{Mg}{1} & $5.108$ & $-1.945$ & $26.75$ & $7.284$\\ 
$5260.387$ & \ion{Ca}{1} & $2.521$ & $-1.719$ & $23.01$ & $6.139$\\ 
$5512.980$ & \ion{Ca}{1} & $2.933$ & $-0.464$ & $63.20$ & $5.966$
\enddata
\tablecomments{This table will be published in its entirety as a machine-readable table.  A portion is shown here for guidance regarding its form and content.} 
\end{deluxetable*}

When possible, we update our elemental abundances derived under the assumptions of LTE to account for departures from LTE (i.e., non-LTE corrections) by interpolating published grids of non-LTE corrections for several elements.  We make use of 3D non-LTE corrections for carbon and oxygen  \citep{amarsi2019}, and 1D non-LTE corrections for sodium \citep{Lind2011}, aluminum \citep{amarsi2020}, silicon \citep{amarsi2017}, potassium \citep{reggiani2019}, calcium \citep{amarsi2020}, iron \citep{amarsi2016}, and barium \citep{amarsi2020}. The NLTE-corrected abundances are listed in Table~\ref{chem_abundances}. 

\begin{deluxetable}{lcccc}
\tablecaption{Chemical Abundances\label{chem_abundances}}
\tablecolumns{21}
%\tabletypesize{\tiny}
\tablewidth{0pt}
\tablehead{
\colhead{Species} & 
\colhead{N} & \colhead{log($\epsilon_X$)} &
\colhead{[X/Fe]} & \colhead{$\sigma_{[X/Fe]}$}}
\startdata
\ion{C}{1} & $4$ & $8.01$ & $-0.15$ & $0.09$  \\ 
\ion{O}{1} & $3$ & $8.65$ & \phs$0.26$ & $0.03$  \\ 
\ion{Na}{1} & $2$ & $5.96$ & \phs$0.04$ & $0.03$  \\ 
\ion{Mg}{1} & $4$ & $7.33$ & \phs$0.08$ & $0.04$  \\ 
\ion{Al}{1} & $5$ & $6.08$ & $-0.05$ & $0.02$  \\ 
\ion{Si}{1} & $11$ & $7.27$ & \phs$0.06$ & $0.03$  \\ 
\ion{S}{1} & $4$ & $6.85$ & \phs$0.03$ & $0.01$  \\ 
\ion{K}{1} & $1$ & $5.18$ & \phs$0.42$ & $0.01$  \\ 
\ion{Ca}{1} & $10$ & $6.06$ & \phs$0.06$ & $0.02$  \\ 
\ion{Sc}{2} & $4$ & $2.94$ & \phs$0.10$ & $0.09$  \\ 
\ion{Ti}{1} & $9$ & $4.60$ & $-0.07$ & $0.03$  \\ 
\ion{Ti}{2} & $10$ & $4.80$ & \phs$0.13$ & $0.04$  \\ 
\ion{V}{1} & $3$ & $3.51$ & $-0.09$ & $0.02$  \\ 
\ion{Cr}{1} & $8$ & $5.26$ & $-0.06$ & $0.02$  \\ 
\ion{Cr}{2} & $5$ & $5.24$ & $-0.07$ & $0.03$  \\ 
\ion{Mn}{1} & $6$ & $4.88$ & $-0.23$ & $0.05$  \\ 
\ion{Fe}{1} & $79$ & $7.16$ & $\cdots$ & $\cdots$   \\ 
\ion{Fe}{2} & $18$ & $7.15$ & $\cdots$ & $\cdots$  \\ 
\ion{Ni}{1} & $14$ & $5.94$ & \phs$0.04$ & $0.03$  \\ 
\ion{Cu}{1} & $2$ & $3.79$ & $-0.09$ & $0.02$  \\ 
\ion{Zn}{1} & $2$ & $4.25$ & $-0.01$ & $0.07$  \\ 
\ion{Sr}{1} & $1$ & $2.41$ & $-0.12$ & $0.00$  \\ 
\ion{Y}{2} & $3$ & $1.82$ & $-0.08$ & $0.05$  \\ 
\ion{Zr}{2} & $1$ & $2.29$ & $-0.00$ & $0.03$  \\ 
\ion{Ba}{2} & $3$ & $2.12$ & \phs$0.15$ & $0.05$  \\ 
\ion{Ce}{2} & $1$ & $1.55$ & \phs$0.27$ & $0.03$  \\ 
\hline
\multicolumn{4}{l}{\textbf{non-LTE Corrected Abundances}}\\
\ion{C}{1} & $4$ & $8.07$ & $-0.04$ & $0.16$  \\ 
\ion{O}{1} & $3$ & $8.57$ & \phs$0.24$ & $0.03$  \\ 
\ion{Na}{1} & $2$ & $5.85$ & $-0.01$ & $0.04$  \\ 
\ion{Al}{1} & $2$ & $6.07$ & \phs$0.00$ & $0.01$  \\
\ion{Si}{1} & $11$ & $7.25$ & \phs$0.10$ & $0.08$  \\ 
\ion{K}{1} & $1$ & $4.74$ & \phs$0.03$ & $\cdots$  \\ 
\ion{Fe}{1} & $79$ & $7.18$ & $\cdots$ & $\cdots$   \\ 
\ion{Fe}{2} & $18$ & $7.22$ & $\cdots$ & $\cdots$   \\ 
\ion{Ba}{2} & $3$ & $1.96$ & \phs$0.06$ & $0.03$
\enddata
\end{deluxetable}

The abundance analysis shows that \bh\ is a mildly $\alpha$-enhanced star ([$\alpha$/Fe]$\sim0.1$), as expected for its metallicity. The iron-peak and neutron-capture elemental abundances are also fully compatible with the typical values for this metallicity range, based on galactic chemical evolution models  \citep[e.g.,][]{kobayashi2020} and large spectroscopic surveys \citep[e.g.,][]{amarsi2020}.  We do not see any sign of the large enhancements in light elements and $\alpha$ elements that have been detected in some X-ray binary companions \citep[e.g.,][]{Orosz2001,Gonzalez-Hernandez2004,Suarez-Andres2015,Casares2017}.

\section{Orbit Fitting}
\label{s:orbits}

Based on the \textit{Gaia} astrometry and our radial velocity measurements, the binary orbit of \bh\ can be determined astrometrically, spectroscopically, or using a combination of both types of data.  In order to test the consistency of the velocities with the \textit{Gaia} astrometric orbit, we consider each of these cases in turn. We fit Keplerian orbits to the RV data using the adaptation of the {\tt orvara} code \citep{Brandt2021AJ} presented in \cite{Lipartito+Bailey+Brandt+etal_2021}.  

In this section, we briefly discuss the Gaia orbital constraints and then present our orbital fits using only RVs.  For joint astrometric+spectroscopic fits, we cannot use {\tt orvara} directly with \textit{Gaia} astrometry because the individual astrometric measurements are not given in DR3.  We have only a set of best-fit parameters and a covariance matrix.  As a result, we incorporate constraints from \textit{Gaia} by conditioning our RV fits on the \textit{Gaia} results.  (An alternate approximate approach would be to determine the times that targets cross the \textit{Gaia} focal plane and use that as an estimate of the times of observation.)  We discuss this approach and our results in Section \ref{s:joint}.    

\subsection{Astrometry Only}
\label{s:astrometry_only}

The \textit{Gaia}~DR3 non\_single\_stars catalog provides orbital parameters for \bh\ based on a fit to the individual astrometric measurements obtained by \textit{Gaia}.  However, the astrometric measurements themselves will not be available until the fourth \textit{Gaia} data release $\sim3$ years from the time of this writing.  The parameters listed directly in the catalog include the period $P$, the time of periastron $T_0$, the eccentricity $e$, the secondary mass $M_{2}$, and the Thiele-Innes (TI) elements $A$, $B$, $F$, and $G$.  Following the method of \citet{Binnendijk1960}, as presented by \citet{Halbwachs2022}, we use the TI elements to compute the semimajor axis $a$, the position angle of the ascending node $\Omega$, the longitude of periastron $\omega$, and the inclination $i$.  The values of these parameters are listed in Table~\ref{tab:gaia_astrom}.  We note that these calculations are based on the results reported in the two body orbit catalog and the binary mass catalog from \citet{GaiaDR3}.

\begin{figure}[h]        
\begin{center}
\includegraphics[width=\linewidth]{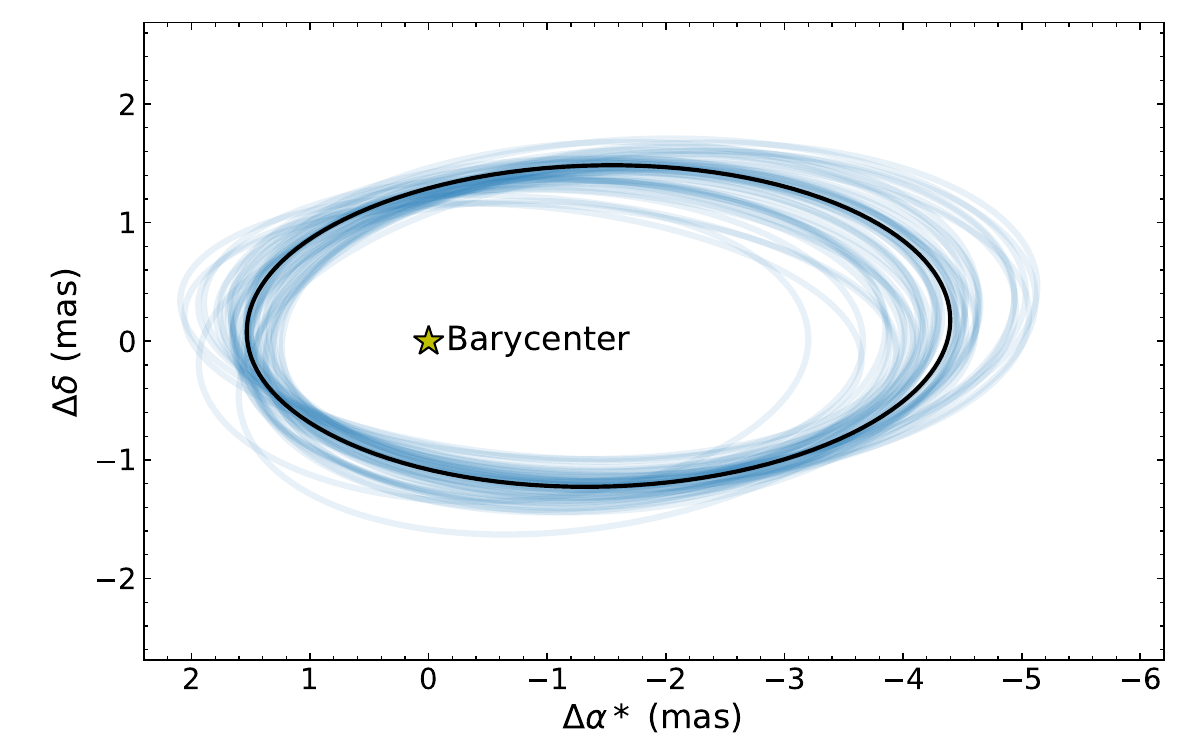}
%\includegraphics[scale=0.6]{astrometric_orbit_GaiaBH4373_Oct3.pdf}
%\hspace*{-0.61in}\includegraphics[scale=0.707]{astrometric_orbit_ourdata.png}
%\hspace*{0.10in}\includegraphics[scale=0.58]{RV_orbit_GaiaBH4373_observedframe.pdf}
\caption{Astrometric orbit determined from the \textit{Gaia} DR3 astrometric orbital solution as described in the text.   The best-fit orbit calculated directly from the \textit{Gaia} parameters is plotted as the thick black curve, and the thin blue curves represent orbits drawn randomly from the posterior distribution.} 
\label{f:astroonly}
\end{center}
\end{figure}

The TI elements, together with the eccentricity, describe an orbit in the plane of the sky.  We generate 50 random realizations of measurements drawn from the \textit{Gaia} best-fit values and covariance matrix.  We then plot the sky paths given by the eccentricities and TI elements.  Figure \ref{f:astroonly} shows these sky paths about the barycenter; the best-fit orbit reported by Gaia DR3 is shown as a thicker black line.
%From the orbital parameters described above, we compute the position angle $\theta$ and separation $\rho$ of the binary as:
The orbit has a semimajor axis of $2.98 \pm 0.22$~mas, with the major axis oriented nearly east--west.\footnote{We note that the perpendicular orientation shown by \cite{ElBadry2022Disc} is a result of an inconsistency in a plotting routine (K. El-Badry 2023, private communication).} 

\begin{deluxetable}{lr}
\tablecaption{Orbital Solutions for \bh for the dataset listed in Table 2} \label{tab:gaia_astrom}
\tablewidth{0pt}
\tablehead{
\colhead{Parameter} & \colhead{Value} }
\startdata
\hspace{0.75in} \bf{Astrometric fit}\\
$P$ & $185.77 \pm 0.31$~d \\ 
$T_{0}$ & $2457377.0 \pm 6.3$~d \\ 
$e$ & $0.49 \pm 0.07$ \\ 
$a$ & $2.98 \pm 0.22$~mas \\ 
$i$ & $121\fdg3 \pm 3.5$ \\ 
$\Omega$ & $89\fdg6 \pm 3.8$ \\ 
$\omega$ & $-10\fdg6 \pm 11.9$ \\
$M_{2}$ & $12.8^{+2.8}_{-2.3}$~\msun \\
\hline
\hspace{0.75in} \bf{RV fit}\\
$P$ & $184.28^{+0.75}_{-0.89}$~d \\ 
$T_{0}$ & $2459802.4 \pm 3.0$~d \\
$e$ & $0.411^{+0.034}_{-0.021}$ \\ 
$\omega$ & $19\fdg1^{+6.0}_{-6.6}$ \\
$K_{1}$ & $60.6^{+10.0}_{-5.8}$~\kms\ \\
$\gamma$ & $47.8^{+2.2}_{-1.4}$~\kms \\
$f$ & $3.2^{+1.6}_{-0.9}$~\msun \\ 
\hline
\hspace{0.75in} \bf{Joint fit}\\ 
$P$ & $185.41_{-0.087}^{+0.101}$ d\\ 
$T_{0}$ & ${2459757.9}_{-1.5}^{+1.2}$~d \\
$e$ & ${0.456} \pm .023$ \\
$a$ & ${1.357}_{-0.061}^{+0.069}$ AU \\ 
%%$a$ & ${3.064}_{-0.035}^{+0.034}$ mas \\ 
$i$ & ${123.6} \pm 1.6$ \\ 
$\Omega$ & ${96\fdg2}_{-2.0}^{+2.3}$ \\ 
$\omega$ & ${10\fdg4}_{-2.0}^{+2.7}$ \\ 
$M_{2}$ & ${11.33}_{-1.32}^{+1.57}$~\msun \\  
\enddata

%\tablecomments{} 
\end{deluxetable}

\begin{deluxetable}{lr}
\tablecaption{Orbital Solutions for \bh ~using all available data}
\label{tab:gaia_astrom_alldata}
\tablewidth{0pt}
\tablehead{
\colhead{Parameter} & \colhead{Value} }
\startdata
\hspace{0.75in} \bf{RV fit}\\
$P$ & $184.52_{-0.80}^{+0.66}$ d\\ 
$T_{0}$ & ${2459801.16}\pm{0.12}$~d \\ %%Double check, could have wrong reference time
$e$ & ${0.4368}_{-0.0037}^{+0.0036}$ \\ 
$\omega$ & ${16\fdg10} \pm 0.63$ \\
$K_{1}$ & $65.58^{+0.31}_{-0.30}$~\kms\  \\
$\gamma$ & $48.48^{+0.36}_{-0.36}$~\kms  \\
$f$ & $3.92^{+0.042}_{-0.042}$~\msun \\
\hline
\hspace{0.75in} \bf{Joint fit}\\
$P$ & $185.52_{-0.08}^{+0.08}$ d\\ 
$T_{0}$ & ${2459759.3} \pm 1.1$ ~d \\ 
$e$ & ${0.439}_{-0.003}^{+0.003}$ \\
$a$ & ${1.258}_{-0.010}^{+0.011}$ AU \\ 
$i$ & ${126\fdg80}_{-0.63}^{+0.62}$ \\ 
$\Omega$ & ${98\fdg71}_{-2.01}^{+1.97}$ \\
$\omega$ & ${15\fdg89} \pm 0.61$ \\ 
$M_{2}$ & ${9.326}_{-0.208}^{+0.216}$~\msun  \\
\enddata
%\tablecomments{} 
\end{deluxetable}

\subsection{Radial Velocity Data Only}

We fit the radial velocity data using the adaptation of {\tt orvara} \citep{Brandt2021AJ} described in \cite{Lipartito+Bailey+Brandt+etal_2021}.  This remains an MCMC-based fitting approach, but differs from the main version of {\tt orvara} in several ways: 
\begin{itemize}
\item it fits only for the RV parameters of period, eccentricity, RV semiamplitude, periastron time, argument of periastron, and RV zero point;
\item the argument of periastron refers to the star (as is the definition for \textit{Gaia} astrometry);
\item only the RV zero point (barycenter RV of the system) is analytically marginalized off.
\end{itemize}

\begin{figure}[h]        
\includegraphics[width=\linewidth]{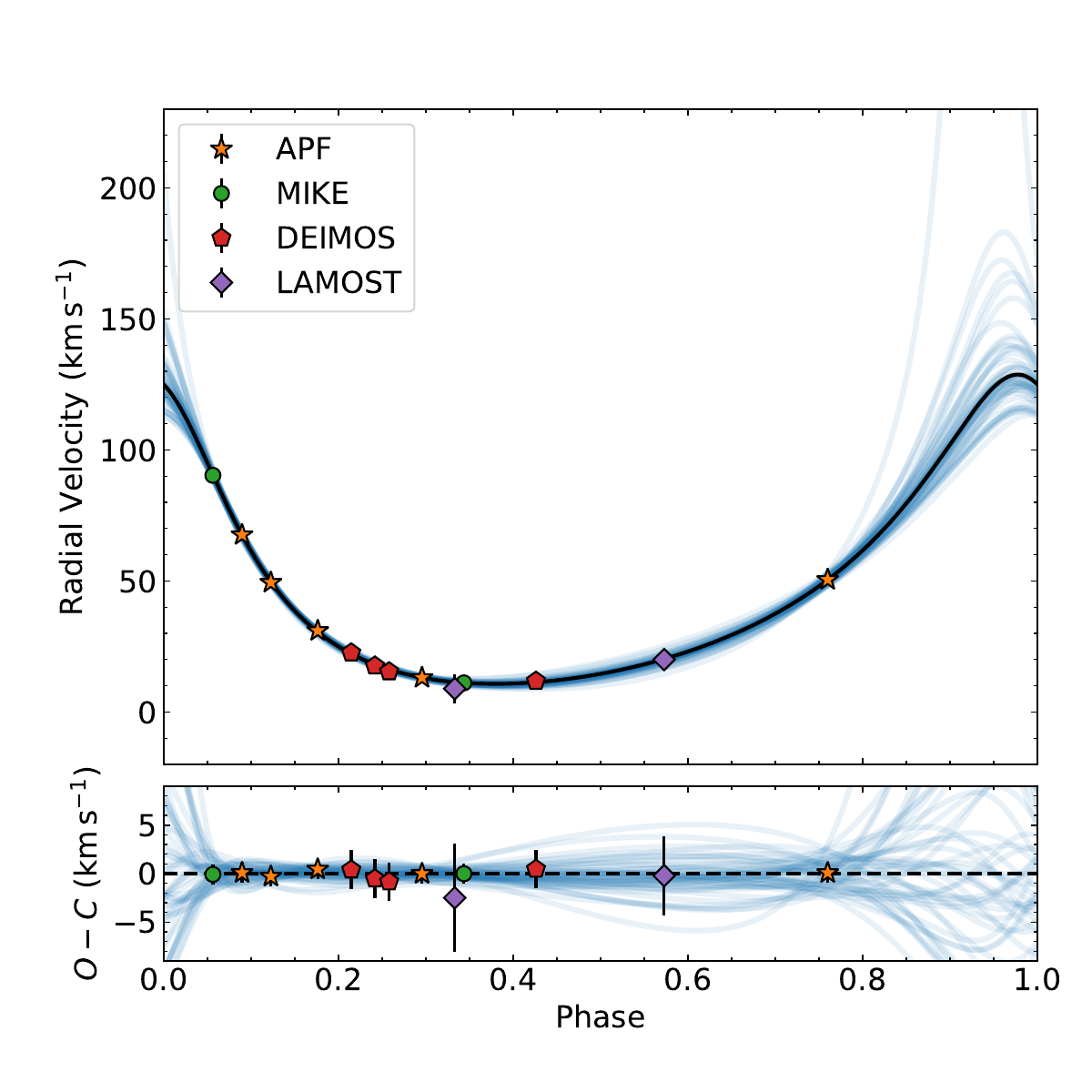}
\includegraphics[width=\linewidth]{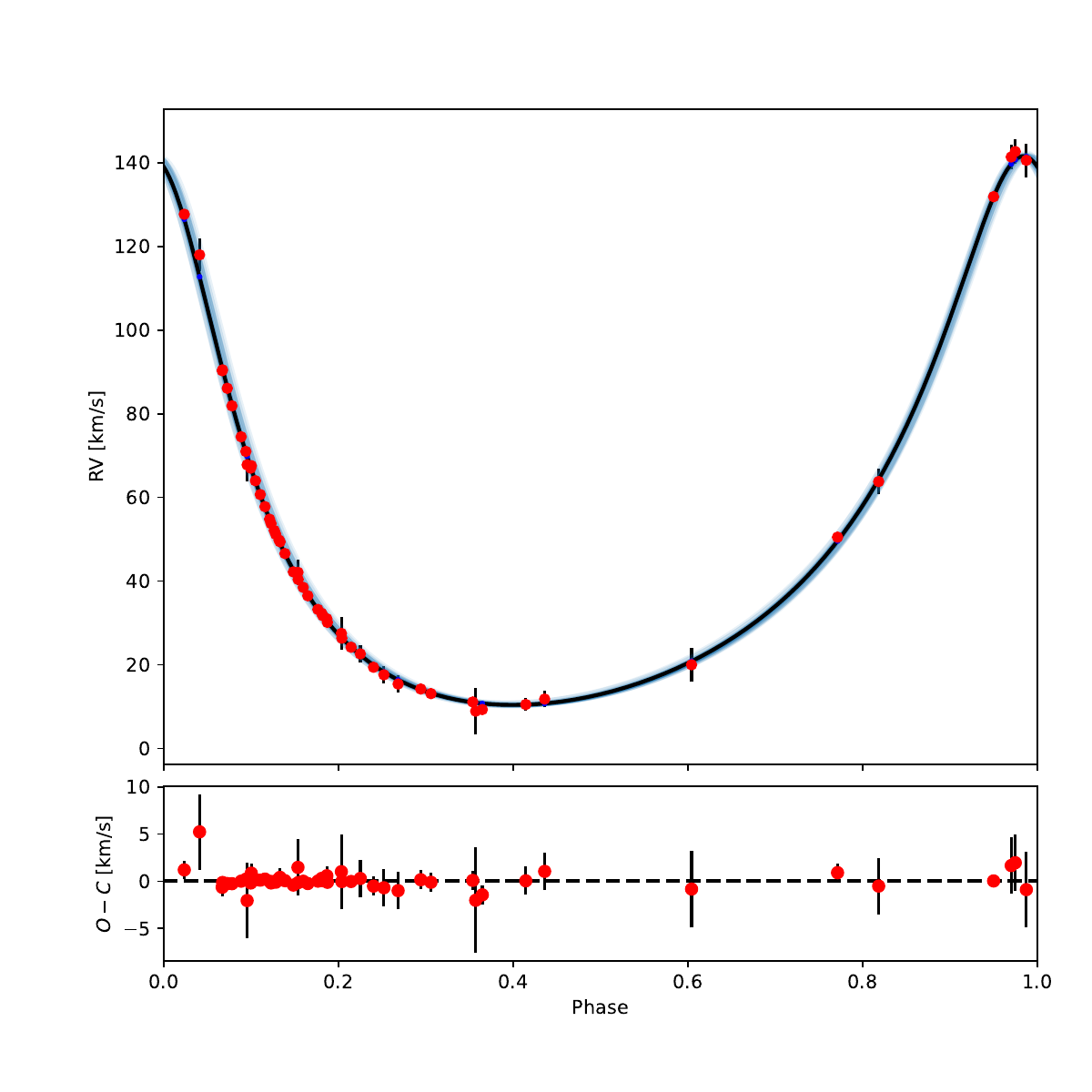}
\caption{(Top) Radial velocity solution for \bh, for the velocity measurements listed in Table~\ref{vel_table}.  The black curve is the most likely MCMC orbital solution determined from the spectroscopic measurements alone, and the light blue curves are a random set of MCMC samples. Many periods are consistent with the RV data. %Here, we impose a prior on the period, similar to the RV-only fit by \cite{ElBadry2022Disc}, which for either our dataset or \cite{ElBadry2022Disc}'s yields a range of periods. 
(Bottom) Radial velocity solution including data from Table 2 and from \cite{ElBadry2022Disc}.  This fit yields a unimodal solution for the period even without incorporating any information from the astrometric solution (0.4\% of the points in the chain are at $P \approx 218\,$d).}
\label{f:RVdataonlynoperiodprior}
\end{figure}

We use uniform priors on all of the RV parameters: period, semiamplitude, eccentricity, periastron time, RV zero point, and (stellar) argument of periastron.  We find a highly multimodal posterior, with orbital solutions at periods of $\sim$142~d, $\sim$164~d, $\sim$184~d, $\sim$220~d, and additional possible periods out to at least 1000 days.  The top panel of Figure \ref{f:RVdataonlynoperiodprior} shows the wide range of RV orbits consistent with our \bh\ velocities.  The fits to the data are satisfactory, with a best-fit $\chi^2$ of 3 for 13 data points with six free parameters, i.e., seven degrees of freedom.  

Because one of our main goals is to determine whether the \textit{Gaia} astrometric orbital solution is correct, we restrict the subsequent analysis of our RV data to solutions with a period range within $\pm10$ days of the Gaia orbital period of $\sim185$~d, discarding those with shorter or longer periods.  A similar procedure was adopted by \cite{ElBadry2022Disc} in their analysis since their RV data (although more extensive than ours) also yields multiple period solutions.  However, as we note below, the combination of our RV data and that of \cite{ElBadry2022Disc} gives a unimodal solution for the period, which is not possible using either data set alone.  This result is illustrated in Figure \ref{f:RVdataonlynoperiodprior}b, where the combined RV data uniquely determine the orbital period without reference to the \textit{Gaia} astrometric solution.  The fit to the combined RV data set remains formally good, with a best-fit $\chi^2$ of 24 for 52 total radial velocity measurements and six free parameters (46 degrees of freedom).  The low $\chi^2$ value suggests that, if anything, the RV errors and, by extension, our derived parameter uncertainties, may be overestimated. 

\begin{figure*}
\epsscale{1.2}
%\plotone{joker_corner_plot2.pdf}
\plotone{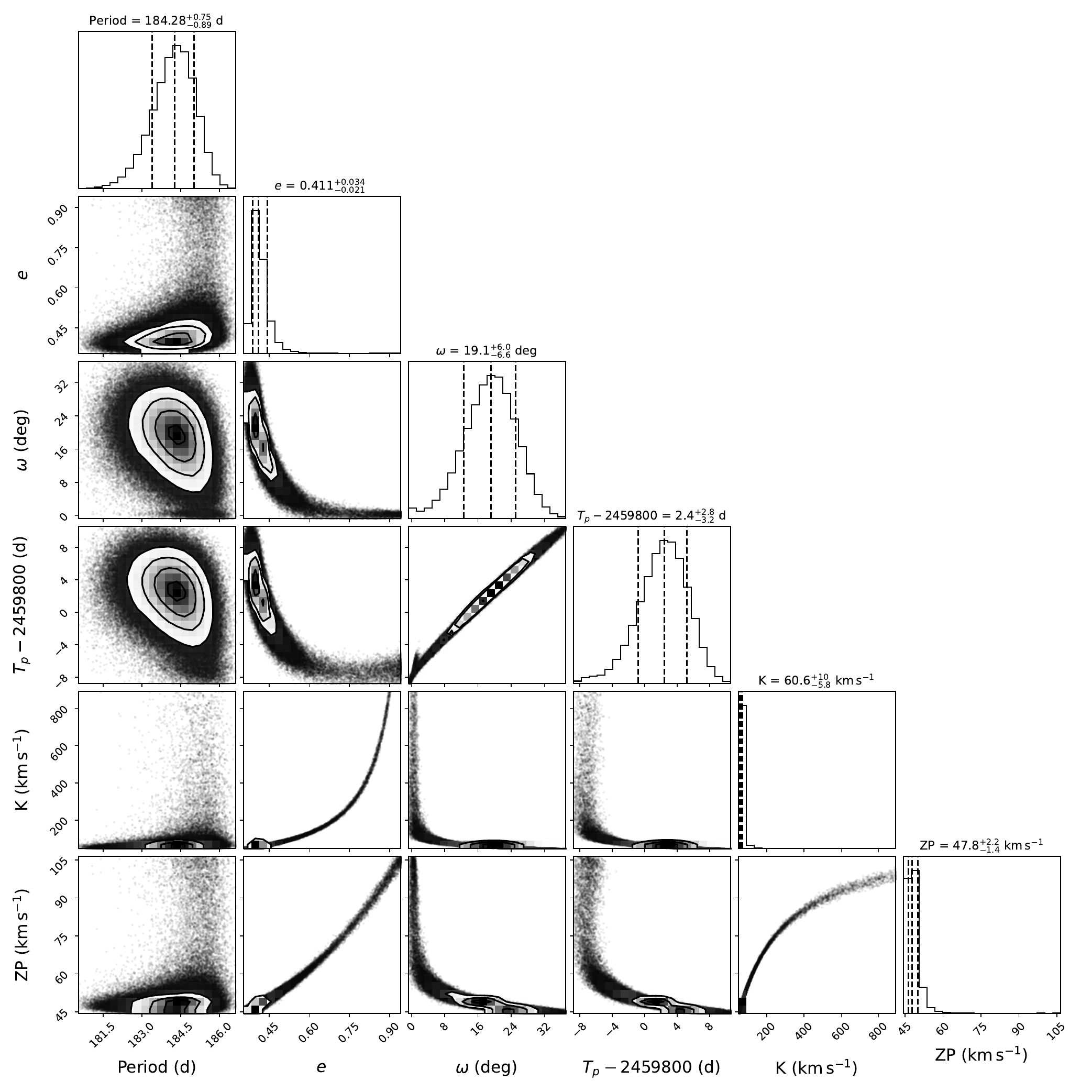}
\caption{MCMC results for the orbital solution determined from the radial velocity data listed in Table 2, showing only elements in the chain with periods between 175 and 195 days.  Although some of the parameters exhibit significant correlations, the overall orbital properties are well constrained by the data, with results similar to those expected from the astrometric solution.  Some parameters, like $K_{1}$, do have significant uncertainties.}
\label{f:OurRVdataonly_orvara}
\end{figure*}

\begin{figure*}
\epsscale{1.2}
%\plotone{joker_corner_plot2.pdf}
\plotone{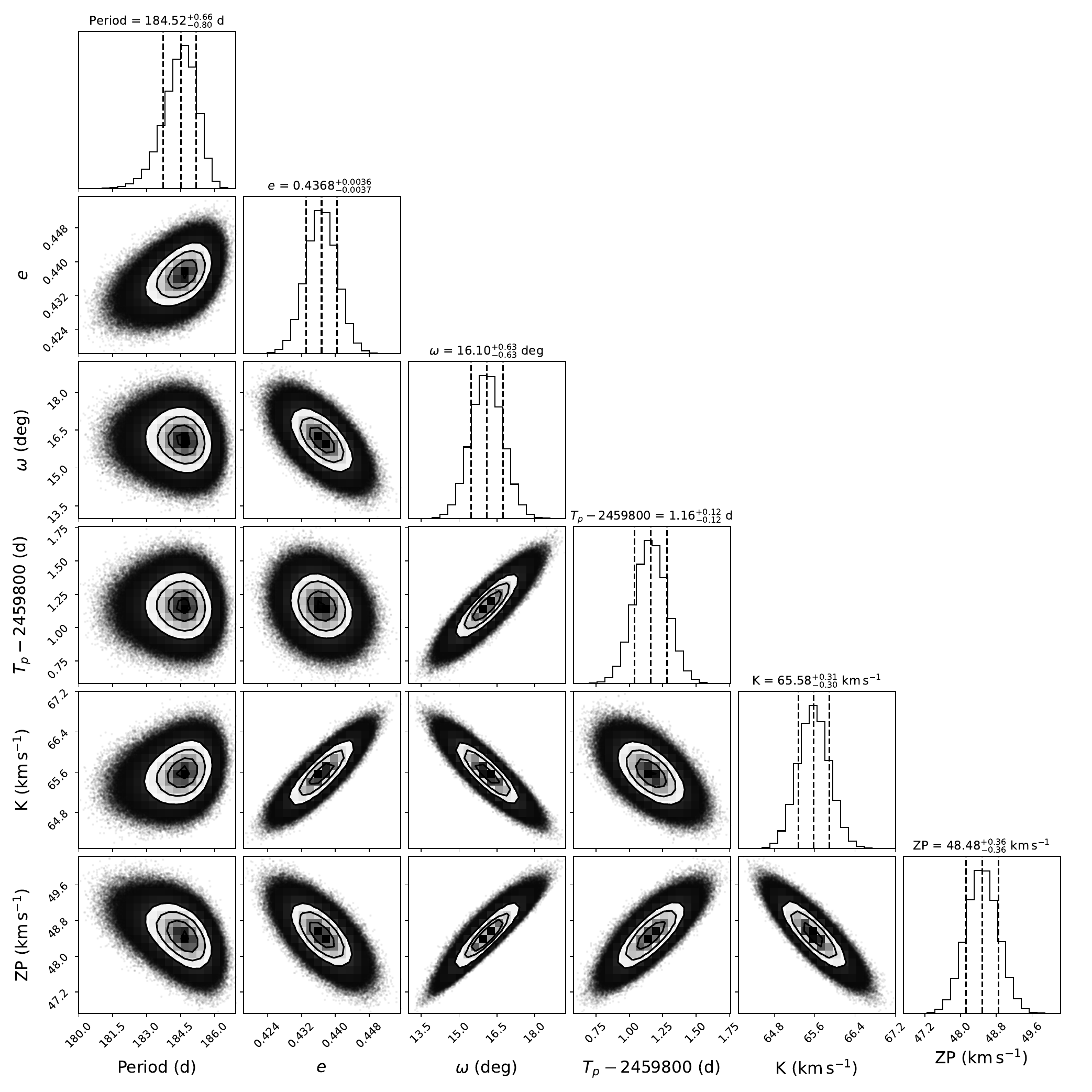}
\caption{MCMC results for the orbital solution determined from all available radial velocity data, including the measurements from \citet{ElBadry2022Disc}.  The orbital solution is much more strongly constrained by the full dataset.  We do not show a much sparser cluster of points at $P\approx218$~d, which contains $\approx$0.4\% of the chain.}
\label{f:AllRVdataonly_orvara}
\end{figure*}

\begin{figure}       
\includegraphics[width=\linewidth]{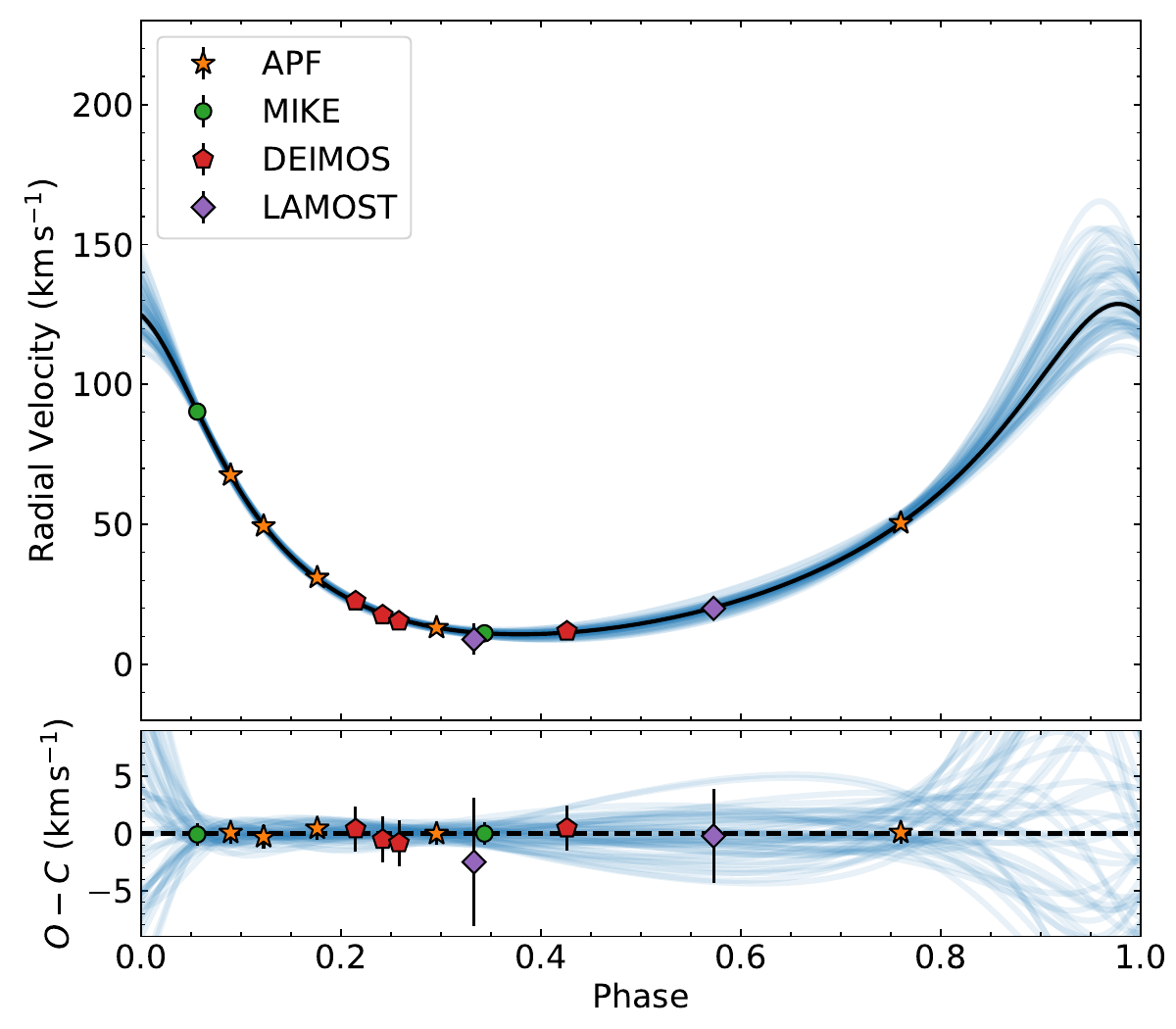}
\includegraphics[width=\linewidth]{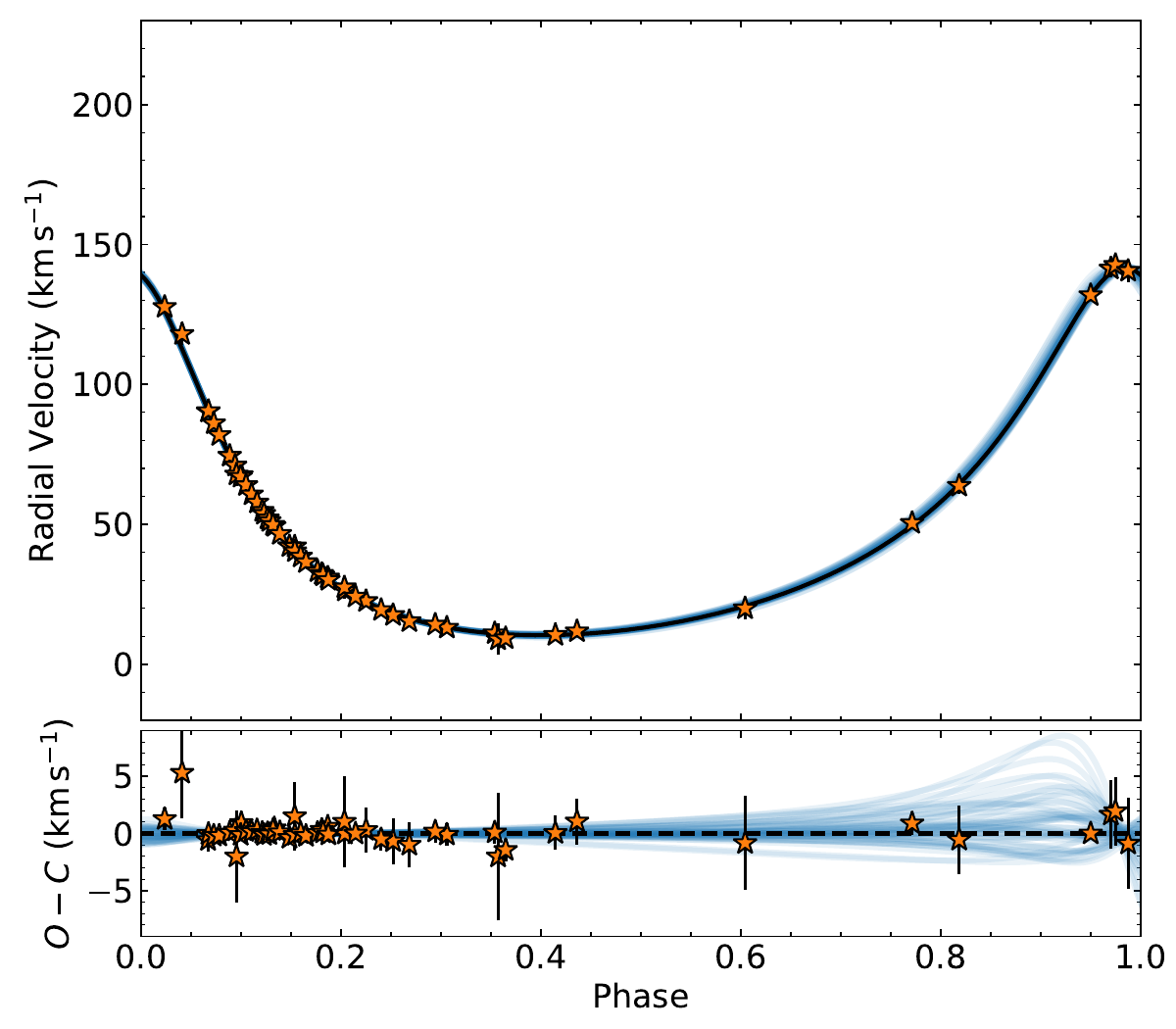}
\caption{Phase-folded radial velocity curve for the data listed in Table 2 (top) and for all available data (bottom).} %, in the frame of the center-of-mass velocity of the system.  
The black line in the top panel represents the most likely orbit, and the light blue curves illustrate a random selection of individual MCMC samples.  The orbits shown in the top panel are restricted to periods between 175 and 195 days.  The residuals, observed minus calculated, are shown in the lower panels. 
%The APF velocities are plotted in blue, the DEIMOS velocities in red, the MIKE velocity in magenta, and the LAMOST velocities in gray.  The apparent offset between the LAMOST data and the best fit results from the offset of 7--11 orbital cycles between those measurements and the remainder of the data and the uncertainty in the time of periastron.  As shown in Figure~\ref{f:RVdataonly}, there are MCMC samples that are in better agreement with the LAMOST measurements, but they may not provide the best match to the more recent measurements.
The agreement is within $\sim5$~km~s$^{-1}$ for all RV measurements, and all $\chi^2$ values indicate satisfactory fits, with $\chi^2$ per degree of freedom below 1 for both data sets.
\label{f:RVphasefolded}
\end{figure}

The spectroscopic orbital parameters from the MCMC fit to our RV data set (the velocities listed in Table~\ref{vel_table}), restricted to periods between 175 and 195 days, are listed in the second section of Table~\ref{tab:gaia_astrom}.  Figure \ref{f:OurRVdataonly_orvara} shows the corner plot of the posterior distribution.  Using all available RVs gives the posterior distributions shown in Figure \ref{f:AllRVdataonly_orvara} and the best-fit parameters listed in the top section of Table~\ref{tab:gaia_astrom_alldata}, whereby the combination of both data sets yields very precise constraints on all parameters that can be determined with RV data.  The radial velocity curve and the phase-folded radial velocity curve with residuals are displayed in Figure~\ref{f:RVphasefolded} for both our RV data alone and all available RV data.  Although our RV data allow a wide range of possible orbits, especially at phases near 0 and 1, the inclusion of data from \cite{ElBadry2022Disc} restricts the range of possible orbits significantly.  Just 0.4\% of the points in the chain lie away from the mode at $P \approx 185$\,d, in a much sparser cluster at $P \approx 218\,$d. 
We obtain very similar results using different orbit-fitting packages such as \textbf{\texttt{TheJoker} \citep{Price-Whelan17}} or \texttt{exoplanet} \citep{exoplanetjoss,exoplanetzenodo}.

The mass function derived from our period-restricted spectroscopic orbital solution is $f=3.2^{+1.6}_{-0.9}$~\msun.  Combined with the primary mass determined from the spectral analysis, the corresponding minimum secondary mass is 4.6~\msun, consistent with the companion being a black hole.  However, 
because our spectroscopic measurements missed the maximum radial velocity at the periastron of the orbit, our constraint on the velocity semi-amplitude $K_{1}$ is not very strong, as is apparent from the definition of the mass function:
\begin{equation}
f = \frac{PK_{1}^{3}}{2\pi G} \left(1-e^{2}\right)^{3/2}= \frac{M_{2}^{3} \sin^{3}{i}}{\left(M_{1} + M_{2}\right)^{2}, \\ 
}
\label{eq:K1}
\end{equation}

\noindent
where $P$ is the orbital period, and $M_{1}$ and $M_{2}$ are the masses of the two stars.
The cubic dependence of the mass function on $K_{1}$ then results in a large uncertainty in the mass function and the mass of the secondary, such that with our radial velocities alone we cannot completely exclude masses below 4~\msun.  %Figure \ref{f:RVdataonly} shows the fit to our RV data using \texttt{orvara}.
The derived constraints on the spectroscopic mass function improve significantly when using all available RV data, combining our measurements and those from \cite{ElBadry2022Disc}.  In this case, we obtain $f = 3.9 \pm 0.04$~\msun, implying a secondary mass of 9.2~\msun\ (when the astrometric inclination is included).

\subsection{Joint Fits}
\label{s:joint}

Our joint fits rely on determining the set of orbits that simultaneously satisfy constraints from both the RV data and the \textit{Gaia} astrometry.
We begin with the MCMC chains derived from RVs alone and condition these chains on the \textit{Gaia} astrometry.  We achieve this by first choosing random values for parameters unconstrained by RV: parallax, inclination, and position angle of the ascending node.  With a random set of these parameters for each step of the chain, we can compute the TI constants and use the \textit{Gaia} DR3 covariance matrix to compute a likelihood.  We then re-weight each step of the chain by its corresponding likelihood.  In practice, we improve our signal-to-noise ratio by choosing many possible inclinations, position angles, and parallaxes for each step of the RV chain, and computing the \textit{Gaia} likelihood for each of them.

Our conditioning of the RV fits on the \textit{Gaia} covariance matrix immediately reveals a problem.  If we adopt the full RV data set --- the union of our data and that of \cite{ElBadry2022Disc} --- the goodness-of-fit is formally good for the RVs ($\chi^2=24$ for 46 degrees of freedom). The step in this chain that agrees best with \textit{Gaia}, even when freely choosing inclination, position angle, and parallax, has a $\chi^2$ with respect to \textit{Gaia} of 11 with five degrees of freedom.  This step is, in turn, a much worse fit to the RVs, with $\Delta \chi^2_{\rm RV} \approx 12$.  The Gaia constraint provides an additional five degrees of freedom but increases the total $\chi^2$ by about 20.  This would be even worse if the RV errors are underestimated, as suggested by the low $\chi^2$
 for the RV-only fit.

\begin{figure*} 
\includegraphics[width=\linewidth]{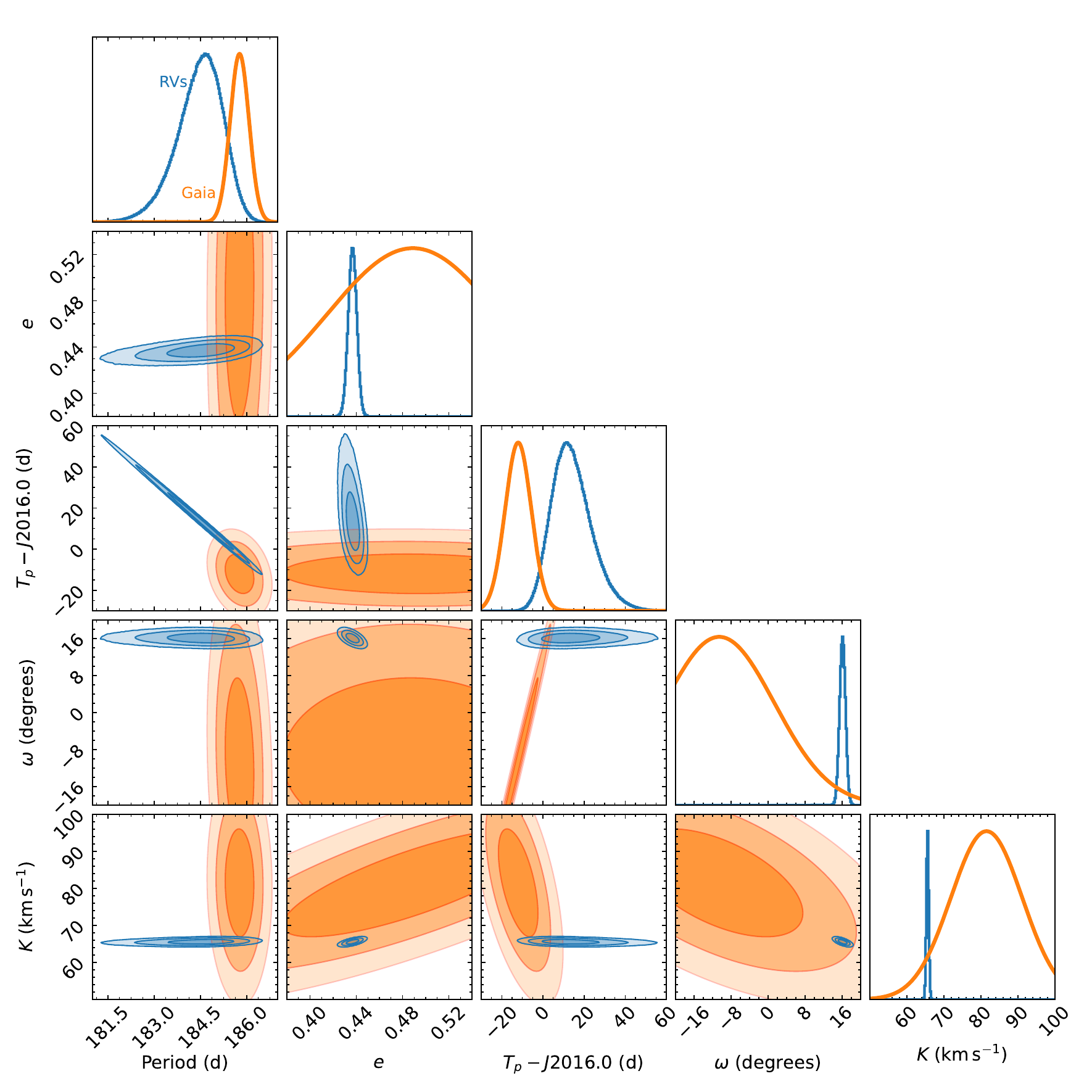}
\caption{1-, 2- and 3-$\sigma$ confidence levels resulting from fits to all available RV data (blue contours), compared to the reported constraints from  \textit{Gaia} astrometry (orange contours).  Although the RV data are qualitatively consistent with the \textit{Gaia} astrometry, most orbital parameters show discrepancies at the 2-3~$\sigma$ level even with the full RV dataset.  Since the $\chi^{2}$ value for RV fits indicates that the velocity uncertainties have not been underestimated, we conclude that the \textit{Gaia} errors must have been underestimated.}
\label{f:confidenceinterval}
%\end{center}
\end{figure*}

 The poor formal agreement of \textit{Gaia} astrometry with the RVs renders any resulting confidence intervals dubious.  The increases in $\chi^2$ from conditioning on Gaia are about a factor of 4 larger what is expected; this suggests that the \textit{Gaia} uncertainties are underestimated by a factor of $\approx$2.  Figure \ref{f:confidenceinterval} shows this tension in the five parameters constrained both by astrometry and radial velocities; the RV confidence intervals are from the entire RV data set.  The $1\sigma$ confidence intervals do not overlap in any parameters apart from eccentricity.  (We emphasize, though, that both the astrometry and RV data sets independently point to \bh\ containing a $\sim10$~\msun\ black hole with a $\sim185$~d period.  The disagreement discussed here relates only to determining the exact orbital parameters of the system.)

Based on the combined evidence from a satisfactory fit to the RV data and an unsatisfactory joint fit, we inflate the Gaia uncertainties by a factor of 2.  We retain the form of the \textit{Gaia} covariance matrix (i.e.,~the correlation coefficients between parameters) and simply multiply this matrix by a factor of 4.  This increase in the uncertainties leads to acceptable joint fits, with $\chi^2$ increasing by $\approx$8 when optimizing the best-fit orbit with the additional five degrees of freedom provided by astrometry. 

We therefore adopt the following prescription to condition the RV fits on the \textit{Gaia} astrometry:
\begin{itemize}
    \item We select many possible sets of random values for parallax, position angle, and inclination for each step in an RV chain;
    \item We compute the TI constants and evaluate 
    \begin{equation}
    \chi^2 = ({\bf p} - {\bf \tilde{p}})^T {\bf C}_{\rm Gaia}^{-1}({\bf p} - {\bf \tilde{p}}),
    \end{equation}
    where ${\bf p}$ is the set of parameters from the RV chain (including the additional parameters above), ${\bf \tilde{p}}$ are the best-fit \textit{Gaia} values, and ${\bf C}_{\rm Gaia}$ is the \textit{Gaia} covariance matrix; and
    \item We weight each of these steps by $e^{(-\chi^2/8)}$, equivalent to multiplying the \textit{Gaia} covariance matrix by 4 (or doubling the uncertainties).  This produces a weighted chain.
\end{itemize}

\begin{figure}[h]        
\begin{center}
\includegraphics[width=\linewidth]{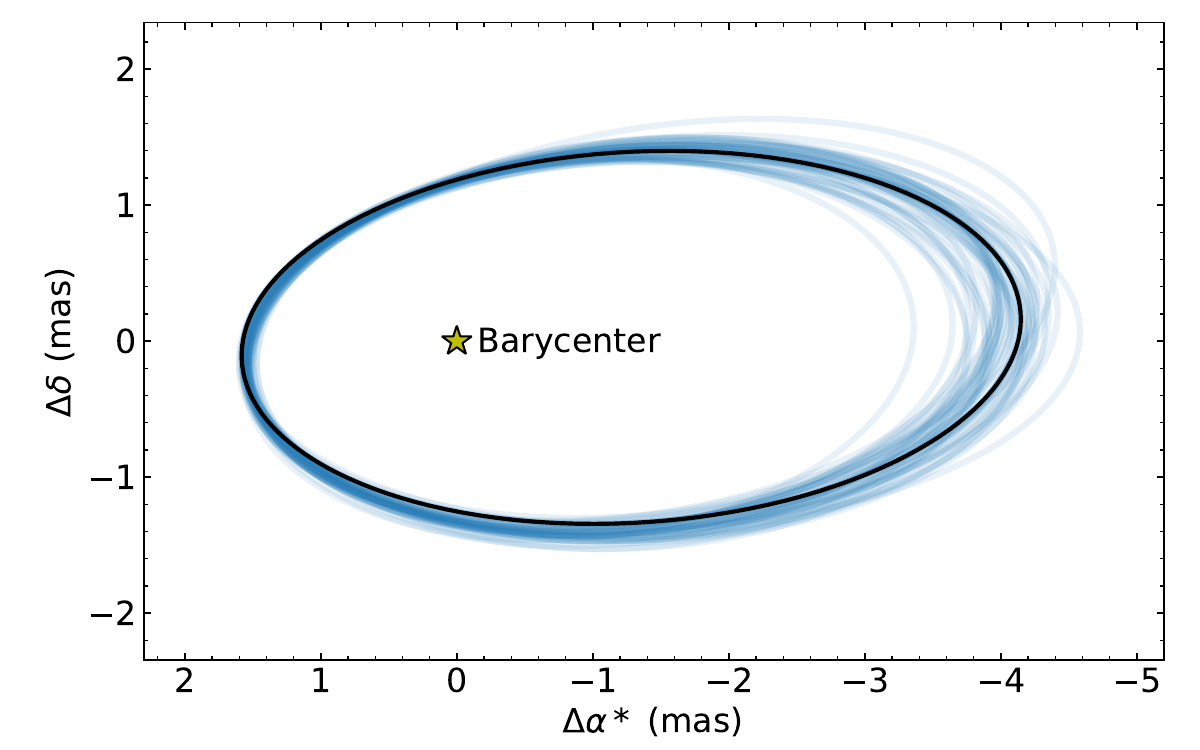}
\includegraphics[width=\linewidth]{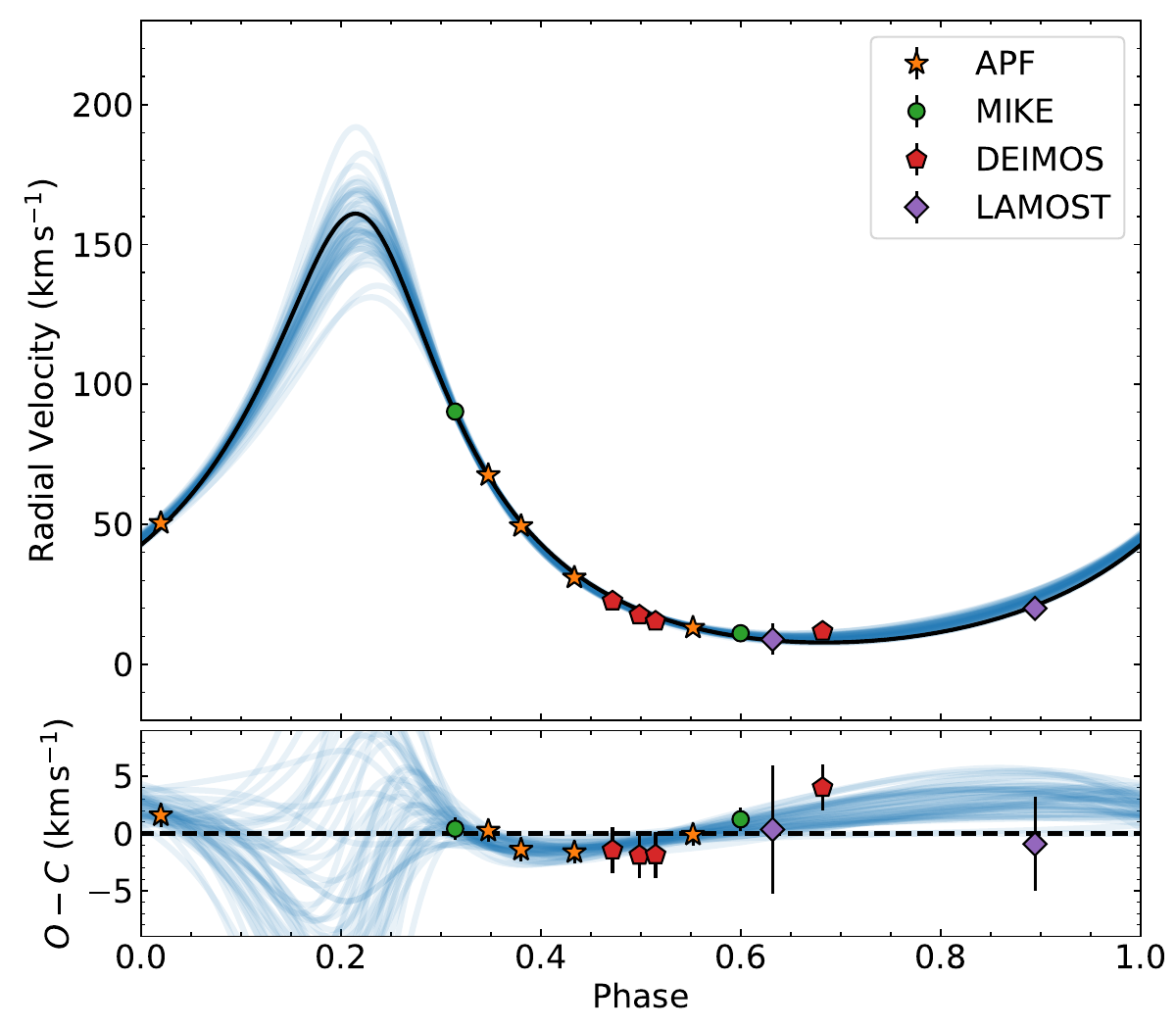}
\includegraphics[width=\linewidth]{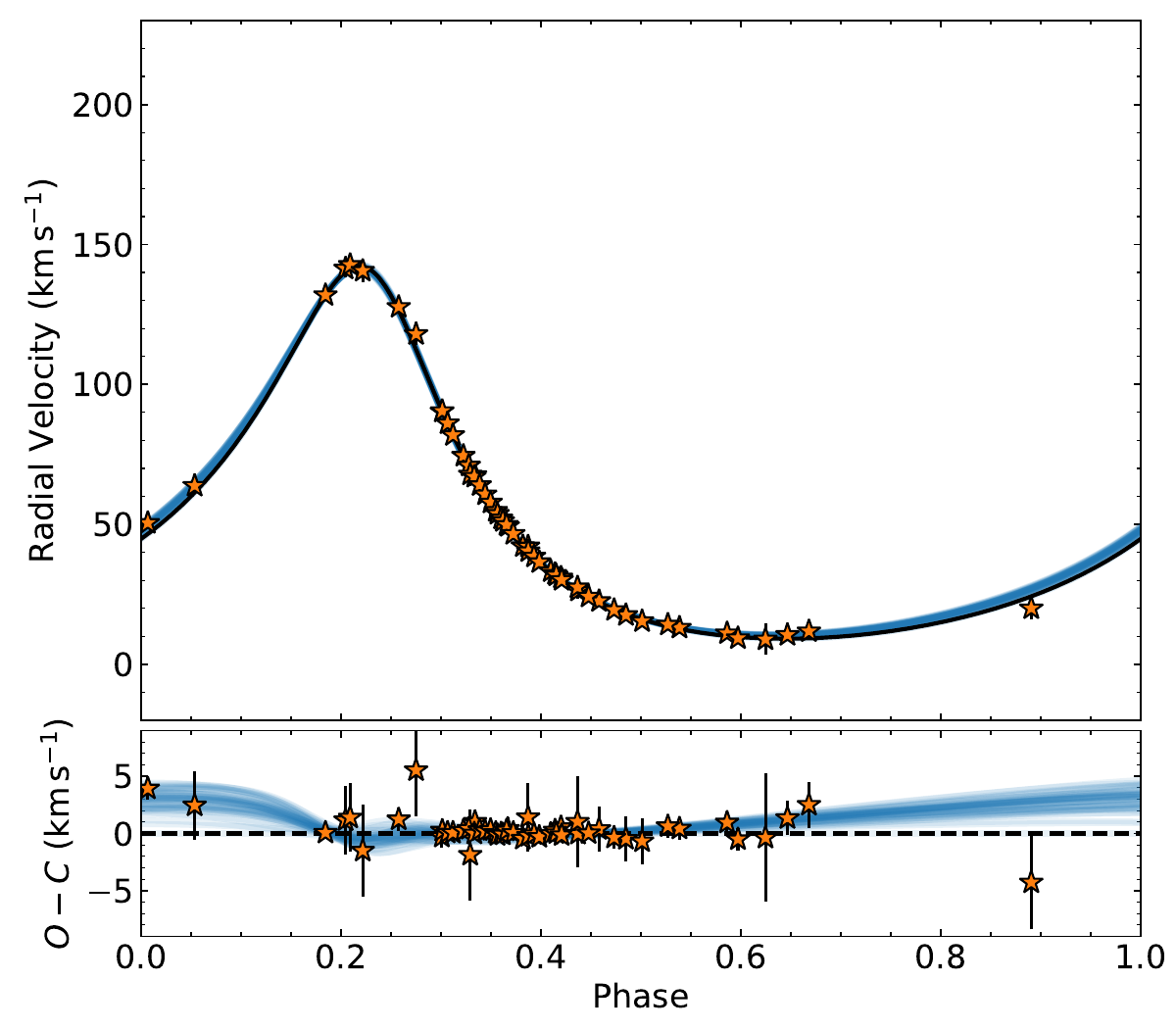}
\caption{Joint orbital fit (see text for the description of the joint fit)} to the astrometry (top) and radial velocities (middle) for the RV data set given in Table 2, and for all available radial velocity data (bottom).  %The astrometric orbit is determined from the \textit{Gaia} DR3 astrometric orbital solution.
The black lines in the lower two panels show the steps in the RV chains most consistent with the Gaia astrometry; these orbits are in tension with the RVs.  The light blue points are drawn from weighted RV chains conditioned on the Gaia data as described in the text. 
\label{f:jointfitourRVdata}
\end{center}
\end{figure}

Figure \ref{f:jointfitourRVdata} presents another way of visualizing the resulting combined constraints and the tension of \textit{Gaia} with the RVs.  The top panel shows a random sampling of astrometric orbits compatible with both the union of the RV data sets and the Gaia covariance matrix.  It is more tightly constrained then the orbits shown in Figure \ref{f:astroonly} but less so than those shown in \cite{ElBadry2022Disc} due to our inflation of the \textit{Gaia} uncertainties.  We note that \cite{ElBadry2022Disc} did perform a check with inflated uncertainties but used the published \textit{Gaia} covariance matrix for their baseline case.  The lower two panels of Figure \ref{f:jointfitourRVdata} show the tension between the astrometry and RVs.  The black lines, denoting the steps in the RV chains that best match the Gaia astrometry, leave systematics in the RV residuals.  This is true both when using only our RVs (middle panel) and 
using all RVs (lower panel).  The light blue lines are randomly drawn from the weighted chains computed as described above.  These RV residuals can be compared to those shown in Fig.~\ref{f:RVphasefolded} for the best fit to the RVs alone.

\begin{figure*} 
\epsscale{1.1}
%\plotone{Corner_GaiaBH4373_Joint_Oct10.pdf}
%\includegraphics[scale=0.7]{Corner_GaiaBH4373_Joint_Oct10.pdf}
%\plotone{CatalogfixedJoint_OurRVdata.pdf}
\plotone{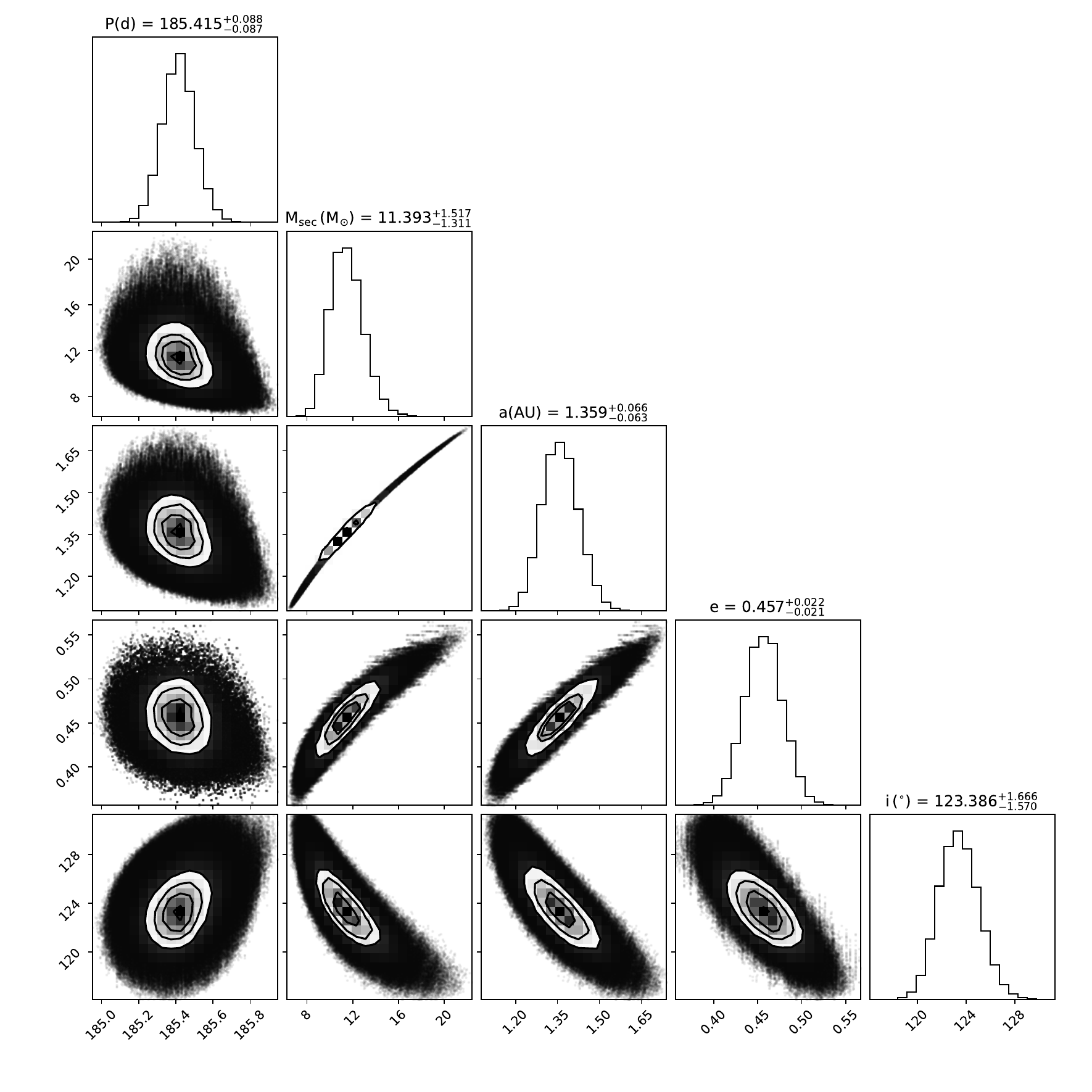}
\caption{Posterior distributions from our joints fits to our velocity data and LAMOST data (as listed in Table 2), and \textit{Gaia} astrometry.  The parameter constraints are tighter and correlations are reduced compared to the fit to the velocities alone.}
\label{f:jointfit_corner_OurRVs}
%\end{center}
\end{figure*}

\begin{figure*} 
\epsscale{1.1}
%\plotone{Corner_GaiaBH4373_Joint_Oct10.pdf}
%\includegraphics[scale=0.7]{Corner_GaiaBH4373_Joint_Oct10.pdf}
%\plotone{CatalogfixedJoint_AllRVdata.pdf}
\plotone{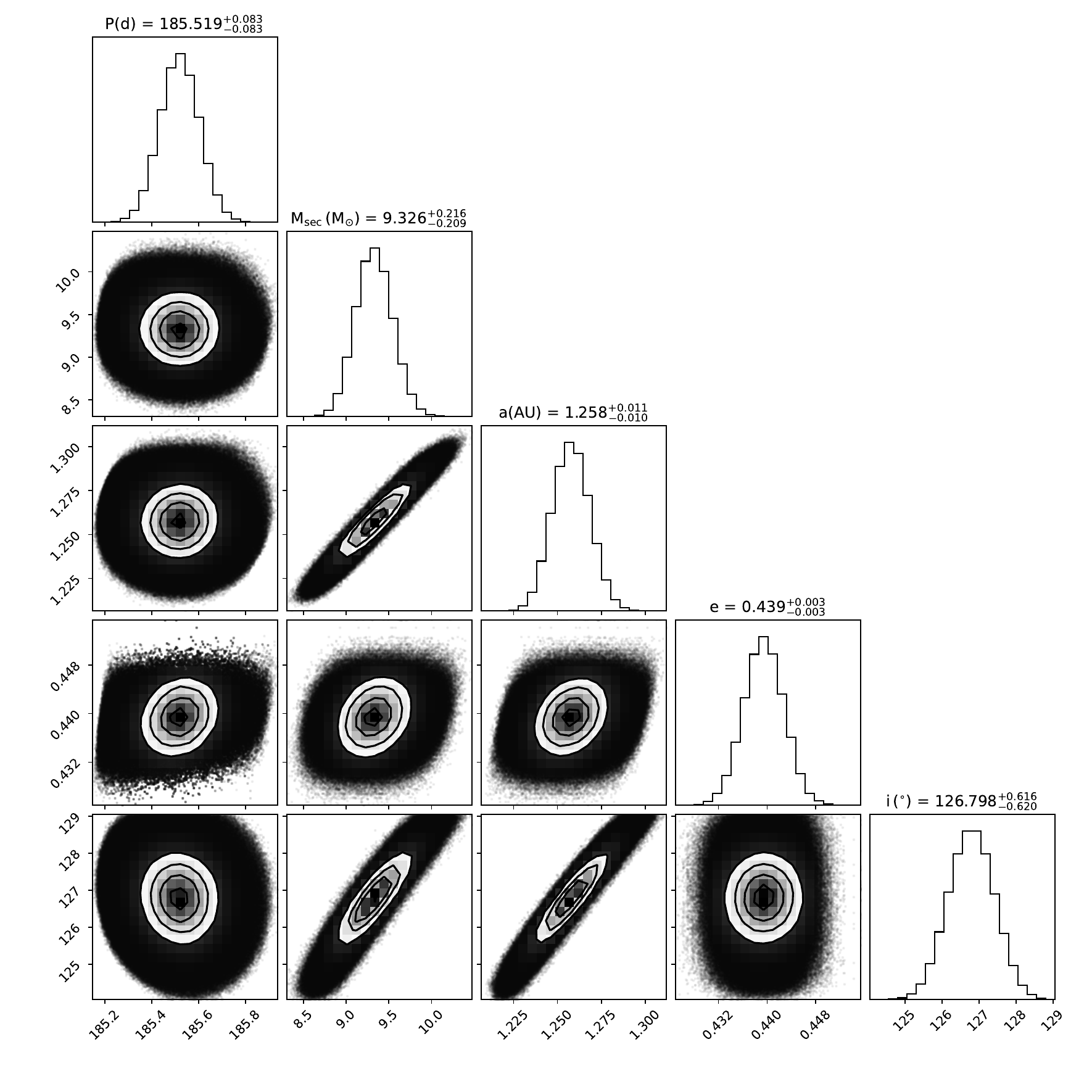}
\caption{Posterior distributions from our joints fits to all available velocity data (including RV data from El-Badry et al. 2023) and \textit{Gaia} astrometry.  The parameter constraints are tighter and correlations are reduced compared to the fit to the velocities alone.}
\label{f:jointfit_corner_AllRVs}
%\end{center}
\end{figure*}

Our joint constraints for \textit{Gaia}~DR3 4373465352415301632 
 using both our radial velocity data and the astrometry 
 are listed in Table 6 (which also provides the astrometric fit and posteriors that result from fitting to the velocities only).  
%The parameters from the joint fits to the astrometry and radial velocities qualitatively agree with the solution reported in \textit{Gaia} DR3.  
Figure \ref{f:jointfit_corner_OurRVs} displays the the resultant corner plot for the derived parameters of the combined fit using the RV data listed in Table 2.  We find that the secondary mass of \bh\ is $11.64^{+1.52}_{-1.31}$~\msun.  Using all of the available radial velocity data gives the posterior distributions shown in Figure \ref{f:jointfit_corner_AllRVs} and parameters listed in the bottom section of Table~\ref{tab:gaia_astrom_alldata}, leading to a smaller secondary mass of ${9.326}_{-0.209}^{+0.216}$~\msun\ that is in closer agreement with \cite{ElBadry2022Disc}. 

%%%%%%%%%%%%%%%
%%%%%Corner plots  - first our data showing results from joint constraints and then joint constraints using all data 

%%%%% edited up to this point

\section{Discussion}
\label{s:discuss}

As noted in Section~\ref{s:intro}, during the preparation of this manuscript, on 2022 September 14 \cite{ElBadry2022Disc} submitted to MNRAS and posted a preprint of an analysis of this same source selected from the \textit{Gaia} DR3 catalog.  %Our goal in this paper is to carry out an independent analysis. 
The overall conclusions of the two studies are in good agreement, that the spectroscopic and astrometric measurements of \bh\ both independently and together demonstrate that the companion is a massive black hole.  \cite{ElBadry2022Disc} have a much more accurate RV-only solution due to their extensive RV coverage.  Based on our joint fits to our RV data and astrometry, we infer a companion mass of $11.64^{+1.52}_{-1.31}$~\msun, whereas \cite{ElBadry2022Disc} report a lower companion mass of $9.62 \pm 0.18 M_{\odot}$.   As noted earlier, our observations did not cover the peak of the radial velocity curve, limiting the precision with which we can determine the velocity semi-amplitude of the primary.  %Secondly, our measurements extend about 3 weeks beyond those of \citet{ElBadry2022Disc}, sampling closer to the minimum of the radial velocity curve.  
Consequently, the best-fit peak-to-trough amplitude of the orbital solution using our RV data is $\sim160$~\kms, compared to $\sim140$~\kms\ in \citet{ElBadry2022Disc}.  As discussed in \S \ref{s:orbits}, the combination of both data sets allows us to derive an unimodal solution for the orbital period, independent of \textit{Gaia}'s astrometric solution.  The joint fit to all available RV data and the astrometry yields the most accurate solution thus far.

\subsection{Comparison with Theoretical Predictions}

Prior to \textit{Gaia} DR3, several studies calculated the expected population of black holes with luminous companions that could be detected with \textit{Gaia}.  The predicted \textit{Gaia} yields range from tens to thousands of such binaries by the end of the mission \citep{Breivik2017,Mashian2017,Wiktorowicz2020,Chawla2022,Shikauchi2022,Wang2022,Janssens2022}.  Although the quality cuts on orbital solutions described by \citet{Halbwachs2022} may have removed more binary systems that contain black holes, the identification of only two confident black hole-luminous companion binaries in DR3 so far (see \cite{ElBadry2023}, as well as the earlier identification of this source by \cite{Tanikawa2022} in their analysis of \textit{Gaia} data) suggests that the upper end of those predictions is likely too optimistic.  The properties of \bh\ are quite similar to the most common binary parameters predicted by, e.g., \citet{Breivik2017}, \citet{Yalinewich2018}, and \citet{Shao2019} in black hole mass, companion mass, period, semi-major axis, and apparent magnitude.  The only deviation from those expectations is the eccentricity, where the observed value differs from the most likely values of $e < 0.1$ or (in the case of significant kick velocities) $e > 0.9$.  %While tidal circularization is expected in binary systems with short periods ($P < 10~\rm d$) \citep{Justesen2021} 
Below we speculate that the observed eccentricity of $e \sim 0.4$ may have resulted from a possible architecture of a hierarchical triple system.  Stellar triples are a common architecture and can affect the dynamical evolution of the system \citep{VignaGomez2022,Tokov2006,Tokov2020,Lennon2022,Chakrabarti2022}.  

\subsection{Formation Scenarios}
\label{s:formation}

Although the characteristics of the \bh\ binary system do not appear to be unusual with respect to population synthesis predictions for binaries detectable by \textit{Gaia}, the particular combination of properties we have determined may require fine-tuning to explain.  To begin with, the mass ratio of the system is quite large, $M_{2}/M_{1} = 10.3 \pm 1.2$.  Although some low-mass companions to massive stars have been identified \citep[e.g.,][]{Karl18,MReggiani2022}, this does not seem to be the most common configuration for OB stars.  The initial mass ratio of the system must have been even larger; the initial-final mass relation of \citet{Raithel2018} suggests that a $\sim$ 10~\msun\ black hole would have had a zero age main sequence mass of $\sim15$--50~\msun.  Moreover, it is not clear that a binary with a separation of $\sim1$~AU should have survived the evolution of the massive component.  Prior to the formation of the black hole, the orbit
must have been tighter because of the mass that the system lost when the black hole formed.  (If the black hole received a natal kick, that also would have heated the system, further reducing the inferred initial separation.)  The maximum radius reached by a 15--50\msun\ star after crossing the Hertzsprung gap is $\gtrsim400$~\rsun\ \citep{MartinsPalacios2013}, indicating that the system could not have avoided a common envelope phase, unless the progenitor evolved quasi-homogeneously (e.g., \citealt{ElBadry22magbraking, Ramachandran2019, Gilkis2021}, suggesting that stars more massive than $\sim 30~M_{\odot}$ may evolve quasi-homogeneously without respect to rotation.)
  Depending on the duration of that period and on the common envelope ejection efficiency, a merger might be expected to follow, making the survival of the G star until the present surprising.
%The low current mass of the primary star indicates that significant mass transfer during the evolution of the system is unlikely; the most it could have accreted from its companion is less than 1~\msun.  

One way to avoid this conundrum is if the system was initially a triple, consisting of an inner binary orbited by the G star.  Massive stars are often found with close massive companions \citep{Sana2012}, so there would be nothing unusual about this architecture.  Interactions between the two inner stars could then have proceeded either via mass loss or a merger in such a way as to prevent either star from expanding enough to engulf their wider companion.  The question posed by this scenario is whether the inner binary ultimately merged, either before or after the formation of the black hole.  If so, that merger could have resulted in the production of the 9.3~\msun\ black hole we have identified.  On the other hand, if the stars did not merge, an inner binary of two compact objects with a combined mass of 9.3~\msun\ could still be present.  Many such combinations are possible, and of course this suggestion is entirely speculative at the moment, but a binary with a combined mass of 9.3~\msun\ would likely involve at least one object residing in the mass gap \citep[e.g.,][]{Ozel2010}, which would be of considerable interest.  In the next section, we discuss whether there would be any observable consequences of the 9.3~\msun\ dark mass being subdivided into a binary.

\subsection{A Third Companion?}
We have investigated with \texttt{orvara} the possibility that \bh\ may be a triple system containing two unseen companions, but our present data set does not allow us to constrain the posterior distributions of a hierarchical triple. Extended radial velocity monitoring could make it possible to study the rich dynamics of a triple system, and we briefly speculate here on some of the possibilities that could be directly observed.  Although the incidence of triples drops dramatically for periods greater than $\sim10$ days \citep{Tokov2006,Tokov2020}, compact triple systems are not uncommon, and indeed in eclipsing binaries, there is a population of compact triples that manifest large eclipse transit variations \citep{Borkovits2015}.  Here, we focus on the more likely possibility of a compact inner binary with the G star as part of the outer system (Section~\ref{s:formation}).  A similar scenario has been investigated in \cite{ElBadry2022Disc}.

%The period drift rate due to the relativistic precession, $\dot{P}_{b}^{GR}$, is described by:
%\begin{equation}
%\dot{P}_{b}^{GR} = \frac {36 \pi e \rm cos \omega} {(1-e^{2})^{1/2} (1+e \rm sin \omega)^{3}} \left(\frac{na}{c}\right)^{4},     
%\label{eq:grprecession}
%\end{equation}
%where $n = 2 \pi /P$ is the orbital frequency, $a$ is the semi-major axis, and $e$ the eccentricity.  For our derived binary parameters, this gives a value of $\dot{P}_{b}^{GR}$ of $3 \times 10^{-13}~\rm s~s^{-1}$, which implies a timescale of $t_{\rm GR}=P_{b}/\dot{P}_{b}^{GR}$ of . 

If this indeed is a hierarchical triple system, the Kozai-Lidov mechanism would allow for an exchange between the eccentricity and the inclination \citep{Chang2009,Naoz2013,Suzuki2019}.  The Kozai-Lidov timescale (Antognini 2015), $t_{\rm KL}$, is given by:
\begin{equation}
t_{\rm KL} \sim P_{\rm in} \frac{m_{1} + m_{2}}{m_{3}} \left({\frac{a_{\rm out}}{a_{\rm in}}} \right)^3 \left(1-e_{\rm out}^{2}\right)^{3/2}   ,
\label{eq:tKL}
\end{equation}

\noindent
where $P_{\rm in}$ is the orbital period of the inner binary, $a_{\rm out}$ and $a_{\rm in}$ are the semi-major axes of the outer and inner systems, $e_{\rm out}$ is the eccentricity of the outer system, and the masses of the three bodies are given by $m_{1}, m_{2}$ and $m_{3}$. For our specific case here, if we consider an inner binary of combined mass $m_{1} + m_{2} \sim 9.3~M_{\odot}$ that has a third body with a semimajor axis that is 3(10) times that of the inner binary, this gives a Kozai-Lidov timescale of $\sim$ 100(5000) years.  However, the shift in the velocity is $\sim 10^{-2}(10^{-4})$ per orbital period. For this specific case, the velocity change in physical units translates to $\sim 1000(10)$~m~s$^{-1}$ per orbit, which could in principle be observed, especially if the inner binary is not too close. This effect (the osculation of the orbit of the inner binary due to the Kozai-Lidov effect) is modified by a general relativistic correction term \citep{Anderson2017}.    This is likely the dominant dynamical effect beyond that of Keplerian motion.  Other dynamical effects for binary systems recently reviewed by \cite{Chakrabarti2022} (including tidal effects and the general relativistic precession) are sub-dominant to this signal.  
%The observed acceleration of this system is $\sim$~km~s$^{-1}$~d$^{-1}$, while the Kozai-Lidov term would contribute (for our supposed case discussed above) $\sim90$~cm~s$^{-1}$~yr$^{-1}$.  
Thus, the effects of 
%an osculating orbit due to 
the Kozai-Lidov effect may become manifest over several orbital periods for this system even for a widely separated hierarchical system.   In this context, it is advantageous that the companion star is a G star, with spectra that indicate that it has low stellar jitter \citep{Wright2005}.  For stars with low stellar jitter, high precision measurements should be able to capture this effect.

\subsection{Comparison with 2MASS J05215658+4359220}

Earlier work on the report of the black hole binary system with a non-interacting giant companion by \citet{Thompson2019} indicated an unusually high [C/N] abundance ratio for its mass and evolutionary state.  If the mass inferred for the red giant is correct, this abundance pattern may have resulted from previous interaction with the black hole progenitor star (we note that \cite{VandenHeuvel2020Sci} argue that if the red giant's mass is $\sim 1 M_{\odot}$, the companion could be a binary with two main-sequence stars).  Low-mass X-ray binaries with black hole companions are also found to be metal rich, with large abundances of $\alpha$ elements \citep{Casares2017}.  For \bh, a comparison with synthetic spectra indicates an approximately solar [C/N] ratio, as expected for a main sequence star of its metallicity.  The [C/N] ratio of \bh\ places it in the typical observed locus of main-sequence stars \citep{Pin2018}, consistent with its small Roche lobe to semi-major axis ratio; we find that the Roche limit to the semi-major axis for the G star is 0.2, indicating that it is stable against tidal disintegration.

\subsection{Estimate for the Number of Systems in the Galaxy}

Given the current uncertainty in the formulation of the common envelope channel, we adopt a simpler approach and estimate the number of such systems that may exist in the Galaxy.  Here, we do not consider  binary (and triple) evolution effects, which may be more significant than IMF dependencies.
We assume a Salpeter IMF:  
\begin{equation}
    \frac{dn}{dm} =\mathcal{N} m^{-2.35},
\end{equation}
where $dn/dm$ is the number of stars per mass bin, and the normalization, $\mathcal{N}$, can be set by integrating the number of stars per mass bin over the limits of the masses of stars in the Salpeter IMF, i.e., where $m_{\rm min} = 0.1~M_{\odot}$ and $m_{\rm max} = 100~M_{\odot}$ are the upper and lower limits, and equating to the observed number of stars in the Galaxy, $N_*$.  The binary fraction of high mass stars is nearly unity ($f_b \sim 1$) \citep{Sana2009,Maiz2016} and the companion masses, $m_c$, are taken from a flat distribution in mass ratio, $q = m_c/m$, e.g., $dP/dq = 1$ \citep{Sana2012,Chulkov2021}.  
\begin{eqnarray}
    N &=& f_b\mathcal{N}\int_{m_{\rm lower}}^{m_{\rm max}} \frac{dn}{dm} dm\int_{q_1}^{q_2} \frac{dP}{dq} dq \nonumber\\
    &=& 0.42f_b \mathcal{N}\left( m_{\rm lower}^{-2.35} - m_{\rm max}^{-2.35}\right)\left(m_2-m_1\right) 
\end{eqnarray}
where $m_{\rm lower}$ is the minimum progenitor mass required to form a $\sim 10~M_{\odot}$ black hole, which we take to be $\sim 20~M_{\odot}$ \citep{Sukhbold2016} and $m_1$ and $m_2$ are the lower and upper mass range of zero-age main sequence masses of G-type stars, which we take to be $0.8~M_{\odot}$ and $1.2~M_{\odot}$. 
This gives a total number of binaries of $\sim 8\times 10^{5}$, which is comparable to the number of binaries with a black hole and a luminous companion in the zero-kick model by \cite{Breivik2017}.
%We can then estimate the number of systems in wide orbits as follows. 
%\begin{equation}
%    N_{*} = \int_{m_{\rm %min}}^{m_{\rm max}}\frac{dn}{dm} %dm = 0.74 \mathcal{N} m_{\rm %min}^{-1.35},
%\end{equation}
%where $N_*$ is the number of %stars in the Galaxy, $m_{\rm %min} = 0.5~M_{\odot}$ and %$m_{\rm max} = 100~M_{\odot}$ %are the limits of the masses of %stars in the Salpeter IMF, 
%Here, $m_{\rm max} \gg m_{\rm min}$ \citep{Salpeter1955}. The number of $M>10$~\msun\ black holes in binaries is then:
%\begin{equation}
%    N_b = f_b \int_{m_{\rm lower}}^{m_{\rm max}} \frac{dn}{dm} dm = 0.74f_b \mathcal{N}\left( m_{\rm lower}^{-1.35} - m_{\rm max}^{-1.35}\right),
%\end{equation}
%where $m_{\rm lower}$ is the minimum progenitor mass required to form a $\sim 10~M_{\odot}$ black hole, which we take to be $\sim 20~M_{\odot}$ \textbf{\citep{Sukhbold2016}}.
%This gives the total number of binaries with a $>10$~\msun\ black hole.  We then determine the number of binaries with a G-star companion :
%\begin{eqnarray}
%    N &=& N_b \int_{m_1}^{m_2} \frac{dn}{dm} dm N_*^{-1} \nonumber\\
%    &=& 0.74f_b \mathcal{N}\left( m_{\rm lower}^{-1.35} - m_{\rm max}^{-1.35}\right)\frac{m_1^{-1.35}-m_2^{-1.35}}{m_{\rm min}^{-1.35}} 
%\end{eqnarray}
%where $m_1$ and $m_2$ are the lower and upper mass range of zero-age main sequence masses of G-type stars, which we take to be $0.8~M_{\odot}$ and $1.2~M_{\odot}$ respectively.  This gives a total number of binaries of $\sim 10^{6}$, which is comparable to the binaries with a black hole and a luminous companion in the zero-kick model by \cite{Breivik2017}.

\subsection{Milky Way Orbit} 

\begin{figure}[h]        
\begin{center}
\includegraphics[scale=0.5]{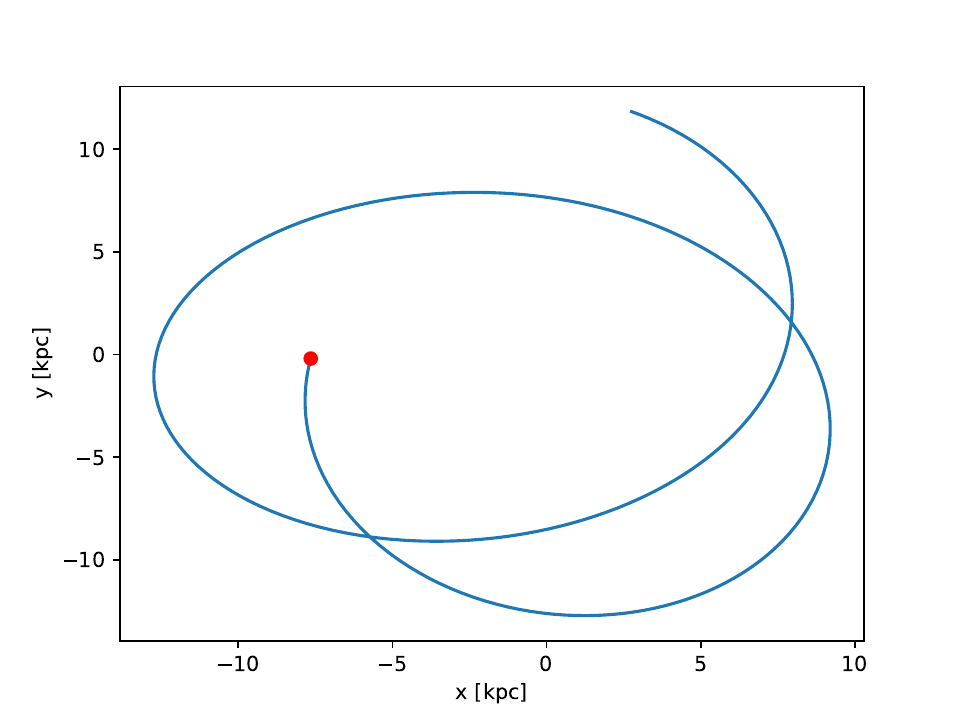}
\includegraphics[scale=0.5]{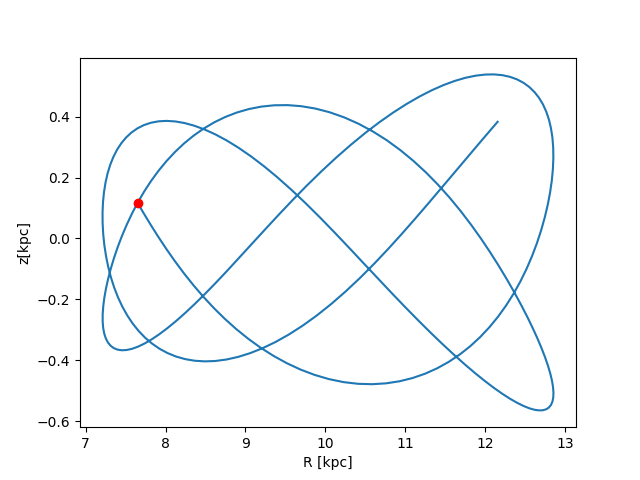}
\includegraphics[scale=0.5]{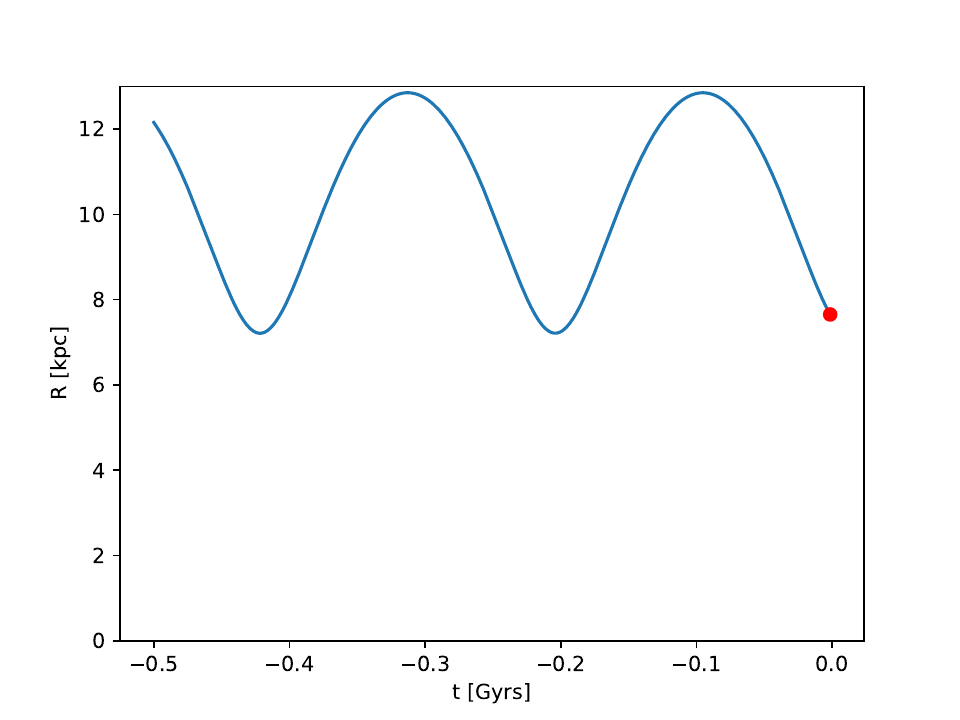}
\caption{Orbit in the Galaxy, integrated backwards from present day for 500~Myr; the red dot marks the present day ($t=0)$.}
\label{f:orbitGalaxy}
\end{center}
\end{figure}

We integrate the orbit of \bh ~backwards in time 500~Myr in the Galactic potential, as shown in Figure \ref{f:orbitGalaxy}.  Here, we use the potential derived from accelerations measured directly from pulsar timing \citep{Chakrabarti2021}, which provides the most direct probe of the Galactic mass distribution.  Using potentials derived from kinematic assumptions \citep[e.g.,][]{Bovy2015} produces similar results to within a factor of $\sim 2$.  The star remains confined within a few hundred pc of the Galactic plane at all times, confirming that it formed as part of the thin disk population.  Unless the natal kick of the black hole happened to be oriented within the disk of the Galaxy, the low scale height of the  orbit indicates that the kick velocity must have been small.  Given the low scale height of the orbit, $\sim$ 0.4 kpc, the vertical kick velocity is of order $\sim$ 10 km/s.  This is similar to the negligible kick velocities for Cyg X-1 and VFTS 243 \citep{Mirabel2003,Shenar2022}.  The orbit of \bh\ around the Milky Way is somewhat eccentric, with a pericenter of $\sim7$~kpc and apocenter of $\sim12$~kpc, reaching substantially larger distances than the Sun does, consistent with its older age.  %The orbit of this source is similar to other thin disk stars, and remains within $\sim$ 500 pc of the Galactic mid-plane.

\section{Conclusion} 
\label{s:conclude}

We summarize below our main results: \\ 

$\bullet$ We searched the \textit{Gaia} DR3 binary catalog to identify possible black hole companions to main sequence stars.  We identified \bh ~as a particularly interesting source on the basis of its large mass ratio and location near the \textit{Gaia} main sequence. \\
$\bullet$ Fitting the SED of \bh\ with single stellar atmosphere models, we find that the photometry is consistent with a modestly metal-poor G star on the main sequence.  Adding a second source does not substantially improve the quality of the SED fits, indicating no evidence for the presence of more than one luminous star. \\ 
$\bullet$ We determine stellar parameters from a high-resolution spectrum, finding $T_{\mathrm{eff}} = 5972 \pm 100$~K, $\log{g} = 4.54 \pm 0.15$, $\mathrm{[Fe/H]} = -0.30 \pm 0.10$, $M = 0.91 \pm 0.10$~\msun, and $\mathrm{age} = 7.1 \pm 3.8$~Gyr, which are generally in good agreement with those inferred from the SED.  The chemical abundance pattern shows no abnormalities compared to other stars with similar metallicities, and the spectrum contains lines from only one star.  \\ 
$\bullet$ We have carried out follow-up radial velocity observations for this source over the four months from its discovery until it went into conjunction with the Sun, and our fits to the velocities give an orbital period of $184.28^{+0.75}_{-0.89}$~d, an eccentricity of $0.411^{+0.034}_{-0.021}$, and a velocity semi-amplitude of $60.6^{+10.0}_{-5.8}$~\kms.  The radial velocity curve predicted from the astrometric orbital solution is in reasonable agreement with the data, confirming that the \textit{Gaia} orbit is not spurious.  We then computed joint fits to the astrometric data and the RVs, which (including the RV measurements from \citealt{ElBadry2022Disc}) give a companion mass of ${9.326}_{-0.209}^{+0.216}$~\msun, an orbital period of $185.52_{-0.08}^{+0.08}$~d, an eccentricity of ${0.439}_{-0.0030}^{+0.0036}$, a semi-major axis of ${1.258}_{-0.011}^{+0.010}$~AU , and an inclination of ${126\fdg79} \pm 0\fdg62$.  \\ 
$\bullet$ The fit to all available RV data gives a best-fit $\chi^{2}$ of 24 for 52 total
radial velocity measurements and six free parameters (46 degrees of freedom). This low $\chi^{2}$ value indicates the RV errors and our derived parameter uncertainties
may be overestimated.  However, there is $\sim2\sigma$ disagreement between the best-fit parameters from the RVs and the \textit{Gaia} astrometry.  The 1-sigma confidence intervals overlap only for the eccentricity.  For our joint fits to the astrometry and RV data, we find that when we condition the RV fits on the \textit{Gaia} data, the increase in $\chi^{2}$ is a factor of four greater than what is expected, which suggests the \textit{Gaia} uncertainties have been underestimated by a factor of $\approx$ 2.

$\bullet$ Given the combination of the large mass of the dark companion and a semi-major axis of \bh\ that is neither very large nor very small, the formation channel for this system is not immediately clear.  However, the most natural scenario may be that the visible G star was originally the outer tertiary component orbiting a close inner binary with two massive stars.  A similar possibility is also discussed in \cite{ElBadry2022Disc}.\\
$\bullet$ The orbit of this system in the Galaxy is consistent with that of thin disk stars, and the low scale height of the integrated Galactic orbit indicates that the kick velocity must have been small.  At a distance of \textbf{$468 \pm 4$}~pc after correcting for the DR3 parallax bias \citep{Lindegren2021b}, as discussed by \citet{ElBadry2022Disc}, \bh\ is closer to the Sun than any black hole X-ray binaries with known distances \citep{Corral-Santana2016} or any of the black holes identified through other techniques \citep{Thompson2019,Lam2022,Sahu2022}. \\ 
$\bullet$ Discoveries of black holes around luminous stars from their measured accelerations now provide a new avenue for understanding the formation channels of black holes in the Galaxy.  Simple estimates suggest that there are $\sim 10^{6}$ similar systems in the Milky Way.  Although our current data cannot constrain the possibility that \bh\ is actually a triple, future RV monitoring may enable us to witness the rich dynamics that can be manifest in hierarchical triple systems.

\bigskip
\bigskip
\bigskip
\begin{acknowledgments}
We thank the anonymous referee for a thorough review that helped improve the paper.
We are grateful to Mirek Brandt for helpful discussions on \texttt{orvara}, Adrian Price-Whelan for assistance with \textit{The Joker}, Brad Holden for his help with APF data and observations, Lieke van Son for discussion of binary evolution, and Alex Ji for advice on MIKE data handling in Python.  SC also acknowledges support from NSF AAG 2009828 and from the RCSA time domain astrophysics program.

This work has made use of data from the European Space Agency (ESA) mission
{\it Gaia} (\url{https://www.cosmos.esa.int/gaia}), processed by the {\it Gaia}
Data Processing and Analysis Consortium (DPAC,
\url{https://www.cosmos.esa.int/web/gaia/dpac/consortium}). Funding for the DPAC
has been provided by national institutions, in particular the institutions
participating in the {\it Gaia} Multilateral Agreement.

This paper includes data gathered with the 6.5 meter Magellan Telescopes located at Las Campanas Observatory, Chile.

Some of the data presented herein were obtained at the W. M. Keck Observatory, which is operated as a scientific partnership among the California Institute of Technology, the University of California and the National Aeronautics and Space Administration. The Observatory was made possible by the generous financial support of the W. M. Keck Foundation. The authors wish to recognize and acknowledge the very significant cultural role and reverence that the summit of Maunakea has always had within the indigenous Hawaiian community.  We are most fortunate to have the opportunity to conduct observations from this mountain.

Guoshoujing Telescope (the Large Sky Area Multi-Object Fiber Spectroscopic Telescope LAMOST) is a National Major Scientific Project built by the Chinese Academy of Sciences. Funding for the project has been provided by the National Development and Reform Commission. LAMOST is operated and managed by the National Astronomical Observatories, Chinese Academy of Sciences.

This research has made use of NASA's Astrophysics Data System
Bibliographic Services.
\end{acknowledgments}

\textit{Facilities:} \facility{\textit{Gaia}, Magellan:II (MIKE), Automated Planet Finder (Levy), Keck:II (DEIMOS)}

\textit{Software:} \software{ARIADNE \citep{Vines_Jenkins2022}, \texttt{SpecMatch-Emp} \citep{Yee17}, \texttt{smhr} \citep{Casey2014}, \texttt{isochrones} \citep{morton2015}, \texttt{MultiNest} \citep{feroz2008,feroz2009,feroz2019}, \texttt{PyMultinest} \citep{buchner2014}, MOOG \citep{sneden1973}, \texttt{orvara} \citep{Brandt2021AJ}, \textit{The Joker} \citep{Price-Whelan17},  \texttt{exoplanet} \citep{exoplanetjoss,exoplanetzenodo}}

\bibliography{bibl}

\end{document}

%% file: 4373_photom_table.tex
\begin{deluxetable}{lccc}
\tablecaption{Photometry for Gaia DR3 4373465352415301632}
\tablewidth{0pt}
\tablehead{
\colhead{Survey} & \colhead{Band} & \colhead{Magnitude} & \colhead{Magnitude} \\
\colhead{} & \colhead{} & \colhead{}
& \colhead{Uncertainty}
}
%COMMENT OUT BANDS WE DECIDE NOT TO USE
\startdata
     GALEX &                  NUV & 20.281 & 0.115 \\
      SDSS &                    u & 16.026 & 0.006 \\
      SDSS &                    g & 14.504 & 0.003 \\
      SDSS &                    r & 13.768 & 0.003 \\
      SDSS &                    i & 13.462 & 0.004 \\
      SDSS &                    z & 13.308 & 0.004 \\
     APASS &                    B & 14.971 & 0.088 \\
     APASS &                    V & 14.091 & 0.045 \\
   GaiaDR3 &  BP\tablenotemark{a} & 14.276 & 0.001 \\
   GaiaDR3 &   G\tablenotemark{a} & 13.772 & 0.001 \\
   GaiaDR3 &  RP\tablenotemark{a} & 13.101 & 0.001 \\
Pan-STARRS &                    g & 14.420 & 0.002 \\
Pan-STARRS &                    r & 13.781 & 0.002 \\
Pan-STARRS &                    i & 13.479 & 0.001 \\
Pan-STARRS &                    z & 13.346 & 0.002 \\
Pan-STARRS &                    y & 13.242 & 0.001 \\
 SkyMapper &                    u & 16.041 & 0.010 \\
 SkyMapper &                    v & 15.654 & 0.008 \\
 SkyMapper &                    g & 14.252 & 0.005 \\
 SkyMapper &                    r & 13.786 & 0.005 \\
 SkyMapper &                    i & 13.450 & 0.004 \\
 SkyMapper &                    z & 13.303 & 0.004 \\
     2MASS &                    J & 12.244 & 0.024 \\
     2MASS &                    H & 11.895 & 0.023 \\
     2MASS &                   Ks & 11.780 & 0.023 \\
      WISE &                   W1 & 11.691 & 0.023 \\
      WISE &                   W2 & 11.717 & 0.022 \\
      WISE &                   W3 & 11.671 & 0.232   
\enddata
\tablenotetext{a}{The formal uncertainties on the \textit{Gaia} DR3 photometry are smaller than listed here, but are rounded to 1~mmag for convenience and because calibration errors between surveys will limit cross-survey comparisons to a coarser level.  }
\label{photom_table}
\end{deluxetable}

%% file: 4373_vel_table.tex
\begin{deluxetable}{lrcr}
\tablecaption{Radial Velocity Measurements for Gaia DR3 4373465352415301632} \label{vel_table}
\tablewidth{0pt}
\tablehead{
\colhead{Julian} & \colhead{Heliocentric} & \colhead{Velocity} & \colhead{Telescope/} \\
\colhead{Date} & \colhead{Velocity} & \colhead{Uncertainty} & \colhead{Instrument} \\
\colhead{} & \colhead{[\kms]} & \colhead{[\kms]} & \colhead{} 
}
\startdata
2457881.29 & 20.0 & 4.1 &        LAMOST \\
2458574.36 &  8.9 & 5.6 &        LAMOST \\
2459758.93 & 50.5 & 1.0 &           APF \\
2459813.53 & 90.3 & 1.0 & Magellan/MIKE \\
2459819.68 & 67.6 & 1.0 &           APF \\
2459825.76 & 49.4 & 1.0 &           APF \\
2459835.65 & 31.0 & 1.0 &           APF \\
2459842.72 & 22.6 & 2.0 &   Keck/DEIMOS \\
2459847.71 & 17.6 & 2.0 &   Keck/DEIMOS \\
2459850.71 & 15.4 & 2.0 &   Keck/DEIMOS \\
2459857.65 & 13.1 & 1.0 &           APF \\
2459866.49 & 11.1 & 1.0 & Magellan/MIKE \\
2459881.69 & 11.8 & 2.0 &   Keck/DEIMOS
\enddata
\end{deluxetable}

%% file: main.bbl
\begin{thebibliography}{}
\expandafter\ifx\csname natexlab\endcsname\relax\def\natexlab#1{#1}\fi
\providecommand{\url}[1]{\href{#1}{#1}}
\providecommand{\dodoi}[1]{doi:~\href{http://doi.org/#1}{\nolinkurl{#1}}}
\providecommand{\doeprint}[1]{\href{http://ascl.net/#1}{\nolinkurl{http://ascl.net/#1}}}
\providecommand{\doarXiv}[1]{\href{https://arxiv.org/abs/#1}{\nolinkurl{https://arxiv.org/abs/#1}}}

\bibitem[{{Abbott} {et~al.}(2016){Abbott}, {Abbott}, {Abbott}, {Abernathy},
  {Acernese}, {Ackley}, {Adams}, {Adams}, {Addesso}, {Adhikari}, {Adya},
  {Affeldt}, {Agathos}, {Agatsuma}, {Aggarwal}, {Aguiar}, {Aiello}, {Ain},
  {Ajith}, {Allen}, {Allocca}, {Altin}, {Anderson}, {Anderson}, {Arai},
  {Araya}, {Arceneaux}, {Areeda}, {Arnaud}, {Arun}, {Ascenzi}, {Ashton}, {Ast},
  {Aston}, {Astone}, {Aufmuth}, {Aulbert}, {Babak}, {Bacon}, {Bader}, {Baker},
  {Baldaccini}, {Ballardin}, {Ballmer}, {Barayoga}, {Barclay}, {Barish},
  {Barker}, {Barone}, {Barr}, {Barsotti}, {Barsuglia}, {Barta}, {Bartlett},
  {Bartos}, {Bassiri}, {Basti}, {Batch}, {Baune}, {Bavigadda}, {Bazzan},
  {Behnke}, {Bejger}, {Bell}, {Bell}, {Berger}, {Bergman}, {Bergmann}, {Berry},
  {Bersanetti}, {Bertolini}, {Betzwieser}, {Bhagwat}, {Bhandare}, {Bilenko},
  {Billingsley}, {Birch}, {Birney}, {Biscans}, {Bisht}, {Bitossi}, {Biwer},
  {Bizouard}, {Blackburn}, {Blair}, {Blair}, {Blair}, {Bloemen}, {Bock},
  {Bodiya}, {Boer}, {Bogaert}, {Bogan}, {Bohe}, {Boh{\'e}mier}, {Bojtos},
  {Bond}, {Bondu}, {Bonnand}, {Boom}, {Bork}, {Boschi}, {Bose}, {Bouffanais},
  {Bozzi}, {Bradaschia}, {Brady}, {Braginsky}, {Branchesi}, {Brau}, {Briant},
  {Brillet}, {Brinkmann}, {Brisson}, {Brockill}, {Brooks}, {Brown}, {Brown},
  {Brown}, {Buchanan}, {Buikema}, {Bulik}, {Bulten}, {Buonanno}, {Buskulic},
  {Buy}, {Byer}, {Cabero}, {Cadonati}, {Cagnoli}, {Cahillane}, {Calder{\'o}n
  Bustillo}, {Callister}, {Calloni}, {Camp}, {Cannon}, {Cao}, {Capano},
  {Capocasa}, {Carbognani}, {Caride}, {Casanueva Diaz}, {Casentini}, {Caudill},
  {Cavagli{\`a}}, {Cavalier}, {Cavalieri}, {Cella}, {Cepeda}, {Cerboni
  Baiardi}, {Cerretani}, {Cesarini}, {Chakraborty}, {Chalermsongsak},
  {Chamberlin}, {Chan}, {Chao}, {Charlton}, {Chassande-Mottin}, {Chen}, {Chen},
  {Cheng}, {Chincarini}, {Chiummo}, {Cho}, {Cho}, {Chow}, {Christensen}, {Chu},
  {Chua}, {Chung}, {Ciani}, {Clara}, {Clark}, {Clayton}, {Cleva}, {Coccia},
  {Cohadon}, {Cokelaer}, {Colla}, {Collette}, {Cominsky}, {Constancio},
  {Conte}, {Conti}, {Cook}, {Corbitt}, {Cornish}, {Corsi}, {Cortese}, {Costa},
  {Coughlin}, {Coughlin}, {Coulon}, {Countryman}, {Couvares}, {Cowan},
  {Coward}, {Cowart}, {Coyne}, {Coyne}, {Craig}, {Creighton}, {Creighton},
  {Cripe}, {Crowder}, {Cumming}, {Cunningham}, {Cuoco}, {Dal Canton},
  {Danilishin}, {D'Antonio}, {Danzmann}, {Darman}, {Dattilo}, {Dave},
  {Daveloza}, {Davier}, {Davies}, {Daw}, {Day}, {De}, {DeBra}, {Debreczeni},
  {Degallaix}, {De Laurentis}, {Del{\'e}glise}, {Del Pozzo}, {Denker}, {Dent},
  {Dereli}, {Dergachev}, {DeRosa}, {De Rosa}, {DeSalvo}, {Dhurandhar},
  {D{\'\i}az}, {Dietz}, {Di Fiore}, {Di Giovanni}, {Di Lieto}, {Di Pace}, {Di
  Palma}, {Di Virgilio}, {Dojcinoski}, {Dolique}, {Donovan}, {Dooley},
  {Doravari}, {Douglas}, {Downes}, {Drago}, {Drever}, {Driggers}, {Du},
  {Ducrot}, {Dwyer}, {Edo}, {Edwards}, {Effler}, {Eggenstein}, {Ehrens},
  {Eichholz}, {Eikenberry}, {Engels}, {Essick}, {Etzel}, {Evans}, {Evans},
  {Everett}, {Factourovich}, {Fafone}, {Fair}, {Fairhurst}, {Fan}, {Fang},
  {Farinon}, {Farr}, {Farr}, {Favata}, {Fays}, {Fehrmann}, {Fejer}, {Ferrante},
  {Ferreira}, {Ferrini}, {Fidecaro}, {Fiori}, {Fiorucci}, {Fisher}, {Flaminio},
  {Fletcher}, {Fotopoulos}, {Fournier}, {Franco}, {Frasca}, {Frasconi}, {Frei},
  {Frei}, {Freise}, {Frey}, {Frey}, {Fricke}, {Fritschel}, {Frolov}, {Fulda},
  {Fyffe}, {Gabbard}, {Gair}, {Gammaitoni}, {Gaonkar}, {Garufi}, {Gatto},
  {Gaur}, {Gehrels}, {Gemme}, {Gendre}, {Genin}, {Gennai}, {George}, {Gergely},
  {Germain}, {Ghosh}, {Ghosh}, {Giaime}, {Giardina}, {Giazotto}, {Gill},
  {Glaefke}, {Goetz}, {Goetz}, {Goggin}, {Gondan}, {Gonz{\'a}lez}, {Gonzalez
  Castro}, {Gopakumar}, {Gordon}, {Gorodetsky}, {Gossan}, {Gosselin}, {Gouaty},
  {Graef}, {Graff}, {Granata}, {Grant}, {Gras}, {Gray}, {Greco}, {Green},
  {Groot}, {Grote}, {Grunewald}, {Guidi}, {Guo}, {Gupta}, {Gupta}, {Gushwa},
  {Gustafson}, {Gustafson}, {Hacker}, {Hall}, {Hall}, {Hammond}, {Haney},
  {Hanke}, {Hanks}, {Hanna}, {Hannam}, {Hanson}, {Hardwick}, {Harms}, {Harry},
  {Harry}, {Hart}, {Hartman}, {Haster}, {Haughian}, {Heidmann}, {Heintze},
  {Heitmann}, {Hello}, {Hemming}, {Hendry}, {Heng}, {Hennig}, {Heptonstall},
  {Heurs}, {Hild}, {Hoak}, {Hodge}, {Hofman}, {Hollitt}, {Holt}, {Holz},
  {Hopkins}, {Hosken}, {Hough}, {Houston}, {Howell}, {Hu}, {Huang}, {Huerta},
  {Huet}, {Hughey}, {Husa}, {Huttner}, {Huynh-Dinh}, {Idrisy}, {Indik},
  {Ingram}, {Inta}, {Isa}, {Isac}, {Isi}, {Islas}, {Isogai}, {Iyer}, {Izumi},
  {Jacqmin}, {Jang}, {Jani}, {Jaranowski}, {Jawahar}, {Jim{\'e}nez-Forteza},
  {Johnson}, {Jones}, {Jones}, {Jones}, {Jonker}, {Ju}, {Haris}, {Kalaghatgi},
  {Kalogera}, {Kandhasamy}, {Kang}, {Kanner}, {Karki}, {Kasprzack},
  {Katsavounidis}, {Katzman}, {Kaufer}, {Kaur}, {Kawabe}, {Kawazoe},
  {K{\'e}f{\'e}lian}, {Kehl}, {Keitel}, {Kelley}, {Kells}, {Keppel}, {Kennedy},
  {Key}, {Khalaidovski}, {Khalili}, {Khan}, {Khan}, {Khan}, {Khazanov},
  {Kijbunchoo}, {Kim}, {Kim}, {Kim}, {Kim}, {Kim}, {Kim}, {King}, {King},
  {Kinzel}, {Kissel}, {Kleybolte}, {Klimenko}, {Koehlenbeck}, {Kokeyama},
  {Koley}, {Kondrashov}, {Kontos}, {Korobko}, {Korth}, {Kowalska}, {Kozak},
  {Kringel}, {Krishnan}, {Kr{\'o}lak}, {Krueger}, {Kuehn}, {Kumar}, {Kuo},
  {Kutynia}, {Lackey}, {Landry}, {Lange}, {Lantz}, {Lasky}, {Lazzarini},
  {Lazzaro}, {Leaci}, {Leavey}, {Lebigot}, {Lee}, {Lee}, {Lee}, {Lee}, {Lenon},
  {Leonardi}, {Leong}, {Leroy}, {Letendre}, {Levin}, {Levine}, {Li}, {Libson},
  {Littenberg}, {Lockerbie}, {Logue}, {Lombardi}, {Lord}, {Lorenzini},
  {Loriette}, {Lormand}, {Losurdo}, {Lough}, {L{\"u}ck}, {Lundgren}, {Luo},
  {Lynch}, {Ma}, {MacDonald}, {Machenschalk}, {MacInnis}, {Macleod},
  {Maga{\~n}a-Sandoval}, {Magee}, {Mageswaran}, {Majorana}, {Maksimovic},
  {Malvezzi}, {Man}, {Mandel}, {Mandic}, {Mangano}, {Mansell}, {Manske},
  {Mantovani}, {Marchesoni}, {Marion}, {M{\'a}rka}, {M{\'a}rka}, {Markosyan},
  {Maros}, {Martelli}, {Martellini}, {Martin}, {Martin}, {Martynov}, {Marx},
  {Mason}, {Masserot}, {Massinger}, {Masso-Reid}, {Matichard}, {Matone},
  {Mavalvala}, {Mazumder}, {Mazzolo}, {McCarthy}, {McClelland}, {McCormick},
  {McGuire}, {McIntyre}, {McIver}, {McKechan}, {McManus}, {McWilliams},
  {Meacher}, {Meadors}, {Meidam}, {Melatos}, {Mendell}, {Mendoza-Gandara},
  {Mercer}, {Merilh}, {Merzougui}, {Meshkov}, {Messaritaki}, {Messenger},
  {Messick}, {Meyers}, {Mezzani}, {Miao}, {Michel}, {Middleton}, {Mikhailov},
  {Milano}, {Miller}, {Millhouse}, {Minenkov}, {Ming}, {Mirshekari}, {Mishra},
  {Mitra}, {Mitrofanov}, {Mitselmakher}, {Mittleman}, {Moggi}, {Mohan},
  {Mohapatra}, {Montani}, {Moore}, {Moore}, {Moraru}, {Moreno}, {Morriss},
  {Mossavi}, {Mours}, {Mow-Lowry}, {Mueller}, {Mueller}, {Muir}, {Mukherjee},
  {Mukherjee}, {Mukherjee}, {Mukund}, {Mullavey}, {Munch}, {Murphy}, {Murray},
  {Mytidis}, {Nardecchia}, {Naticchioni}, {Nayak}, {Necula}, {Nedkova},
  {Nelemans}, {Neri}, {Neunzert}, {Newton}, {Nguyen}, {Nielsen}, {Nissanke},
  {Nitz}, {Nocera}, {Nolting}, {Normandin}, {Nuttall}, {Oberling}, {Ochsner},
  {O'Dell}, {Oelker}, {Ogin}, {Oh}, {Oh}, {Ohme}, {Oliver}, {Oppermann},
  {Oram}, {O'Reilly}, {O'Shaughnessy}, {Ottaway}, {Ottens}, {Overmier}, {Owen},
  {Pai}, {Pai}, {Palamos}, {Palashov}, {Palomba}, {Pal-Singh}, {Pan}, {Pan},
  {Pankow}, {Pannarale}, {Pant}, {Paoletti}, {Paoli}, {Papa}, {Paris},
  {Parker}, {Pascucci}, {Pasqualetti}, {Passaquieti}, {Passuello},
  {Patricelli}, {Patrick}, {Pearlstone}, {Pedraza}, {Pedurand}, {Pekowsky},
  {Pele}, {Penn}, {Perreca}, {Phelps}, {Piccinni}, {Pichot}, {Piergiovanni},
  {Pierro}, {Pillant}, {Pinard}, {Pinto}, {Pitkin}, {Poggiani}, {Popolizio},
  {Post}, {Powell}, {Prasad}, {Predoi}, {Premachandra}, {Prestegard}, {Price},
  {Prijatelj}, {Principe}, {Privitera}, {Prodi}, {Prokhorov}, {Puncken},
  {Punturo}, {Puppo}, {P{\"u}rrer}, {Qi}, {Qin}, {Quetschke}, {Quintero},
  {Quitzow-James}, {Raab}, {Rabeling}, {Radkins}, {Raffai}, {Raja},
  {Rakhmanov}, {Rapagnani}, {Raymond}, {Razzano}, {Re}, {Read}, {Reed},
  {Regimbau}, {Rei}, {Reid}, {Reitze}, {Rew}, {Reyes}, {Ricci}, {Riles},
  {Robertson}, {Robie}, {Robinet}, {Robinson}, {Rocchi}, {Rodriguez},
  {Rolland}, {Rollins}, {Roma}, {Romano}, {Romanov}, {Romie}, {Rosi{\'n}ska},
  {Rowan}, {R{\"u}diger}, {Ruggi}, {Ryan}, {Sachdev}, {Sadecki}, {Sadeghian},
  {Salconi}, {Saleem}, {Salemi}, {Samajdar}, {Sammut}, {Sanchez}, {Sandberg},
  {Sandeen}, {Sanders}, {Santamar{\'\i}a}, {Sassolas}, {Sathyaprakash},
  {Saulson}, {Sauter}, {Savage}, {Sawadsky}, {Schale}, {Schilling}, {Schmidt},
  {Schmidt}, {Schnabel}, {Schofield}, {Sch{\"o}nbeck}, {Schreiber}, {Schuette},
  {Schutz}, {Scott}, {Scott}, {Sellers}, {Sengupta}, {Sentenac}, {Sequino},
  {Sergeev}, {Serna}, {Setyawati}, {Sevigny}, {Shaddock}, {Shah}, {Shahriar},
  {Shaltev}, {Shao}, {Shapiro}, {Shawhan}, {Sheperd}, {Shoemaker}, {Shoemaker},
  {Siellez}, {Siemens}, {Sigg}, {Silva}, {Simakov}, {Singer}, {Singer},
  {Singh}, {Singh}, {Singhal}, {Sintes}, {Slagmolen}, {Smith}, {Smith},
  {Smith}, {Son}, {Sorazu}, {Sorrentino}, {Souradeep}, {Srivastava}, {Staley},
  {Steinke}, {Steinlechner}, {Steinlechner}, {Steinmeyer}, {Stephens}, {Stone},
  {Strain}, {Straniero}, {Stratta}, {Strauss}, {Strigin}, {Sturani}, {Stuver},
  {Summerscales}, {Sun}, {Sutton}, {Swinkels}, {Szczepa{\'n}czyk}, {Tacca},
  {Talukder}, {Tanner}, {T{\'a}pai}, {Tarabrin}, {Taracchini}, {Taylor},
  {Theeg}, {Thirugnanasambandam}, {Thomas}, {Thomas}, {Thomas}, {Thorne},
  {Thorne}, {Thrane}, {Tiwari}, {Tiwari}, {Tokmakov}, {Tomlinson}, {Tonelli},
  {Torres}, {Torrie}, {T{\"o}yr{\"a}}, {Travasso}, {Traylor}, {Trifir{\`o}},
  {Tringali}, {Trozzo}, {Tse}, {Turconi}, {Tuyenbayev}, {Ugolini},
  {Unnikrishnan}, {Urban}, {Usman}, {Vahlbruch}, {Vajente}, {Valdes}, {van
  Bakel}, {van Beuzekom}, {van den Brand}, {Van Den Broeck}, {Vander-Hyde},
  {van der Schaaf}, {van Heijningen}, {van Veggel}, {Vardaro}, {Vass},
  {Vas{\'u}th}, {Vaulin}, {Vecchio}, {Vedovato}, {Veitch}, {Veitch},
  {Venkateswara}, {Verkindt}, {Vetrano}, {Vicer{\'e}}, {Vinciguerra}, {Vine},
  {Vinet}, {Vitale}, {Vo}, {Vocca}, {Vorvick}, {Voss}, {Vousden}, {Vyatchanin},
  {Wade}, {Wade}, {Wade}, {Walker}, {Wallace}, {Walsh}, {Wang}, {Wang}, {Wang},
  {Wang}, {Wang}, {Ward}, {Warner}, {Was}, {Weaver}, {Wei}, {Weinert},
  {Weinstein}, {Weiss}, {Welborn}, {Wen}, {We{\ss}els}, {West}, {Westphal},
  {Wette}, {Whelan}, {White}, {Whiting}, {Wiesner}, {Williams}, {Williamson},
  {Willis}, {Willke}, {Wimmer}, {Winkler}, {Wipf}, {Wiseman}, {Wittel}, {Woan},
  {Worden}, {Wright}, {Wu}, {Yablon}, {Yam}, {Yamamoto}, {Yancey}, {Yap}, {Yu},
  {Yvert}, {Zadro{\.Z}ny}, {Zangrando}, {Zanolin}, {Zendri}, {Zevin}, {Zhang},
  {Zhang}, {Zhang}, {Zhang}, {Zhao}, {Zhou}, {Zhou}, {Zhu}, {Zucker}, {Zuraw},
  {Zweizig}, {LIGO Scientific Collaboration}, \& {Virgo
  Collaboration}}]{abbott2016}
{Abbott}, B.~P., {Abbott}, R., {Abbott}, T.~D., {et~al.} 2016, \prd, 93,
  122003, \dodoi{10.1103/PhysRevD.93.122003}

\bibitem[{{Abbott} {et~al.}(2019){Abbott}, {Abbott}, {Abbott}, {Abraham},
  {Acernese}, {Ackley}, {Adams}, {Adhikari}, {Adya}, {Affeldt}, {Agathos},
  {Agatsuma}, {Aggarwal}, {Aguiar}, {Aiello}, {Ain}, {Ajith}, {Allen},
  {Allocca}, {Aloy}, {Altin}, {Amato}, {Ananyeva}, {Anderson}, {Anderson},
  {Angelova}, {Antier}, {Appert}, {Arai}, {Araya}, {Areeda}, {Ar{\`e}ne},
  {Arnaud}, {Arun}, {Ascenzi}, {Ashton}, {Aston}, {Astone}, {Aubin}, {Aufmuth},
  {AultONeal}, {Austin}, {Avendano}, {Avila-Alvarez}, {Babak}, {Bacon},
  {Badaracco}, {Bader}, {Bae}, {Baker}, {Baldaccini}, {Ballardin}, {Ballmer},
  {Banagiri}, {Barayoga}, {Barclay}, {Barish}, {Barker}, {Barkett}, {Barnum},
  {Barone}, {Barr}, {Barsotti}, {Barsuglia}, {Barta}, {Bartlett}, {Bartos},
  {Bassiri}, {Basti}, {Bawaj}, {Bayley}, {Bazzan}, {B{\'e}csy}, {Bejger},
  {Belahcene}, {Bell}, {Beniwal}, {Berger}, {Bergmann}, {Bernuzzi}, {Bero},
  {Berry}, {Bersanetti}, {Bertolini}, {Betzwieser}, {Bhandare}, {Bidler},
  {Bilenko}, {Bilgili}, {Billingsley}, {Birch}, {Birney}, {Birnholtz},
  {Biscans}, {Biscoveanu}, {Bisht}, {Bitossi}, {Bizouard}, {Blackburn},
  {Blair}, {Blair}, {Blair}, {Bloemen}, {Bode}, {Boer}, {Boetzel}, {Bogaert},
  {Bondu}, {Bonilla}, {Bonnand}, {Booker}, {Boom}, {Booth}, {Bork}, {Boschi},
  {Bose}, {Bossie}, {Bossilkov}, {Bosveld}, {Bouffanais}, {Bozzi},
  {Bradaschia}, {Brady}, {Bramley}, {Branchesi}, {Brau}, {Briant}, {Briggs},
  {Brighenti}, {Brillet}, {Brinkmann}, {Brisson}, {Brockill}, {Brooks},
  {Brown}, {Brunett}, {Buikema}, {Bulik}, {Bulten}, {Buonanno}, {Buscicchio},
  {Buskulic}, {Buy}, {Byer}, {Cabero}, {Cadonati}, {Cagnoli}, {Cahillane},
  {Calder{\'o}n Bustillo}, {Callister}, {Calloni}, {Camp}, {Campbell},
  {Canepa}, {Cannon}, {Cao}, {Cao}, {Capocasa}, {Carbognani}, {Caride},
  {Carney}, {Carullo}, {Casanueva Diaz}, {Casentini}, {Caudill},
  {Cavagli{\`a}}, {Cavalier}, {Cavalieri}, {Cella}, {Cerd{\'a}-Dur{\'a}n},
  {Cerretani}, {Cesarini}, {Chaibi}, {Chakravarti}, {Chamberlin}, {Chan},
  {Chao}, {Charlton}, {Chase}, {Chassande-Mottin}, {Chatterjee}, {Chaturvedi},
  {Chatziioannou}, {Cheeseboro}, {Chen}, {Chen}, {Chen}, {Cheng}, {Cheong},
  {Chia}, {Chincarini}, {Chiummo}, {Cho}, {Cho}, {Cho}, {Christensen}, {Chu},
  {Chua}, {Chung}, {Chung}, {Ciani}, {Ciobanu}, {Ciolfi}, {Cipriano}, {Cirone},
  {Clara}, {Clark}, {Clearwater}, {Cleva}, {Cocchieri}, {Coccia}, {Cohadon},
  {Cohen}, {Colgan}, {Colleoni}, {Collette}, {Collins}, {Cominsky},
  {Constancio}, {Conti}, {Cooper}, {Corban}, {Corbitt}, {Cordero-Carri{\'o}n},
  {Corley}, {Cornish}, {Corsi}, {Cortese}, {Costa}, {Cotesta}, {Coughlin},
  {Coughlin}, {Coulon}, {Countryman}, {Couvares}, {Covas}, {Cowan}, {Coward},
  {Cowart}, {Coyne}, {Coyne}, {Creighton}, {Creighton}, {Cripe}, {Croquette},
  {Crowder}, {Cullen}, {Cumming}, {Cunningham}, {Cuoco}, {Dal Canton},
  {D{\'a}lya}, {Danilishin}, {D'Antonio}, {Danzmann}, {Dasgupta}, {Da Silva
  Costa}, {Datrier}, {Dattilo}, {Dave}, {Davier}, {Davis}, {Daw}, {DeBra},
  {Deenadayalan}, {Degallaix}, {De Laurentis}, {Del{\'e}glise}, {Del Pozzo},
  {DeMarchi}, {Demos}, {Dent}, {De Pietri}, {Derby}, {De Rosa}, {De Rossi},
  {DeSalvo}, {de Varona}, {Dhurandhar}, {D{\'\i}az}, {Dietrich}, {Di Fiore},
  {Di Giovanni}, {Di Girolamo}, {Di Lieto}, {Ding}, {Di Pace}, {Di Palma}, {Di
  Renzo}, {Dmitriev}, {Doctor}, {Donovan}, {Dooley}, {Doravari}, {Dorrington},
  {Downes}, {Drago}, {Driggers}, {Du}, {Ducoin}, {Dupej}, {Dwyer}, {Easter},
  {Edo}, {Edwards}, {Effler}, {Ehrens}, {Eichholz}, {Eikenberry}, {Eisenmann},
  {Eisenstein}, {Essick}, {Estelles}, {Estevez}, {Etienne}, {Etzel}, {Evans},
  {Evans}, {Fafone}, {Fair}, {Fairhurst}, {Fan}, {Farinon}, {Farr}, {Farr},
  {Fauchon-Jones}, {Favata}, {Fays}, {Fazio}, {Fee}, {Feicht}, {Fejer}, {Feng},
  {Fernandez-Galiana}, {Ferrante}, {Ferreira}, {Ferreira}, {Ferrini},
  {Fidecaro}, {Fiori}, {Fiorucci}, {Fishbach}, {Fisher}, {Fishner},
  {Fitz-Axen}, {Flaminio}, {Fletcher}, {Flynn}, {Fong}, {Font}, {Forsyth},
  {Fournier}, {Frasca}, {Frasconi}, {Frei}, {Freise}, {Frey}, {Frey},
  {Fritschel}, {Frolov}, {Fulda}, {Fyffe}, {Gabbard}, {Gadre}, {Gaebel},
  {Gair}, {Gammaitoni}, {Ganija}, {Gaonkar}, {Garcia},
  {Garc{\'\i}a-Quir{\'o}s}, {Garufi}, {Gateley}, {Gaudio}, {Gaur}, {Gayathri},
  {Gemme}, {Genin}, {Gennai}, {George}, {George}, {Gergely}, {Germain},
  {Ghonge}, {Ghosh}, {Ghosh}, {Ghosh}, {Giacomazzo}, {Giaime}, {Giardina},
  {Giazotto}, {Gill}, {Giordano}, {Glover}, {Godwin}, {Goetz}, {Goetz},
  {Goncharov}, {Gonz{\'a}lez}, {Gonzalez Castro}, {Gopakumar}, {Gorodetsky},
  {Gossan}, {Gosselin}, {Gouaty}, {Grado}, {Graef}, {Granata}, {Grant}, {Gras},
  {Grassia}, {Gray}, {Gray}, {Greco}, {Green}, {Green}, {Gretarsson}, {Groot},
  {Grote}, {Grunewald}, {Gruning}, {Guidi}, {Gulati}, {Guo}, {Gupta}, {Gupta},
  {Gustafson}, {Gustafson}, {Haegel}, {Halim}, {Hall}, {Hall}, {Hamilton},
  {Hammond}, {Haney}, {Hanke}, {Hanks}, {Hanna}, {Hannam}, {Hannuksela},
  {Hanson}, {Hardwick}, {Haris}, {Harms}, {Harry}, {Harry}, {Haster},
  {Haughian}, {Hayes}, {Healy}, {Heidmann}, {Heintze}, {Heitmann}, {Hello},
  {Hemming}, {Hendry}, {Heng}, {Hennig}, {Heptonstall}, {Hernandez Vivanco},
  {Heurs}, {Hild}, {Hinderer}, {Hoak}, {Hochheim}, {Hofman}, {Holgado},
  {Holland}, {Holt}, {Holz}, {Hopkins}, {Horst}, {Hough}, {Howell}, {Hoy},
  {Hreibi}, {Huerta}, {Huet}, {Hughey}, {Hulko}, {Husa}, {Huttner},
  {Huynh-Dinh}, {Idzkowski}, {Iess}, {Ingram}, {Inta}, {Intini}, {Irwin},
  {Isa}, {Isac}, {Isi}, {Iyer}, {Izumi}, {Jacqmin}, {Jadhav}, {Jani},
  {Janthalur}, {Jaranowski}, {Jenkins}, {Jiang}, {Johnson}, {Jones}, {Jones},
  {Jones}, {Jonker}, {Ju}, {Junker}, {Kalaghatgi}, {Kalogera}, {Kamai},
  {Kandhasamy}, {Kang}, {Kanner}, {Kapadia}, {Karki}, {Karvinen}, {Kashyap},
  {Kasprzack}, {Katsanevas}, {Katsavounidis}, {Katzman}, {Kaufer}, {Kawabe},
  {Keerthana}, {K{\'e}f{\'e}lian}, {Keitel}, {Kennedy}, {Key}, {Khalili},
  {Khan}, {Khan}, {Khan}, {Khan}, {Khazanov}, {Khursheed}, {Kijbunchoo}, {Kim},
  {Kim}, {Kim}, {Kim}, {Kim}, {Kim}, {Kimball}, {King}, {King},
  {Kinley-Hanlon}, {Kirchhoff}, {Kissel}, {Kleybolte}, {Klika}, {Klimenko},
  {Knowles}, {Koch}, {Koehlenbeck}, {Koekoek}, {Koley}, {Kondrashov}, {Kontos},
  {Koper}, {Korobko}, {Korth}, {Kowalska}, {Kozak}, {Kringel}, {Krishnendu},
  {Kr{\'o}lak}, {Kuehn}, {Kumar}, {Kumar}, {Kumar}, {Kumar}, {Kuo}, {Kutynia},
  {Kwang}, {Lackey}, {Lai}, {Lam}, {Landry}, {Lane}, {Lang}, {Lange}, {Lantz},
  {Lanza}, {Lartaux-Vollard}, {Lasky}, {Laxen}, {Lazzarini}, {Lazzaro},
  {Leaci}, {Leavey}, {Lecoeuche}, {Lee}, {Lee}, {Lee}, {Lee}, {Lee}, {Lee},
  {Lehmann}, {Lenon}, {Leroy}, {Letendre}, {Levin}, {Li}, {Li}, {Li}, {Li},
  {Lin}, {Linde}, {Linker}, {Littenberg}, {Liu}, {Liu}, {Lo}, {Lockerbie},
  {London}, {Longo}, {Lorenzini}, {Loriette}, {Lormand}, {Losurdo}, {Lough},
  {Lousto}, {Lovelace}, {Lower}, {L{\"u}ck}, {Lumaca}, {Lundgren}, {Lynch},
  {Ma}, {Macas}, {Macfoy}, {MacInnis}, {Macleod}, {Macquet},
  {Maga{\~n}a-Sandoval}, {Maga{\~n}a Zertuche}, {Magee}, {Majorana},
  {Maksimovic}, {Malik}, {Man}, {Mandic}, {Mangano}, {Mansell}, {Manske},
  {Mantovani}, {Mapelli}, {Marchesoni}, {Marion}, {M{\'a}rka}, {M{\'a}rka},
  {Markakis}, {Markosyan}, {Markowitz}, {Maros}, {Marquina}, {Marsat},
  {Martelli}, {Martin}, {Martin}, {Martynov}, {Mason}, {Massera}, {Masserot},
  {Massinger}, {Masso-Reid}, {Mastrogiovanni}, {Matas}, {Matichard}, {Matone},
  {Mavalvala}, {Mazumder}, {McCann}, {McCarthy}, {McClelland}, {McCormick},
  {McCuller}, {McGuire}, {McIver}, {McManus}, {McRae}, {McWilliams}, {Meacher},
  {Meadors}, {Mehmet}, {Mehta}, {Meidam}, {Melatos}, {Mendell}, {Mercer},
  {Mereni}, {Merilh}, {Merzougui}, {Meshkov}, {Messenger}, {Messick},
  {Metzdorff}, {Meyers}, {Miao}, {Michel}, {Middleton}, {Mikhailov}, {Milano},
  {Miller}, {Miller}, {Millhouse}, {Mills}, {Milovich-Goff}, {Minazzoli},
  {Minenkov}, {Mishkin}, {Mishra}, {Mistry}, {Mitra}, {Mitrofanov},
  {Mitselmakher}, {Mittleman}, {Mo}, {Moffa}, {Mogushi}, {Mohapatra},
  {Montani}, {Moore}, {Moraru}, {Moreno}, {Morisaki}, {Mours}, {Mow-Lowry},
  {Mukherjee}, {Mukherjee}, {Mukherjee}, {Mukund}, {Mullavey}, {Munch},
  {Mu{\~n}iz}, {Muratore}, {Murray}, {Nagar}, {Nardecchia}, {Naticchioni},
  {Nayak}, {Neilson}, {Nelemans}, {Nelson}, {Nery}, {Neunzert}, {Ng}, {Ng},
  {Nguyen}, {Nichols}, {Nissanke}, {Nocera}, {North}, {Nuttall},
  {Obergaulinger}, {Oberling}, {O'Brien}, {O'Dea}, {Ogin}, {Oh}, {Oh}, {Ohme},
  {Ohta}, {Okada}, {Oliver}, {Oppermann}, {Oram}, {O'Reilly}, {Ormiston},
  {Ortega}, {O'Shaughnessy}, {Ossokine}, {Ottaway}, {Overmier}, {Owen}, {Pace},
  {Pagano}, {Page}, {Pai}, {Pai}, {Palamos}, {Palashov}, {Palomba},
  {Pal-Singh}, {Pan}, {Pang}, {Pang}, {Pankow}, {Pannarale}, {Pant},
  {Paoletti}, {Paoli}, {Parida}, {Parker}, {Pascucci}, {Pasqualetti},
  {Passaquieti}, {Passuello}, {Patil}, {Patricelli}, {Pearlstone}, {Pedersen},
  {Pedraza}, {Pedurand}, {Pele}, {Penn}, {Perez}, {Perreca}, {Pfeiffer},
  {Phelps}, {Phukon}, {Piccinni}, {Pichot}, {Piergiovanni}, {Pillant},
  {Pinard}, {Pirello}, {Pitkin}, {Poggiani}, {Pong}, {Ponrathnam}, {Popolizio},
  {Porter}, {Powell}, {Prajapati}, {Prasad}, {Prasai}, {Prasanna}, {Pratten},
  {Prestegard}, {Privitera}, {Prodi}, {Prokhorov}, {Puncken}, {Punturo},
  {Puppo}, {P{\"u}rrer}, {Qi}, {Quetschke}, {Quinonez}, {Quintero},
  {Quitzow-James}, {Raab}, {Radkins}, {Radulescu}, {Raffai}, {Raja}, {Rajan},
  {Rajbhandari}, {Rakhmanov}, {Ramirez}, {Ramos-Buades}, {Rana}, {Rao},
  {Rapagnani}, {Raymond}, {Razzano}, {Read}, {Regimbau}, {Rei}, {Reid},
  {Reitze}, {Ren}, {Ricci}, {Richardson}, {Richardson}, {Ricker}, {Riles},
  {Rizzo}, {Robertson}, {Robie}, {Robinet}, {Rocchi}, {Rolland}, {Rollins},
  {Roma}, {Romanelli}, {Romano}, {Romel}, {Romie}, {Rose}, {Rosi{\'n}ska},
  {Rosofsky}, {Ross}, {Rowan}, {R{\"u}diger}, {Ruggi}, {Rutins}, {Ryan},
  {Sachdev}, {Sadecki}, {Sakellariadou}, {Salconi}, {Saleem}, {Samajdar},
  {Sammut}, {Sanchez}, {Sanchez}, {Sanchis-Gual}, {Sandberg}, {Sanders},
  {Santiago}, {Sarin}, {Sassolas}, {Sathyaprakash}, {Saulson}, {Sauter},
  {Savage}, {Schale}, {Scheel}, {Scheuer}, {Schmidt}, {Schnabel}, {Schofield},
  {Sch{\"o}nbeck}, {Schreiber}, {Schulte}, {Schutz}, {Schwalbe}, {Scott},
  {Scott}, {Seidel}, {Sellers}, {Sengupta}, {Sennett}, {Sentenac}, {Sequino},
  {Sergeev}, {Setyawati}, {Shaddock}, {Shaffer}, {Shahriar}, {Shaner}, {Shao},
  {Sharma}, {Shawhan}, {Shen}, {Shink}, {Shoemaker}, {Shoemaker},
  {ShyamSundar}, {Siellez}, {Sieniawska}, {Sigg}, {Silva}, {Singer}, {Singh},
  {Singhal}, {Sintes}, {Sitmukhambetov}, {Skliris}, {Slagmolen},
  {Slaven-Blair}, {Smith}, {Smith}, {Somala}, {Son}, {Sorazu}, {Sorrentino},
  {Souradeep}, {Sowell}, {Spencer}, {Spera}, {Srivastava}, {Srivastava},
  {Staats}, {Stachie}, {Standke}, {Steer}, {Steinke}, {Steinlechner},
  {Steinlechner}, {Steinmeyer}, {Stevenson}, {Stocks}, {Stone}, {Stops},
  {Strain}, {Stratta}, {Strigin}, {Strunk}, {Sturani}, {Stuver}, {Sudhir},
  {Summerscales}, {Sun}, {Sunil}, {Suresh}, {Sutton}, {Swinkels},
  {Szczepa{\'n}czyk}, {Tacca}, {Tait}, {Talbot}, {Talukder}, {Tanner},
  {T{\'a}pai}, {Taracchini}, {Tasson}, {Taylor}, {Thies}, {Thomas}, {Thomas},
  {Thondapu}, {Thorne}, {Thrane}, {Tiwari}, {Tiwari}, {Tiwari}, {Toland},
  {Tonelli}, {Tornasi}, {Torres-Forn{\'e}}, {Torrie}, {T{\"o}yr{\"a}},
  {Travasso}, {Traylor}, {Tringali}, {Trovato}, {Trozzo}, {Trudeau}, {Tsang},
  {Tse}, {Tso}, {Tsukada}, {Tsuna}, {Tuyenbayev}, {Ueno}, {Ugolini},
  {Unnikrishnan}, {Urban}, {Usman}, {Vahlbruch}, {Vajente}, {Valdes}, {van
  Bakel}, {van Beuzekom}, {van den Brand}, {Van Den Broeck}, {Vander-Hyde},
  {van der Schaaf}, {van Heijningen}, {van Veggel}, {Vardaro}, {Varma}, {Vass},
  {Vas{\'u}th}, {Vecchio}, {Vedovato}, {Veitch}, {Veitch}, {Venkateswara},
  {Venugopalan}, {Verkindt}, {Vetrano}, {Vicer{\'e}}, {Viets}, {Vine}, {Vinet},
  {Vitale}, {Vo}, {Vocca}, {Vorvick}, {Vyatchanin}, {Wade}, {Wade}, {Wade},
  {Walet}, {Walker}, {Wallace}, {Walsh}, {Wang}, {Wang}, {Wang}, {Wang},
  {Wang}, {Ward}, {Warden}, {Warner}, {Was}, {Watchi}, {Weaver}, {Wei},
  {Weinert}, {Weinstein}, {Weiss}, {Wellmann}, {Wen}, {Wessel}, {We{\ss}els},
  {Westhouse}, {Wette}, {Whelan}, {Whiting}, {Whittle}, {Wilken}, {Williams},
  {Williamson}, {Willis}, {Willke}, {Wimmer}, {Winkler}, {Wipf}, {Wittel},
  {Woan}, {Woehler}, {Wofford}, {Worden}, {Wright}, {Wu}, {Wysocki}, {Xiao},
  {Yamamoto}, {Yancey}, {Yang}, {Yap}, {Yazback}, {Yeeles}, {Yu}, {Yu}, {Yuen},
  {Yvert}, {Zadro{\.z}ny}, {Zanolin}, {Zelenova}, {Zendri}, {Zevin}, {Zhang},
  {Zhang}, {Zhang}, {Zhao}, {Zhou}, {Zhou}, {Zhu}, {Zimmerman}, {Zlochower},
  {Zucker}, {Zweizig}, {LIGO Scientific Collaboration}, \& {Virgo
  Collaboration}}]{abbott2019}
---. 2019, \apjl, 882, L24, \dodoi{10.3847/2041-8213/ab3800}

\bibitem[{{Abbott} {et~al.}(2021){Abbott}, {Abbott}, {Abraham}, {Acernese},
  {Ackley}, {Adams}, {Adams}, {Adhikari}, {Adya}, {Affeldt}, {Agathos},
  {Agatsuma}, {Aggarwal}, {Aguiar}, {Aiello}, {Ain}, {Ajith}, {Akcay}, {Allen},
  {Allocca}, {Altin}, {Amato}, {Anand}, {Ananyeva}, {Anderson}, {Anderson},
  {Angelova}, {Ansoldi}, {Antelis}, {Antier}, {Appert}, {Arai}, {Araya},
  {Areeda}, {Ar{\`e}ne}, {Arnaud}, {Aronson}, {Arun}, {Asali}, {Ascenzi},
  {Ashton}, {Aston}, {Astone}, {Aubin}, {Aufmuth}, {AultONeal}, {Austin},
  {Avendano}, {Babak}, {Badaracco}, {Bader}, {Bae}, {Baer}, {Bagnasco},
  {Baird}, {Ball}, {Ballardin}, {Ballmer}, {Bals}, {Balsamo}, {Baltus},
  {Banagiri}, {Bankar}, {Bankar}, {Barayoga}, {Barbieri}, {Barish}, {Barker},
  {Barneo}, {Barnum}, {Barone}, {Barr}, {Barsotti}, {Barsuglia}, {Barta},
  {Bartlett}, {Bartos}, {Bassiri}, {Basti}, {Bawaj}, {Bayley}, {Bazzan},
  {Becher}, {B{\'e}csy}, {Bedakihale}, {Bejger}, {Belahcene}, {Beniwal},
  {Benjamin}, {Bennett}, {Bentley}, {Bergamin}, {Berger}, {Bergmann},
  {Bernuzzi}, {Berry}, {Bersanetti}, {Bertolini}, {Betzwieser}, {Bhandare},
  {Bhandari}, {Bhattacharjee}, {Bidler}, {Bilenko}, {Billingsley}, {Birney},
  {Birnholtz}, {Biscans}, {Bischi}, {Biscoveanu}, {Bisht}, {Bitossi},
  {Bizouard}, {Blackburn}, {Blackman}, {Blair}, {Blair}, {Blair}, {Blanch},
  {Bobba}, {Bode}, {Boer}, {Boetzel}, {Bogaert}, {Boldrini}, {Bondu},
  {Bonilla}, {Bonnand}, {Booker}, {Boom}, {Bork}, {Boschi}, {Bose},
  {Bossilkov}, {Boudart}, {Bouffanais}, {Bozzi}, {Bradaschia}, {Brady},
  {Bramley}, {Branchesi}, {Brau}, {Breschi}, {Briant}, {Briggs}, {Brighenti},
  {Brillet}, {Brinkmann}, {Brockill}, {Brooks}, {Brooks}, {Brown}, {Brunett},
  {Bruno}, {Bruntz}, {Buikema}, {Bulik}, {Bulten}, {Buonanno}, {Buscicchio},
  {Buskulic}, {Byer}, {Cabero}, {Cadonati}, {Caesar}, {Cagnoli}, {Cahillane},
  {Calder{\'o}n Bustillo}, {Callaghan}, {Callister}, {Calloni}, {Camp},
  {Canepa}, {Cannon}, {Cao}, {Cao}, {Carapella}, {Carbognani}, {Carney},
  {Carpinelli}, {Carullo}, {Carver}, {Casanueva Diaz}, {Casentini}, {Caudill},
  {Cavagli{\`a}}, {Cavalier}, {Cavalieri}, {Cella}, {Cerd{\'a}-Dur{\'a}n},
  {Cesarini}, {Chaibi}, {Chakravarti}, {Chan}, {Chan}, {Chandra}, {Chanial},
  {Chao}, {Charlton}, {Chase}, {Chassande-Mottin}, {Chatterjee},
  {Chattopadhyay}, {Chaturvedi}, {Chatziioannou}, {Chen}, {Chen}, {Chen},
  {Chen}, {Cheng}, {Cheong}, {Chia}, {Chiadini}, {Chierici}, {Chincarini},
  {Chiummo}, {Cho}, {Cho}, {Cho}, {Choate}, {Christensen}, {Chu}, {Chua},
  {Chung}, {Chung}, {Ciani}, {Ciecielag}, {Cie{\'s}lar}, {Cifaldi}, {Ciobanu},
  {Ciolfi}, {Cipriano}, {Cirone}, {Clara}, {Clark}, {Clark}, {Clarke},
  {Clearwater}, {Clesse}, {Cleva}, {Coccia}, {Cohadon}, {Cohen}, {Colleoni},
  {Collette}, {Collins}, {Colpi}, {Constancio}, {Conti}, {Cooper}, {Corban},
  {Corbitt}, {Cordero-Carri{\'o}n}, {Corezzi}, {Corley}, {Cornish}, {Corre},
  {Corsi}, {Cortese}, {Costa}, {Cotesta}, {Coughlin}, {Coughlin}, {Coulon},
  {Countryman}, {Cousins}, {Couvares}, {Covas}, {Coward}, {Cowart}, {Coyne},
  {Coyne}, {Creighton}, {Creighton}, {Croquette}, {Crowder}, {Cudell},
  {Cullen}, {Cumming}, {Cummings}, {Cunningham}, {Cuoco}, {Cury{\l}o},
  {Canton}, {D{\'a}lya}, {Dana}, {DaneshgaranBajastani}, {D'Angelo}, {Danila},
  {Danilishin}, {D'Antonio}, {Danzmann}, {Darsow-Fromm}, {Dasgupta}, {Datrier},
  {Dattilo}, {Dave}, {Davier}, {Davies}, {Davis}, {Daw}, {Dean}, {DeBra},
  {Deenadayalan}, {Degallaix}, {De Laurentis}, {Del{\'e}glise}, {Del Favero},
  {De Lillo}, {De Lillo}, {Del Pozzo}, {DeMarchi}, {De Matteis}, {D'Emilio},
  {Demos}, {Denker}, {Dent}, {Depasse}, {De Pietri}, {De Rosa}, {De Rossi},
  {DeSalvo}, {de Varona}, {Dhurandhar}, {D{\'\i}az}, {Diaz-Ortiz}, {Didio},
  {Dietrich}, {Di Fiore}, {DiFronzo}, {Di Giorgio}, {Di Giovanni}, {Di
  Giovanni}, {Di Girolamo}, {Di Lieto}, {Ding}, {Di Pace}, {Di Palma}, {Di
  Renzo}, {Divakarla}, {Dmitriev}, {Doctor}, {D'Onofrio}, {Donovan}, {Dooley},
  {Doravari}, {Dorrington}, {Downes}, {Drago}, {Driggers}, {Du}, {Ducoin},
  {Dupej}, {Durante}, {D'Urso}, {Duverne}, {Dwyer}, {Easter}, {Eddolls},
  {Edelman}, {Edo}, {Edy}, {Effler}, {Eichholz}, {Eikenberry}, {Eisenmann},
  {Eisenstein}, {Ejlli}, {Errico}, {Essick}, {Estell{\'e}s}, {Estevez},
  {Etienne}, {Etzel}, {Evans}, {Evans}, {Ewing}, {Fafone}, {Fair}, {Fairhurst},
  {Fan}, {Farah}, {Farinon}, {Farr}, {Farr}, {Fauchon-Jones}, {Favata}, {Fays},
  {Fazio}, {Feicht}, {Fejer}, {Feng}, {Fenyvesi}, {Ferguson},
  {Fernandez-Galiana}, {Ferrante}, {Ferreira}, {Fidecaro}, {Figura}, {Fiori},
  {Fiorucci}, {Fishbach}, {Fisher}, {Fishner}, {Fittipaldi}, {Fitz-Axen},
  {Fiumara}, {Flaminio}, {Floden}, {Flynn}, {Fong}, {Font}, {Forsyth},
  {Fournier}, {Frasca}, {Frasconi}, {Frei}, {Freise}, {Frey}, {Frey},
  {Fritschel}, {Frolov}, {Fronz{\'e}}, {Fulda}, {Fyffe}, {Gabbard}, {Gadre},
  {Gaebel}, {Gair}, {Gais}, {Galaudage}, {Gamba}, {Ganapathy}, {Ganguly},
  {Gaonkar}, {Garaventa}, {Garc{\'\i}a-Quir{\'o}s}, {Garufi}, {Gateley},
  {Gaudio}, {Gayathri}, {Gemme}, {Gennai}, {George}, {George}, {George},
  {Gergely}, {Ghonge}, {Ghosh}, {Ghosh}, {Ghosh}, {Giacomazzo}, {Giacoppo},
  {Giaime}, {Giardina}, {Gibson}, {Gier}, {Gill}, {Giri}, {Glanzer}, {Gleckl},
  {Godwin}, {Goetz}, {Goetz}, {Gohlke}, {Goncharov}, {Gonz{\'a}lez},
  {Gopakumar}, {Gossan}, {Gosselin}, {Gouaty}, {Grace}, {Grado}, {Granata},
  {Granata}, {Grant}, {Gras}, {Grassia}, {Gray}, {Gray}, {Greco}, {Green},
  {Green}, {Gretarsson}, {Griggs}, {Grignani}, {Grimaldi}, {Grimes}, {Grimm},
  {Grote}, {Grunewald}, {Gruning}, {Guerrero}, {Guidi}, {Guimaraes},
  {Guix{\'e}}, {Gulati}, {Guo}, {Gupta}, {Gupta}, {Gupta}, {Gustafson},
  {Gustafson}, {Guzman}, {Haegel}, {Halim}, {Hall}, {Hamilton}, {Hammond},
  {Haney}, {Hanke}, {Hanks}, {Hanna}, {Hannam}, {Hannuksela}, {Hannuksela},
  {Hansen}, {Hansen}, {Hanson}, {Harder}, {Hardwick}, {Haris}, {Harms},
  {Harry}, {Harry}, {Hartwig}, {Hasskew}, {Haster}, {Haughian}, {Hayes},
  {Healy}, {Heidmann}, {Heintze}, {Heinze}, {Heinzel}, {Heitmann}, {Hellman},
  {Hello}, {Helmling-Cornell}, {Hemming}, {Hendry}, {Heng}, {Hennes}, {Hennig},
  {Hennig}, {Hernandez Vivanco}, {Heurs}, {Hild}, {Hill}, {Hines}, {Hochheim},
  {Hofgard}, {Hofman}, {Hohmann}, {Holgado}, {Holland}, {Hollows}, {Holmes},
  {Holt}, {Holz}, {Hopkins}, {Horst}, {Hough}, {Howell}, {Hoy}, {Hoyland},
  {Huang}, {H{\"u}bner}, {Huddart}, {Huerta}, {Hughey}, {Hui}, {Husa},
  {Huttner}, {Hutzler}, {Huxford}, {Huynh-Dinh}, {Idzkowski}, {Iess},
  {Imperato}, {Inchauspe}, {Ingram}, {Intini}, {Isi}, {Iyer},
  {JaberianHamedan}, {Jacqmin}, {Jadhav}, {Jadhav}, {James}, {Jani},
  {Janssens}, {Janthalur}, {Jaranowski}, {Jariwala}, {Jaume}, {Jenkins},
  {Jeunon}, {Jiang}, {Johns}, {Johnson-McDaniel}, {Jones}, {Jones}, {Jones},
  {Jones}, {Jones}, {Jonker}, {Ju}, {Junker}, {Kalaghatgi}, {Kalogera},
  {Kamai}, {Kandhasamy}, {Kang}, {Kanner}, {Kapadia}, {Kapasi}, {Karathanasis},
  {Karki}, {Kashyap}, {Kasprzack}, {Kastaun}, {Katsanevas}, {Katsavounidis},
  {Katzman}, {Kawabe}, {K{\'e}f{\'e}lian}, {Keitel}, {Key}, {Khadka},
  {Khalili}, {Khan}, {Khan}, {Khazanov}, {Khetan}, {Khursheed}, {Kijbunchoo},
  {Kim}, {Kim}, {Kim}, {Kim}, {Kim}, {Kim}, {Kimball}, {King}, {Kinley-Hanlon},
  {Kirchhoff}, {Kissel}, {Kleybolte}, {Klimenko}, {Knowles}, {Knyazev}, {Koch},
  {Koehlenbeck}, {Koekoek}, {Koley}, {Kolstein}, {Komori}, {Kondrashov},
  {Kontos}, {Koper}, {Korobko}, {Korth}, {Kovalam}, {Kozak}, {Kr{\"a}mer},
  {Kringel}, {Krishnendu}, {Kr{\'o}lak}, {Kuehn}, {Kumar}, {Kumar}, {Kumar},
  {Kumar}, {Kuns}, {Kwang}, {Lackey}, {Laghi}, {Lalande}, {Lam}, {Lamberts},
  {Landry}, {Lane}, {Lang}, {Lange}, {Lantz}, {Lanza}, {La Rosa},
  {Lartaux-Vollard}, {Lasky}, {Laxen}, {Lazzarini}, {Lazzaro}, {Leaci},
  {Leavey}, {Lecoeuche}, {Lee}, {Lee}, {Lee}, {Lee}, {Lehmann}, {Leon},
  {Leroy}, {Letendre}, {Levin}, {Li}, {Li}, {Li}, {Li}, {Li}, {Linde},
  {Linker}, {Linley}, {Littenberg}, {Liu}, {Liu}, {Llorens-Monteagudo}, {Lo},
  {Lockwood}, {London}, {Longo}, {Lorenzini}, {Loriette}, {Lormand}, {Losurdo},
  {Lough}, {Lousto}, {Lovelace}, {L{\"u}ck}, {Lumaca}, {Lundgren}, {Ma},
  {Macas}, {MacInnis}, {Macleod}, {MacMillan}, {Macquet}, {Maga{\~n}a
  Hernandez}, {Maga{\~n}a-Sandoval}, {Magazz{\~A}{\textonesuperior}}, {Magee},
  {Majorana}, {Maksimovic}, {Maliakal}, {Malik}, {Man}, {Mandic}, {Mangano},
  {Mansell}, {Manske}, {Mantovani}, {Mapelli}, {Marchesoni}, {Marion},
  {M{\'a}rka}, {M{\'a}rka}, {Markakis}, {Markosyan}, {Markowitz}, {Maros},
  {Marquina}, {Marsat}, {Martelli}, {Martin}, {Martin}, {Martinez}, {Martinez},
  {Martynov}, {Masalehdan}, {Mason}, {Massera}, {Masserot}, {Massinger},
  {Masso-Reid}, {Mastrogiovanni}, {Matas}, {Mateu-Lucena}, {Matichard},
  {Matiushechkina}, {Mavalvala}, {Maynard}, {McCann}, {McCarthy}, {McClelland},
  {McCormick}, {McCuller}, {McGuire}, {McIsaac}, {McIver}, {McManus}, {McRae},
  {McWilliams}, {Meacher}, {Meadors}, {Mehmet}, {Mehta}, {Melatos}, {Melchor},
  {Mendell}, {Menendez-Vazquez}, {Mercer}, {Mereni}, {Merfeld}, {Merilh},
  {Merritt}, {Merzougui}, {Meshkov}, {Messenger}, {Messick}, {Metzdorff},
  {Meyers}, {Meylahn}, {Mhaske}, {Miani}, {Miao}, {Michaloliakos}, {Michel},
  {Middleton}, {Milano}, {Miller}, {Millhouse}, {Mills}, {Milotti},
  {Milovich-Goff}, {Minazzoli}, {Minenkov}, {Mir}, {Mishkin}, {Mishra},
  {Mistry}, {Mitra}, {Mitrofanov}, {Mitselmakher}, {Mittleman}, {Mo},
  {Mogushi}, {Mohapatra}, {Mohite}, {Molina}, {Molina-Ruiz}, {Mondin},
  {Montani}, {Moore}, {Moraru}, {Morawski}, {Moreno}, {Morisaki}, {Mours},
  {Mow-Lowry}, {Mozzon}, {Muciaccia}, {Mukherjee}, {Mukherjee}, {Mukherjee},
  {Mukherjee}, {Mukund}, {Mullavey}, {Munch}, {Mu{\~n}iz}, {Murray}, {Nadji},
  {Nagar}, {Nardecchia}, {Naticchioni}, {Nayak}, {Neil}, {Neilson}, {Nelemans},
  {Nelson}, {Nery}, {Neunzert}, {Nitz}, {Ng}, {Ng}, {Nguyen}, {Nguyen},
  {Nguyen}, {Nichols}, {Nissanke}, {Nocera}, {Noh}, {North}, {Nothard},
  {Nuttall}, {Oberling}, {O'Brien}, {O'Dell}, {Oganesyan}, {Ogin}, {Oh}, {Oh},
  {Ohme}, {Ohta}, {Okada}, {Olivetto}, {Oppermann}, {Oram}, {O'Reilly},
  {Ormiston}, {Ortega}, {O'Shaughnessy}, {Ossokine}, {Osthelder}, {Ottaway},
  {Overmier}, {Owen}, {Pace}, {Pagano}, {Page}, {Pagliaroli}, {Pai}, {Pai},
  {Palamos}, {Palashov}, {Palomba}, {Pan}, {Panda}, {Pang}, {Pankow},
  {Pannarale}, {Pant}, {Paoletti}, {Paoli}, {Paolone}, {Parker}, {Pascucci},
  {Pasqualetti}, {Passaquieti}, {Passuello}, {Patel}, {Patricelli}, {Payne},
  {Pechsiri}, {Pedraza}, {Pegoraro}, {Pele}, {Penn}, {Perego}, {Perez},
  {P{\'e}rigois}, {Perreca}, {Perri{\`e}s}, {Petermann}, {Petterson},
  {Pfeiffer}, {Pham}, {Phukon}, {Piccinni}, {Pichot}, {Piendibene},
  {Piergiovanni}, {Pierini}, {Pierro}, {Pillant}, {Pilo}, {Pinard}, {Pinto},
  {Piotrzkowski}, {Pirello}, {Pitkin}, {Placidi}, {Plastino}, {Pluchar},
  {Poggiani}, {Polini}, {Pong}, {Ponrathnam}, {Popolizio}, {Porter},
  {Poverman}, {Powell}, {Pracchia}, {Prajapati}, {Prasai}, {Prasanna},
  {Pratten}, {Prestegard}, {Principe}, {Prodi}, {Prokhorov}, {Prosposito},
  {Prudenzi}, {Puecher}, {Punturo}, {Puosi}, {Puppo}, {P{\"u}rrer}, {Qi},
  {Quetschke}, {Quinonez}, {Quitzow-James}, {Raab}, {Raaijmakers}, {Radkins},
  {Radulesco}, {Raffai}, {Rafferty}, {Rail}, {Raja}, {Rajan}, {Rajbhandari},
  {Rakhmanov}, {Ramirez}, {Ramirez}, {Ramos-Buades}, {Rana}, {Rao},
  {Rapagnani}, {Rapol}, {Ratto}, {Raymond}, {Razzano}, {Read}, {Regimbau},
  {Rei}, {Reid}, {Reitze}, {Rettegno}, {Ricci}, {Richardson}, {Richardson},
  {Richardson}, {Ricker}, {Riemenschneider}, {Riles}, {Rizzo}, {Robertson},
  {Robinet}, {Rocchi}, {Rocha}, {Rodriguez}, {Rodriguez-Soto}, {Rolland},
  {Rollins}, {Roma}, {Romanelli}, {Romano}, {Romel}, {Romero}, {Romero-Shaw},
  {Romie}, {Ronchini}, {Rose}, {Rose}, {Rose}, {Rosell}, {Rosi{\'n}ska},
  {Rosofsky}, {Ross}, {Rowan}, {Rowlinson}, {Roy}, {Roy}, {Ruggi}, {Ryan},
  {Sachdev}, {Sadecki}, {Sadiq}, {Sakellariadou}, {Salafia}, {Salconi},
  {Saleem}, {Samajdar}, {Sanchez}, {Sanchez}, {Sanchez}, {Sanchis-Gual},
  {Sanders}, {Sandles}, {Santiago}, {Santos}, {Saravanan}, {Sarin}, {Sassolas},
  {Sathyaprakash}, {Sauter}, {Savage}, {Savant}, {Sawant}, {Sayah}, {Schaetzl},
  {Schale}, {Scheel}, {Scheuer}, {Schindler-Tyka}, {Schmidt}, {Schnabel},
  {Schofield}, {Sch{\"o}nbeck}, {Schreiber}, {Schulte}, {Schutz}, {Schwarm},
  {Schwartz}, {Scott}, {Scott}, {Seglar-Arroyo}, {Seidel}, {Sellers},
  {Sengupta}, {Sennett}, {Sentenac}, {Sequino}, {Sergeev}, {Setyawati},
  {Shaffer}, {Shahriar}, {Sharifi}, {Sharma}, {Sharma}, {Shawhan}, {Shen},
  {Shikauchi}, {Shink}, {Shoemaker}, {Shoemaker}, {Shukla}, {ShyamSundar},
  {Sieniawska}, {Sigg}, {Singer}, {Singh}, {Singh}, {Singha}, {Singhal},
  {Sintes}, {Sipala}, {Skliris}, {Slagmolen}, {Slaven-Blair}, {Smetana},
  {Smith}, {Smith}, {Somala}, {Son}, {Soni}, {Soni}, {Sorazu}, {Sordini},
  {Sorrentino}, {Sorrentino}, {Soulard}, {Souradeep}, {Sowell}, {Spencer},
  {Spera}, {Srivastava}, {Srivastava}, {Staats}, {Stachie}, {Steer},
  {Steinhoff}, {Steinke}, {Steinlechner}, {Steinlechner}, {Steinmeyer},
  {Stevenson}, {Stolle-McAllister}, {Stops}, {Stover}, {Strain}, {Stratta},
  {Strunk}, {Sturani}, {Stuver}, {S{\"u}dbeck}, {Sudhagar}, {Sudhir}, {Suh},
  {Summerscales}, {Sun}, {Sun}, {Sunil}, {Sur}, {Suresh}, {Sutton}, {Swinkels},
  {Szczepa{\'n}czyk}, {Tacca}, {Tait}, {Talbot}, {Tanasijczuk}, {Tanner},
  {Tao}, {Tapia}, {Tapia San Martin}, {Tasson}, {Taylor}, {Tenorio},
  {Terkowski}, {Thirugnanasambandam}, {Thomas}, {Thomas}, {Thomas}, {Thompson},
  {Thondapu}, {Thorne}, {Thrane}, {Tiwari}, {Tiwari}, {Tiwari}, {Toland},
  {Tolley}, {Tonelli}, {Tornasi}, {Torres-Forn{\'e}}, {Torrie}, {e Melo},
  {T{\"o}yr{\"a}}, {Tran}, {Trapananti}, {Travasso}, {Traylor}, {Tringali},
  {Tripathee}, {Trovato}, {Trudeau}, {Tsai}, {Tsang}, {Tse}, {Tso}, {Tsukada},
  {Tsuna}, {Tsutsui}, {Turconi}, {Ubhi}, {Udall}, {Ueno}, {Ugolini},
  {Unnikrishnan}, {Urban}, {Usman}, {Utina}, {Vahlbruch}, {Vajente}, {Vajpeyi},
  {Valdes}, {Valentini}, {Valsan}, {van Bakel}, {van Beuzekom}, {van den
  Brand}, {Van Den Broeck}, {Vander-Hyde}, {van der Schaaf}, {van Heijningen},
  {Vardaro}, {Vargas}, {Varma}, {Vass}, {Vas{\'u}th}, {Vecchio}, {Vedovato},
  {Veitch}, {Veitch}, {Venkateswara}, {Venneberg}, {Venugopalan}, {Verkindt},
  {Verma}, {Veske}, {Vetrano}, {Vicer{\'e}}, {Viets}, {Vijaykumar},
  {Villa-Ortega}, {Vinet}, {Vitale}, {Vo}, {Vocca}, {Vorvick}, {Vyatchanin},
  {Wade}, {Wade}, {Wade}, {Walet}, {Walker}, {Wallace}, {Wallace}, {Walsh},
  {Wang}, {Wang}, {Wang}, {Wang}, {Ward}, {Warner}, {Was}, {Washington},
  {Watchi}, {Weaver}, {Wei}, {Weinert}, {Weinstein}, {Weiss}, {Wellmann},
  {Wen}, {We{\ss}els}, {Westhouse}, {Wette}, {Whelan}, {White}, {White},
  {Whiting}, {Whittle}, {Wilken}, {Williams}, {Williams}, {Williamson},
  {Willis}, {Willke}, {Wilson}, {Wimmer}, {Winkler}, {Wipf}, {Woan}, {Woehler},
  {Wofford}, {Wong}, {Wrangel}, {Wright}, {Wu}, {Wysocki}, {Xiao}, {Yamamoto},
  {Yang}, {Yang}, {Yang}, {Yap}, {Yeeles}, {Yoon}, {Yu}, {Yu}, {Yuen},
  {Zadro{\.Z}ny}, {Zanolin}, {Zelenova}, {Zendri}, {Zevin}, {Zhang}, {Zhang},
  {Zhang}, {Zhang}, {Zhao}, {Zhao}, {Zheng}, {Zhou}, {Zhou}, {Zhu},
  {Zimmerman}, {Zlochower}, {Zucker}, {Zweizig}, {LIGO Scientific
  Collaboration}, \& {Virgo Collaboration}}]{abbott2021}
{Abbott}, R., {Abbott}, T.~D., {Abraham}, S., {et~al.} 2021, Physical Review X,
  11, 021053, \dodoi{10.1103/PhysRevX.11.021053}

\bibitem[{{Abdul-Masih} {et~al.}(2020){Abdul-Masih}, {Banyard}, {Bodensteiner},
  {Bordier}, {Bowman}, {Dsilva}, {Fabry}, {Hawcroft}, {Mahy}, {Marchant},
  {Raskin}, {Reggiani}, {Shenar}, {Tkachenko}, {Van Winckel}, {Vermeylen}, \&
  {Sana}}]{Abdul-Masih2020}
{Abdul-Masih}, M., {Banyard}, G., {Bodensteiner}, J., {et~al.} 2020, \nat, 580,
  E11, \dodoi{10.1038/s41586-020-2216-x}

\bibitem[{{Ahumada} {et~al.}(2020){Ahumada}, {Prieto}, {Almeida}, {Anders},
  {Anderson}, {Andrews}, {Anguiano}, {Arcodia}, {Armengaud}, {Aubert}, {Avila},
  {Avila-Reese}, {Badenes}, {Balland}, {Barger}, {Barrera-Ballesteros}, {Basu},
  {Bautista}, {Beaton}, {Beers}, {Benavides}, {Bender}, {Bernardi}, {Bershady},
  {Beutler}, {Bidin}, {Bird}, {Bizyaev}, {Blanc}, {Blanton}, {Boquien},
  {Borissova}, {Bovy}, {Brandt}, {Brinkmann}, {Brownstein}, {Bundy}, {Bureau},
  {Burgasser}, {Burtin}, {Cano-D{\'\i}az}, {Capasso}, {Cappellari}, {Carrera},
  {Chabanier}, {Chaplin}, {Chapman}, {Cherinka}, {Chiappini}, {Doohyun Choi},
  {Chojnowski}, {Chung}, {Clerc}, {Coffey}, {Comerford}, {Comparat}, {da
  Costa}, {Cousinou}, {Covey}, {Crane}, {Cunha}, {Ilha}, {Dai}, {Damsted},
  {Darling}, {Davidson}, {Davies}, {Dawson}, {De}, {de la Macorra}, {De Lee},
  {Queiroz}, {Deconto Machado}, {de la Torre}, {Dell'Agli}, {du Mas des
  Bourboux}, {Diamond-Stanic}, {Dillon}, {Donor}, {Drory}, {Duckworth},
  {Dwelly}, {Ebelke}, {Eftekharzadeh}, {Davis Eigenbrot}, {Elsworth},
  {Eracleous}, {Erfanianfar}, {Escoffier}, {Fan}, {Farr},
  {Fern{\'a}ndez-Trincado}, {Feuillet}, {Finoguenov}, {Fofie},
  {Fraser-McKelvie}, {Frinchaboy}, {Fromenteau}, {Fu}, {Galbany}, {Garcia},
  {Garc{\'\i}a-Hern{\'a}ndez}, {Oehmichen}, {Ge}, {Maia}, {Geisler}, {Gelfand},
  {Goddy}, {Gonzalez-Perez}, {Grabowski}, {Green}, {Grier}, {Guo}, {Guy},
  {Harding}, {Hasselquist}, {Hawken}, {Hayes}, {Hearty}, {Hekker}, {Hogg},
  {Holtzman}, {Horta}, {Hou}, {Hsieh}, {Huber}, {Hunt}, {Chitham}, {Imig},
  {Jaber}, {Angel}, {Johnson}, {Jones}, {J{\"o}nsson}, {Jullo}, {Kim},
  {Kinemuchi}, {Kirkpatrick}, {Kite}, {Klaene}, {Kneib}, {Kollmeier}, {Kong},
  {Kounkel}, {Krishnarao}, {Lacerna}, {Lan}, {Lane}, {Law}, {Le Goff}, {Leung},
  {Lewis}, {Li}, {Lian}, {Lin}, {Long}, {Longa-Pe{\~n}a}, {Lundgren}, {Lyke},
  {Ted Mackereth}, {MacLeod}, {Majewski}, {Manchado}, {Maraston}, {Martini},
  {Masseron}, {Masters}, {Mathur}, {McDermid}, {Merloni}, {Merrifield},
  {M{\'e}sz{\'a}ros}, {Miglio}, {Minniti}, {Minsley}, {Miyaji}, {Mohammad},
  {Mosser}, {Mueller}, {Muna}, {Mu{\~n}oz-Guti{\'e}rrez}, {Myers}, {Nadathur},
  {Nair}, {Nandra}, {do Nascimento}, {Nevin}, {Newman}, {Nidever}, {Nitschelm},
  {Noterdaeme}, {O'Connell}, {Olmstead}, {Oravetz}, {Oravetz}, {Osorio},
  {Pace}, {Padilla}, {Palanque-Delabrouille}, {Palicio}, {Pan}, {Pan},
  {Parker}, {Paviot}, {Peirani}, {Ram{\'r}ez}, {Penny}, {Percival},
  {Perez-Fournon}, {P{\'e}rez-R{\`a}fols}, {Petitjean}, {Pieri},
  {Pinsonneault}, {Poovelil}, {Povick}, {Prakash}, {Price-Whelan}, {Raddick},
  {Raichoor}, {Ray}, {Rembold}, {Rezaie}, {Riffel}, {Riffel}, {Rix}, {Robin},
  {Roman-Lopes}, {Rom{\'a}n-Z{\'u}{\~n}iga}, {Rose}, {Ross}, {Rossi},
  {Rowlands}, {Rubin}, {Salvato}, {S{\'a}nchez}, {S{\'a}nchez-Menguiano},
  {S{\'a}nchez-Gallego}, {Sayres}, {Schaefer}, {Schiavon}, {Schimoia},
  {Schlafly}, {Schlegel}, {Schneider}, {Schultheis}, {Schwope}, {Seo},
  {Serenelli}, {Shafieloo}, {Shamsi}, {Shao}, {Shen}, {Shetrone}, {Shirley},
  {Aguirre}, {Simon}, {Skrutskie}, {Slosar}, {Smethurst}, {Sobeck}, {Sodi},
  {Souto}, {Stark}, {Stassun}, {Steinmetz}, {Stello}, {Stermer},
  {Storchi-Bergmann}, {Streblyanska}, {Stringfellow}, {Stutz}, {Su{\'a}rez},
  {Sun}, {Taghizadeh-Popp}, {Talbot}, {Tayar}, {Thakar}, {Theriault}, {Thomas},
  {Thomas}, {Tinker}, {Tojeiro}, {Toledo}, {Tremonti}, {Troup}, {Tuttle},
  {Unda-Sanzana}, {Valentini}, {Vargas-Gonz{\'a}lez}, {Vargas-Maga{\~n}a},
  {V{\'a}zquez-Mata}, {Vivek}, {Wake}, {Wang}, {Weaver}, {Weijmans}, {Wild},
  {Wilson}, {Wilson}, {Wolthuis}, {Wood-Vasey}, {Yan}, {Yang}, {Y{\`e}che},
  {Zamora}, {Zarrouk}, {Zasowski}, {Zhang}, {Zhao}, {Zhao}, {Zheng}, {Zheng},
  {Zhu}, \& {Zou}}]{Ahumada20}
{Ahumada}, R., {Prieto}, C.~A., {Almeida}, A., {et~al.} 2020, \apjs, 249, 3,
  \dodoi{10.3847/1538-4365/ab929e}

\bibitem[{{Amarsi} \& {Asplund}(2017)}]{amarsi2017}
{Amarsi}, A.~M., \& {Asplund}, M. 2017, \mnras, 464, 264,
  \dodoi{10.1093/mnras/stw2445}

\bibitem[{{Amarsi} {et~al.}(2016){Amarsi}, {Lind}, {Asplund}, {Barklem}, \&
  {Collet}}]{amarsi2016}
{Amarsi}, A.~M., {Lind}, K., {Asplund}, M., {Barklem}, P.~S., \& {Collet}, R.
  2016, \mnras, 463, 1518, \dodoi{10.1093/mnras/stw2077}

\bibitem[{{Amarsi} {et~al.}(2019){Amarsi}, {Nissen}, \&
  {Sk{\'u}lad{\'o}ttir}}]{amarsi2019}
{Amarsi}, A.~M., {Nissen}, P.~E., \& {Sk{\'u}lad{\'o}ttir}, {\'A}. 2019, \aap,
  630, A104, \dodoi{10.1051/0004-6361/201936265}

\bibitem[{{Amarsi} {et~al.}(2020){Amarsi}, {Lind}, {Osorio}, {Nordlander},
  {Bergemann}, {Reggiani}, {Wang}, {Buder}, {Asplund}, {Barklem}, {Wehrhahn},
  {Sk{\'u}lad{\'o}ttir}, {Kobayashi}, {Karakas}, {Gao}, {Bland-Hawthorn}, {de
  Silva}, {Kos}, {Lewis}, {Martell}, {Sharma}, {Simpson}, {Zucker},
  {{\v{C}}otar}, {Horner}, \& {Galah Collaboration}}]{amarsi2020}
{Amarsi}, A.~M., {Lind}, K., {Osorio}, Y., {et~al.} 2020, \aap, 642, A62,
  \dodoi{10.1051/0004-6361/202038650}

\bibitem[{{Anderson} {et~al.}(2017){Anderson}, {Lai}, \&
  {Storch}}]{Anderson2017}
{Anderson}, K.~R., {Lai}, D., \& {Storch}, N.~I. 2017, \mnras, 467, 3066,
  \dodoi{10.1093/mnras/stx293}

\bibitem[{{Andrews} {et~al.}(2022){Andrews}, {Taggart}, \&
  {Foley}}]{Andrews2022}
{Andrews}, J.~J., {Taggart}, K., \& {Foley}, R. 2022, arXiv e-prints,
  arXiv:2207.00680.
\newblock \doarXiv{2207.00680}

\bibitem[{{Antonini} \& {Rasio}(2016)}]{Antonini_Rasio2016}
{Antonini}, F., \& {Rasio}, F.~A. 2016, \apj, 831, 187,
  \dodoi{10.3847/0004-637X/831/2/187}

\bibitem[{{Asplund} {et~al.}(2021){Asplund}, {Amarsi}, \&
  {Grevesse}}]{asplund2021}
{Asplund}, M., {Amarsi}, A.~M., \& {Grevesse}, N. 2021, \aap, 653, A141,
  \dodoi{10.1051/0004-6361/202140445}

\bibitem[{{Bavera} {et~al.}(2020){Bavera}, {Fragos}, {Qin}, {Zapartas},
  {Neijssel}, {Mandel}, {Batta}, {Gaebel}, {Kimball}, \&
  {Stevenson}}]{Bavera2020}
{Bavera}, S.~S., {Fragos}, T., {Qin}, Y., {et~al.} 2020, \aap, 635, A97,
  \dodoi{10.1051/0004-6361/201936204}

\bibitem[{{Belczynski} {et~al.}(2016){Belczynski}, {Holz}, {Bulik}, \&
  {O'Shaughnessy}}]{Belczynski2016}
{Belczynski}, K., {Holz}, D.~E., {Bulik}, T., \& {O'Shaughnessy}, R. 2016,
  \nat, 534, 512, \dodoi{10.1038/nature18322}

\bibitem[{{Berg} {et~al.}(2020){Berg}, {Pogge}, {Skillman}, {Croxall},
  {Moustakas}, {Rogers}, \& {Sun}}]{Berg2020}
{Berg}, D.~A., {Pogge}, R.~W., {Skillman}, E.~D., {et~al.} 2020, \apj, 893, 96,
  \dodoi{10.3847/1538-4357/ab7eab}

\bibitem[{{Bernstein} {et~al.}(2003){Bernstein}, {Shectman}, {Gunnels},
  {Mochnacki}, \& {Athey}}]{Bernstein2003}
{Bernstein}, R., {Shectman}, S.~A., {Gunnels}, S.~M., {Mochnacki}, S., \&
  {Athey}, A.~E. 2003, in Society of Photo-Optical Instrumentation Engineers
  (SPIE) Conference Series, Vol. 4841, Instrument Design and Performance for
  Optical/Infrared Ground-based Telescopes, ed. M.~{Iye} \& A.~F.~M.
  {Moorwood}, 1694--1704, \dodoi{10.1117/12.461502}

\bibitem[{{Binnendijk}(1960)}]{Binnendijk1960}
{Binnendijk}, L. 1960, {Properties of double stars; a survey of parallaxes and
  orbits.}

\bibitem[{{Blackwell-Whitehead} {et~al.}(2005){Blackwell-Whitehead}, {Toner},
  {Hibbert}, {Webb}, \& {Ivarsson}}]{Blackwell-Whitehead2005}
{Blackwell-Whitehead}, R.~J., {Toner}, A., {Hibbert}, A., {Webb}, J., \&
  {Ivarsson}, S. 2005, \mnras, 364, 705,
  \dodoi{10.1111/j.1365-2966.2005.09597.x}

\bibitem[{{Bodensteiner} {et~al.}(2020){Bodensteiner}, {Shenar}, {Mahy},
  {Fabry}, {Marchant}, {Abdul-Masih}, {Banyard}, {Bowman}, {Dsilva}, {Frost},
  {Hawcroft}, {Reggiani}, \& {Sana}}]{Bodensteiner2020}
{Bodensteiner}, J., {Shenar}, T., {Mahy}, L., {et~al.} 2020, \aap, 641, A43,
  \dodoi{10.1051/0004-6361/202038682}

\bibitem[{{Borkovits} {et~al.}(2015){Borkovits}, {Rappaport}, {Hajdu}, \&
  {Sztakovics}}]{Borkovits2015}
{Borkovits}, T., {Rappaport}, S., {Hajdu}, T., \& {Sztakovics}, J. 2015,
  \mnras, 448, 946, \dodoi{10.1093/mnras/stv015}

\bibitem[{{Bovy}(2015)}]{Bovy2015}
{Bovy}, J. 2015, \apjs, 216, 29, \dodoi{10.1088/0067-0049/216/2/29}

\bibitem[{{Brandt} {et~al.}(2021){Brandt}, {Dupuy}, {Li}, {Brandt}, {Zeng},
  {Michalik}, {Bardalez Gagliuffi}, \& {Raposo-Pulido}}]{Brandt2021AJ}
{Brandt}, T.~D., {Dupuy}, T.~J., {Li}, Y., {et~al.} 2021, \aj, 162, 186,
  \dodoi{10.3847/1538-3881/ac042e}

\bibitem[{{Breivik} {et~al.}(2017){Breivik}, {Chatterjee}, \&
  {Larson}}]{Breivik2017}
{Breivik}, K., {Chatterjee}, S., \& {Larson}, S.~L. 2017, \apjl, 850, L13,
  \dodoi{10.3847/2041-8213/aa97d5}

\bibitem[{{Bresolin} {et~al.}(2012){Bresolin}, {Kennicutt}, \&
  {Ryan-Weber}}]{Bresolin2012}
{Bresolin}, F., {Kennicutt}, R.~C., \& {Ryan-Weber}, E. 2012, \apj, 750, 122,
  \dodoi{10.1088/0004-637X/750/2/122}

\bibitem[{{Brewer} {et~al.}(2016){Brewer}, {Fischer}, {Valenti}, \&
  {Piskunov}}]{Brewer16}
{Brewer}, J.~M., {Fischer}, D.~A., {Valenti}, J.~A., \& {Piskunov}, N. 2016,
  \apjs, 225, 32, \dodoi{10.3847/0067-0049/225/2/32}

\bibitem[{{Buchner} {et~al.}(2014){Buchner}, {Georgakakis}, {Nandra}, {Hsu},
  {Rangel}, {Brightman}, {Merloni}, {Salvato}, {Donley}, \&
  {Kocevski}}]{buchner2014}
{Buchner}, J., {Georgakakis}, A., {Nandra}, K., {et~al.} 2014, \aap, 564, A125,
  \dodoi{10.1051/0004-6361/201322971}

\bibitem[{{Butler} {et~al.}(1996){Butler}, {Marcy}, {Williams}, {McCarthy},
  {Dosanjh}, \& {Vogt}}]{Butler96}
{Butler}, R.~P., {Marcy}, G.~W., {Williams}, E., {et~al.} 1996, \pasp, 108,
  500, \dodoi{10.1086/133755}

\bibitem[{{Capitanio} {et~al.}(2017){Capitanio}, {Lallement}, {Vergely},
  {Elyajouri}, \& {Monreal-Ibero}}]{capitanio2017}
{Capitanio}, L., {Lallement}, R., {Vergely}, J.~L., {Elyajouri}, M., \&
  {Monreal-Ibero}, A. 2017, \aap, 606, A65, \dodoi{10.1051/0004-6361/201730831}

\bibitem[{{Casali} {et~al.}(2020){Casali}, {Spina}, {Magrini}, {Karakas},
  {Kobayashi}, {Casey}, {Feltzing}, {Van der Swaelmen}, {Tsantaki},
  {Jofr{\'e}}, {Bragaglia}, {Feuillet}, {Bensby}, {Biazzo}, {Gonneau},
  {Tautvai{\v{s}}ien{\.{e}}}, {Baratella}, {Roccatagliata}, {Pancino}, {Sousa},
  {Adibekyan}, {Martell}, {Bayo}, {Jackson}, {Jeffries}, {Gilmore}, {Randich},
  {Alfaro}, {Koposov}, {Korn}, {Recio-Blanco}, {Smiljanic}, {Franciosini},
  {Hourihane}, {Monaco}, {Morbidelli}, {Sacco}, {Worley}, \&
  {Zaggia}}]{Casali20}
{Casali}, G., {Spina}, L., {Magrini}, L., {et~al.} 2020, \aap, 639, A127,
  \dodoi{10.1051/0004-6361/202038055}

\bibitem[{{Casares} {et~al.}(2017){Casares}, {Jonker}, \&
  {Israelian}}]{Casares2017}
{Casares}, J., {Jonker}, P.~G., \& {Israelian}, G. 2017, in Handbook of
  Supernovae, ed. A.~W. {Alsabti} \& P.~{Murdin}, 1499,
  \dodoi{10.1007/978-3-319-21846-5_111}

\bibitem[{{Casey}(2014)}]{Casey2014}
{Casey}, A.~R. 2014, PhD thesis, Australian National University, Canberra

\bibitem[{{Castelli} \& {Kurucz}(2003)}]{Castelli03}
{Castelli}, F., \& {Kurucz}, R.~L. 2003, in Modelling of Stellar Atmospheres,
  ed. N.~{Piskunov}, W.~W. {Weiss}, \& D.~F. {Gray}, Vol. 210, A20.
\newblock \doarXiv{astro-ph/0405087}

\bibitem[{{Chakrabarti} {et~al.}(2021){Chakrabarti}, {Chang}, {Lam},
  {Vigeland}, \& {Quillen}}]{Chakrabarti2021}
{Chakrabarti}, S., {Chang}, P., {Lam}, M.~T., {Vigeland}, S.~J., \& {Quillen},
  A.~C. 2021, \apjl, 907, L26, \dodoi{10.3847/2041-8213/abd635}

\bibitem[{{Chakrabarti} {et~al.}(2017){Chakrabarti}, {Chang}, {O'Shaughnessy},
  {Brooks}, {Shen}, {Bellovary}, {Gladysz}, \& {Belczynski}}]{Chakrabarti2017}
{Chakrabarti}, S., {Chang}, P., {O'Shaughnessy}, R., {et~al.} 2017, \apjl, 850,
  L4, \dodoi{10.3847/2041-8213/aa9655}

\bibitem[{{Chakrabarti} {et~al.}(2018){Chakrabarti}, {Dell}, {Graur},
  {Filippenko}, {Lewis}, \& {McKee}}]{Chakrabarti2018}
{Chakrabarti}, S., {Dell}, B., {Graur}, O., {et~al.} 2018, \apjl, 863, L1,
  \dodoi{10.3847/2041-8213/aad0a4}

\bibitem[{{Chakrabarti} {et~al.}(2022){Chakrabarti}, {Stevens}, {Wright},
  {Rafikov}, {Chang}, {Beatty}, \& {Huber}}]{Chakrabarti2022}
{Chakrabarti}, S., {Stevens}, D.~J., {Wright}, J., {et~al.} 2022, \apjl, 928,
  L17, \dodoi{10.3847/2041-8213/ac5c43}

\bibitem[{{Chang}(2009)}]{Chang2009}
{Chang}, P. 2009, \mnras, 393, 224, \dodoi{10.1111/j.1365-2966.2008.14202.x}

\bibitem[{{Chawla} {et~al.}(2022){Chawla}, {Chatterjee}, {Breivik}, {Moorthy},
  {Andrews}, \& {Sanderson}}]{Chawla2022}
{Chawla}, C., {Chatterjee}, S., {Breivik}, K., {et~al.} 2022, \apj, 931, 107,
  \dodoi{10.3847/1538-4357/ac60a5}

\bibitem[{{Choi} {et~al.}(2016){Choi}, {Dotter}, {Conroy}, {Cantiello},
  {Paxton}, \& {Johnson}}]{choi2016}
{Choi}, J., {Dotter}, A., {Conroy}, C., {et~al.} 2016, \apj, 823, 102,
  \dodoi{10.3847/0004-637X/823/2/102}

\bibitem[{{Chubak} {et~al.}(2012){Chubak}, {Marcy}, {Fischer}, {Howard},
  {Isaacson}, {Johnson}, \& {Wright}}]{Chubak12}
{Chubak}, C., {Marcy}, G., {Fischer}, D.~A., {et~al.} 2012, arXiv e-prints,
  arXiv:1207.6212.
\newblock \doarXiv{1207.6212}

\bibitem[{{Chulkov}(2021)}]{Chulkov2021}
{Chulkov}, D. 2021, \mnras, 501, 769, \dodoi{10.1093/mnras/staa3601}

\bibitem[{{Cooper} {et~al.}(2012){Cooper}, {Newman}, {Davis}, {Finkbeiner}, \&
  {Gerke}}]{Cooper12}
{Cooper}, M.~C., {Newman}, J.~A., {Davis}, M., {Finkbeiner}, D.~P., \& {Gerke},
  B.~F. 2012, {spec2d: DEEP2 DEIMOS Spectral Pipeline}, Astrophysics Source
  Code Library, record ascl:1203.003.
\newblock \doeprint{1203.003}

\bibitem[{{Corral-Santana} {et~al.}(2016){Corral-Santana}, {Casares},
  {Mu{\~n}oz-Darias}, {Bauer}, {Mart{\'\i}nez-Pais}, \&
  {Russell}}]{Corral-Santana2016}
{Corral-Santana}, J.~M., {Casares}, J., {Mu{\~n}oz-Darias}, T., {et~al.} 2016,
  \aap, 587, A61, \dodoi{10.1051/0004-6361/201527130}

\bibitem[{{Corral-Santana} {et~al.}(2013){Corral-Santana}, {Casares},
  {Mu{\~n}oz-Darias}, {Rodr{\'\i}guez-Gil}, {Shahbaz}, {Torres}, {Zurita}, \&
  {Tyndall}}]{Corral2013}
---. 2013, Science, 339, 1048, \dodoi{10.1126/science.1228222}

\bibitem[{{Cui} {et~al.}(2012){Cui}, {Zhao}, {Chu}, {Li}, {Li}, {Zhang}, {Su},
  {Yao}, {Wang}, {Xing}, {Li}, {Zhu}, {Wang}, {Gu}, {Luo}, {Xu}, {Zhang},
  {Liu}, {Zhang}, {Yang}, {Cao}, {Chen}, {Chen}, {Chen}, {Chen}, {Chu}, {Feng},
  {Gong}, {Hou}, {Hu}, {Hu}, {Hu}, {Jia}, {Jiang}, {Jiang}, {Jiang}, {Jin},
  {Li}, {Li}, {Li}, {Liu}, {Liu}, {Lu}, {Mao}, {Men}, {Qi}, {Qi}, {Shi},
  {Tang}, {Tao}, {Wang}, {Wang}, {Wang}, {Wang}, {Wang}, {Wang}, {Wang},
  {Wang}, {Wang}, {Wang}, {Wang}, {Wang}, {Xu}, {Xu}, {Yang}, {Yu}, {Yuan},
  {Yuan}, {Zhai}, {Zhang}, {Zhang}, {Zhang}, {Zhao}, {Zhou}, {Zhou}, {Zhu}, \&
  {Zou}}]{Cui12}
{Cui}, X.-Q., {Zhao}, Y.-H., {Chu}, Y.-Q., {et~al.} 2012, Research in Astronomy
  and Astrophysics, 12, 1197, \dodoi{10.1088/1674-4527/12/9/003}

\bibitem[{{Cutri} {et~al.}(2003){Cutri}, {Skrutskie}, {van Dyk}, {Beichman},
  {Carpenter}, {Chester}, {Cambresy}, {Evans}, {Fowler}, {Gizis}, {Howard},
  {Huchra}, {Jarrett}, {Kopan}, {Kirkpatrick}, {Light}, {Marsh}, {McCallon},
  {Schneider}, {Stiening}, {Sykes}, {Weinberg}, {Wheaton}, {Wheelock}, \&
  {Zacarias}}]{Cutri03}
{Cutri}, R.~M., {Skrutskie}, M.~F., {van Dyk}, S., {et~al.} 2003, {2MASS All
  Sky Catalog of point sources.}

\bibitem[{{Cutri} {et~al.}(2021){Cutri}, {Wright}, {Conrow}, {Fowler},
  {Eisenhardt}, {Grillmair}, {Kirkpatrick}, {Masci}, {McCallon}, {Wheelock},
  {Fajardo-Acosta}, {Yan}, {Benford}, {Harbut}, {Jarrett}, {Lake}, {Leisawitz},
  {Ressler}, {Stanford}, {Tsai}, {Liu}, {Helou}, {Mainzer}, {Gettngs},
  {Gonzalez}, {Hoffman}, {Marsh}, {Padgett}, {Skrutskie}, {Beck}, {Papin}, \&
  {Wittman}}]{Cutri14}
{Cutri}, R.~M., {Wright}, E.~L., {Conrow}, T., {et~al.} 2021, VizieR Online
  Data Catalog, II/328

\bibitem[{{de Mink} \& {Mandel}(2016)}]{DeMink_Mandel2016}
{de Mink}, S.~E., \& {Mandel}, I. 2016, \mnras, 460, 3545,
  \dodoi{10.1093/mnras/stw1219}

\bibitem[{{Dotter}(2016)}]{dotter2016}
{Dotter}, A. 2016, \apjs, 222, 8, \dodoi{10.3847/0067-0049/222/1/8}

\bibitem[{{El-Badry} \& {Burdge}(2022)}]{El-BadryBurdge2022}
{El-Badry}, K., \& {Burdge}, K.~B. 2022, \mnras, 511, 24,
  \dodoi{10.1093/mnrasl/slab135}

\bibitem[{{El-Badry} {et~al.}(2022{\natexlab{a}}){El-Badry}, {Burdge}, \&
  {Mr{\'o}z}}]{El-BadryBurdgeMroz2022}
{El-Badry}, K., {Burdge}, K.~B., \& {Mr{\'o}z}, P. 2022{\natexlab{a}}, \mnras,
  511, 3089, \dodoi{10.1093/mnras/stac274}

\bibitem[{{El-Badry} {et~al.}(2022{\natexlab{b}}){El-Badry}, {Conroy},
  {Fuller}, {Kiman}, {van Roestel}, {Rodriguez}, \&
  {Burdge}}]{ElBadry22magbraking}
{El-Badry}, K., {Conroy}, C., {Fuller}, J., {et~al.} 2022{\natexlab{b}},
  \mnras, 517, 4916, \dodoi{10.1093/mnras/stac2945}

\bibitem[{{El-Badry} \& {Quataert}(2020)}]{El-Badry_Quataert20202}
{El-Badry}, K., \& {Quataert}, E. 2020, \mnras, 493, L22,
  \dodoi{10.1093/mnrasl/slaa004}

\bibitem[{{El-Badry} \& {Quataert}(2021)}]{El-Badry2021}
---. 2021, \mnras, 502, 3436, \dodoi{10.1093/mnras/stab285}

\bibitem[{{El-Badry} \& {Rix}(2022)}]{ElBadryRix2022}
{El-Badry}, K., \& {Rix}, H.-W. 2022, \mnras, 515, 1266,
  \dodoi{10.1093/mnras/stac1797}

\bibitem[{{El-Badry} {et~al.}(2022{\natexlab{c}}){El-Badry}, {Seeburger},
  {Jayasinghe}, {Rix}, {Almada}, {Conroy}, {Price-Whelan}, \&
  {Burdge}}]{ElBadry2022}
{El-Badry}, K., {Seeburger}, R., {Jayasinghe}, T., {et~al.} 2022{\natexlab{c}},
  \mnras, 512, 5620, \dodoi{10.1093/mnras/stac815}

\bibitem[{{El-Badry} {et~al.}(2023{\natexlab{a}}){El-Badry}, {Rix}, {Quataert},
  {Howard}, {Isaacson}, {Fuller}, {Hawkins}, {Breivik}, {Wong}, {Rodriguez},
  {Conroy}, {Shahaf}, {Mazeh}, {Arenou}, {Burdge}, {Bashi}, {Faigler}, {Weisz},
  {Seeburger}, {Almada Monter}, \& {Wojno}}]{ElBadry2022Disc}
{El-Badry}, K., {Rix}, H.-W., {Quataert}, E., {et~al.} 2023{\natexlab{a}},
  \mnras, 518, 1057, \dodoi{10.1093/mnras/stac3140}

\bibitem[{{El-Badry} {et~al.}(2023{\natexlab{b}}){El-Badry}, {Rix}, {Cendes},
  {Rodriguez}, {Conroy}, {Quataert}, {Hawkins}, {Zari}, {Hobson}, {Breivik},
  {Rau}, {Berger}, {Shahaf}, {Seeburger}, {Burdge}, {Latham}, {Buchhave},
  {Bieryla}, {Bashi}, {Mazeh}, \& {Faigler}}]{ElBadry2023}
{El-Badry}, K., {Rix}, H.-W., {Cendes}, Y., {et~al.} 2023{\natexlab{b}}, arXiv
  e-prints, arXiv:2302.07880, \dodoi{10.48550/arXiv.2302.07880}

\bibitem[{{Eldridge} {et~al.}(2020){Eldridge}, {Stanway}, {Breivik}, {Casey},
  {Steeghs}, \& {Stevance}}]{Eldridge2020}
{Eldridge}, J.~J., {Stanway}, E.~R., {Breivik}, K., {et~al.} 2020, \mnras, 495,
  2786, \dodoi{10.1093/mnras/staa1324}

\bibitem[{{Faber} {et~al.}(2003){Faber}, {Phillips}, {Kibrick}, {Alcott},
  {Allen}, {Burrous}, {Cantrall}, {Clarke}, {Coil}, {Cowley}, {Davis}, {Deich},
  {Dietsch}, {Gilmore}, {Harper}, {Hilyard}, {Lewis}, {McVeigh}, {Newman},
  {Osborne}, {Schiavon}, {Stover}, {Tucker}, {Wallace}, {Wei}, {Wirth}, \&
  {Wright}}]{Faber03}
{Faber}, S.~M., {Phillips}, A.~C., {Kibrick}, R.~I., {et~al.} 2003, in Society
  of Photo-Optical Instrumentation Engineers (SPIE) Conference Series, Vol.
  4841, Instrument Design and Performance for Optical/Infrared Ground-based
  Telescopes, ed. M.~{Iye} \& A.~F.~M. {Moorwood}, 1657--1669,
  \dodoi{10.1117/12.460346}

\bibitem[{{Feroz} \& {Hobson}(2008)}]{feroz2008}
{Feroz}, F., \& {Hobson}, M.~P. 2008, \mnras, 384, 449,
  \dodoi{10.1111/j.1365-2966.2007.12353.x}

\bibitem[{{Feroz} {et~al.}(2009){Feroz}, {Hobson}, \& {Bridges}}]{feroz2009}
{Feroz}, F., {Hobson}, M.~P., \& {Bridges}, M. 2009, \mnras, 398, 1601,
  \dodoi{10.1111/j.1365-2966.2009.14548.x}

\bibitem[{{Feroz} {et~al.}(2019){Feroz}, {Hobson}, {Cameron}, \&
  {Pettitt}}]{feroz2019}
{Feroz}, F., {Hobson}, M.~P., {Cameron}, E., \& {Pettitt}, A.~N. 2019, The Open
  Journal of Astrophysics, 2, 10, \dodoi{10.21105/astro.1306.2144}

\bibitem[{{Fitzpatrick}(1999)}]{Fitzpatrick99}
{Fitzpatrick}, E.~L. 1999, \pasp, 111, 63, \dodoi{10.1086/316293}

\bibitem[{{Flewelling} {et~al.}(2020){Flewelling}, {Magnier}, {Chambers},
  {Heasley}, {Holmberg}, {Huber}, {Sweeney}, {Waters}, {Calamida}, {Casertano},
  {Chen}, {Farrow}, {Hasinger}, {Henderson}, {Long}, {Metcalfe}, {Narayan},
  {Nieto-Santisteban}, {Norberg}, {Rest}, {Saglia}, {Szalay}, {Thakar},
  {Tonry}, {Valenti}, {Werner}, {White}, {Denneau}, {Draper}, {Hodapp},
  {Jedicke}, {Kaiser}, {Kudritzki}, {Price}, {Wainscoat}, {Chastel}, {McLean},
  {Postman}, \& {Shiao}}]{Flewelling20}
{Flewelling}, H.~A., {Magnier}, E.~A., {Chambers}, K.~C., {et~al.} 2020, \apjs,
  251, 7, \dodoi{10.3847/1538-4365/abb82d}

\bibitem[{{Foreman-Mackey} {et~al.}(2021{\natexlab{a}}){Foreman-Mackey},
  {Luger}, {Agol}, {Barclay}, {Bouma}, {Brandt}, {Czekala}, {David}, {Dong},
  {Gilbert}, {Gordon}, {Hedges}, {Hey}, {Morris}, {Price-Whelan}, \&
  {Savel}}]{exoplanetjoss}
{Foreman-Mackey}, D., {Luger}, R., {Agol}, E., {et~al.} 2021{\natexlab{a}}, The
  Journal of Open Source Software, 6, 3285, \dodoi{10.21105/joss.03285}

\bibitem[{{Foreman-Mackey} {et~al.}(2021{\natexlab{b}}){Foreman-Mackey},
  {Luger}, {Agol}, {Barclay}, {Bouma}, {Brandt}, {Czekala}, {David}, {Dong},
  {Gilbert}, {Gordon}, {Hedges}, {Hey}, {Morris}, {Price-Whelan}, \&
  {Savel}}]{exoplanetzenodo}
---. 2021{\natexlab{b}}, {exoplanet: Gradient-based probabilistic inference for
  exoplanet data \& other astronomical time series}, 0.5.1, Zenodo,  Zenodo,
  \dodoi{10.5281/zenodo.1998447}

\bibitem[{{Frost} {et~al.}(2022){Frost}, {Bodensteiner}, {Rivinius}, {Baade},
  {Merand}, {Selman}, {Abdul-Masih}, {Banyard}, {Bordier}, {Dsilva},
  {Hawcroft}, {Mahy}, {Reggiani}, {Shenar}, {Cabezas}, {Hadrava}, {Heida},
  {Klement}, \& {Sana}}]{Frost2022}
{Frost}, A.~J., {Bodensteiner}, J., {Rivinius}, T., {et~al.} 2022, \aap, 659,
  L3, \dodoi{10.1051/0004-6361/202143004}

\bibitem[{{Fukugita} {et~al.}(1996){Fukugita}, {Ichikawa}, {Gunn}, {Doi},
  {Shimasaku}, \& {Schneider}}]{Fukugita1996}
{Fukugita}, M., {Ichikawa}, T., {Gunn}, J.~E., {et~al.} 1996, \aj, 111, 1748,
  \dodoi{10.1086/117915}

\bibitem[{{Fulton} {et~al.}(2015){Fulton}, {Weiss}, {Sinukoff}, {Isaacson},
  {Howard}, {Marcy}, {Henry}, {Holden}, \& {Kibrick}}]{Fulton15}
{Fulton}, B.~J., {Weiss}, L.~M., {Sinukoff}, E., {et~al.} 2015, \apj, 805, 175,
  \dodoi{10.1088/0004-637X/805/2/175}

\bibitem[{{Gaia Collaboration} {et~al.}(2016){Gaia Collaboration}, {Prusti},
  {de Bruijne}, {Brown}, {Vallenari}, {Babusiaux}, {Bailer-Jones}, {Bastian},
  {Biermann}, {Evans}, {Eyer}, {Jansen}, {Jordi}, {Klioner}, {Lammers},
  {Lindegren}, {Luri}, {Mignard}, {Milligan}, {Panem}, {Poinsignon},
  {Pourbaix}, {Randich}, {Sarri}, {Sartoretti}, {Siddiqui}, {Soubiran},
  {Valette}, {van Leeuwen}, {Walton}, {Aerts}, {Arenou}, {Cropper}, {Drimmel},
  {H{\o}g}, {Katz}, {Lattanzi}, {O'Mullane}, {Grebel}, {Holland}, {Huc},
  {Passot}, {Bramante}, {Cacciari}, {Casta{\~n}eda}, {Chaoul}, {Cheek}, {De
  Angeli}, {Fabricius}, {Guerra}, {Hern{\'a}ndez}, {Jean-Antoine-Piccolo},
  {Masana}, {Messineo}, {Mowlavi}, {Nienartowicz}, {Ord{\'o}{\~n}ez-Blanco},
  {Panuzzo}, {Portell}, {Richards}, {Riello}, {Seabroke}, {Tanga},
  {Th{\'e}venin}, {Torra}, {Els}, {Gracia-Abril}, {Comoretto},
  {Garcia-Reinaldos}, {Lock}, {Mercier}, {Altmann}, {Andrae}, {Astraatmadja},
  {Bellas-Velidis}, {Benson}, {Berthier}, {Blomme}, {Busso}, {Carry},
  {Cellino}, {Clementini}, {Cowell}, {Creevey}, {Cuypers}, {Davidson}, {De
  Ridder}, {de Torres}, {Delchambre}, {Dell'Oro}, {Ducourant}, {Fr{\'e}mat},
  {Garc{\'\i}a-Torres}, {Gosset}, {Halbwachs}, {Hambly}, {Harrison}, {Hauser},
  {Hestroffer}, {Hodgkin}, {Huckle}, {Hutton}, {Jasniewicz}, {Jordan},
  {Kontizas}, {Korn}, {Lanzafame}, {Manteiga}, {Moitinho}, {Muinonen},
  {Osinde}, {Pancino}, {Pauwels}, {Petit}, {Recio-Blanco}, {Robin}, {Sarro},
  {Siopis}, {Smith}, {Smith}, {Sozzetti}, {Thuillot}, {van Reeven}, {Viala},
  {Abbas}, {Abreu Aramburu}, {Accart}, {Aguado}, {Allan}, {Allasia},
  {Altavilla}, {{\'A}lvarez}, {Alves}, {Anderson}, {Andrei}, {Anglada Varela},
  {Antiche}, {Antoja}, {Ant{\'o}n}, {Arcay}, {Atzei}, {Ayache}, {Bach},
  {Baker}, {Balaguer-N{\'u}{\~n}ez}, {Barache}, {Barata}, {Barbier}, {Barblan},
  {Baroni}, {Barrado y Navascu{\'e}s}, {Barros}, {Barstow}, {Becciani},
  {Bellazzini}, {Bellei}, {Bello Garc{\'\i}a}, {Belokurov}, {Bendjoya},
  {Berihuete}, {Bianchi}, {Bienaym{\'e}}, {Billebaud}, {Blagorodnova},
  {Blanco-Cuaresma}, {Boch}, {Bombrun}, {Borrachero}, {Bouquillon}, {Bourda},
  {Bouy}, {Bragaglia}, {Breddels}, {Brouillet}, {Br{\"u}semeister},
  {Bucciarelli}, {Budnik}, {Burgess}, {Burgon}, {Burlacu}, {Busonero}, {Buzzi},
  {Caffau}, {Cambras}, {Campbell}, {Cancelliere}, {Cantat-Gaudin}, {Carlucci},
  {Carrasco}, {Castellani}, {Charlot}, {Charnas}, {Charvet}, {Chassat},
  {Chiavassa}, {Clotet}, {Cocozza}, {Collins}, {Collins}, {Costigan}, {Crifo},
  {Cross}, {Crosta}, {Crowley}, {Dafonte}, {Damerdji}, {Dapergolas}, {David},
  {David}, {De Cat}, {de Felice}, {de Laverny}, {De Luise}, {De March}, {de
  Martino}, {de Souza}, {Debosscher}, {del Pozo}, {Delbo}, {Delgado},
  {Delgado}, {di Marco}, {Di Matteo}, {Diakite}, {Distefano}, {Dolding}, {Dos
  Anjos}, {Drazinos}, {Dur{\'a}n}, {Dzigan}, {Ecale}, {Edvardsson}, {Enke},
  {Erdmann}, {Escolar}, {Espina}, {Evans}, {Eynard Bontemps}, {Fabre},
  {Fabrizio}, {Faigler}, {Falc{\~a}o}, {Farr{\`a}s Casas}, {Faye}, {Federici},
  {Fedorets}, {Fern{\'a}ndez-Hern{\'a}ndez}, {Fernique}, {Fienga}, {Figueras},
  {Filippi}, {Findeisen}, {Fonti}, {Fouesneau}, {Fraile}, {Fraser}, {Fuchs},
  {Furnell}, {Gai}, {Galleti}, {Galluccio}, {Garabato}, {Garc{\'\i}a-Sedano},
  {Gar{\'e}}, {Garofalo}, {Garralda}, {Gavras}, {Gerssen}, {Geyer}, {Gilmore},
  {Girona}, {Giuffrida}, {Gomes}, {Gonz{\'a}lez-Marcos},
  {Gonz{\'a}lez-N{\'u}{\~n}ez}, {Gonz{\'a}lez-Vidal}, {Granvik}, {Guerrier},
  {Guillout}, {Guiraud}, {G{\'u}rpide}, {Guti{\'e}rrez-S{\'a}nchez}, {Guy},
  {Haigron}, {Hatzidimitriou}, {Haywood}, {Heiter}, {Helmi}, {Hobbs},
  {Hofmann}, {Holl}, {Holland }, {Hunt}, {Hypki}, {Icardi}, {Irwin}, {Jevardat
  de Fombelle}, {Jofr{\'e}}, {Jonker}, {Jorissen}, {Julbe}, {Karampelas},
  {Kochoska}, {Kohley}, {Kolenberg}, {Kontizas}, {Koposov}, {Kordopatis},
  {Koubsky}, {Kowalczyk}, {Krone-Martins}, {Kudryashova}, {Kull}, {Bachchan},
  {Lacoste-Seris}, {Lanza}, {Lavigne}, {Le Poncin-Lafitte}, {Lebreton},
  {Lebzelter}, {Leccia}, {Leclerc}, {Lecoeur-Taibi}, {Lemaitre}, {Lenhardt},
  {Leroux}, {Liao}, {Licata}, {Lindstr{\o}m}, {Lister}, {Livanou}, {Lobel},
  {L{\"o}ffler}, {L{\'o}pez}, {Lopez-Lozano}, {Lorenz}, {Loureiro},
  {MacDonald}, {Magalh{\~a}es Fernandes}, {Managau}, {Mann}, {Mantelet},
  {Marchal}, {Marchant}, {Marconi}, {Marie}, {Marinoni}, {Marrese},
  {Marschalk{\'o}}, {Marshall}, {Mart{\'\i}n-Fleitas}, {Martino}, {Mary},
  {Matijevi{\v{c}}}, {Mazeh}, {McMillan}, {Messina}, {Mestre}, {Michalik},
  {Millar}, {Miranda}, {Molina}, {Molinaro}, {Molinaro}, {Moln{\'a}r},
  {Moniez}, {Montegriffo}, {Monteiro}, {Mor}, {Mora}, {Morbidelli}, {Morel},
  {Morgenthaler}, {Morley}, {Morris}, {Mulone}, {Muraveva}, {Musella},
  {Narbonne}, {Nelemans}, {Nicastro}, {Noval}, {Ord{\'e}novic},
  {Ordieres-Mer{\'e}}, {Osborne}, {Pagani}, {Pagano}, {Pailler}, {Palacin},
  {Palaversa}, {Parsons}, {Paulsen}, {Pecoraro}, {Pedrosa}, {Pentik{\"a}inen},
  {Pereira}, {Pichon}, {Piersimoni}, {Pineau}, {Plachy}, {Plum}, {Poujoulet},
  {Pr{\v{s}}a}, {Pulone}, {Ragaini}, {Rago}, {Rambaux}, {Ramos-Lerate},
  {Ranalli}, {Rauw}, {Read}, {Regibo}, {Renk}, {Reyl{\'e}}, {Ribeiro},
  {Rimoldini}, {Ripepi}, {Riva}, {Rixon}, {Roelens}, {Romero-G{\'o}mez},
  {Rowell}, {Royer}, {Rudolph}, {Ruiz-Dern}, {Sadowski}, {Sagrist{\`a}
  Sell{\'e}s}, {Sahlmann}, {Salgado}, {Salguero}, {Sarasso}, {Savietto},
  {Schnorhk}, {Schultheis}, {Sciacca}, {Segol}, {Segovia}, {Segransan},
  {Serpell}, {Shih}, {Smareglia}, {Smart}, {Smith}, {Solano}, {Solitro},
  {Sordo}, {Soria Nieto}, {Souchay}, {Spagna}, {Spoto}, {Stampa}, {Steele},
  {Steidelm{\"u}ller}, {Stephenson}, {Stoev}, {Suess}, {S{\"u}veges}, {Surdej},
  {Szabados}, {Szegedi-Elek}, {Tapiador}, {Taris}, {Tauran}, {Taylor},
  {Teixeira}, {Terrett}, {Tingley}, {Trager}, {Turon}, {Ulla}, {Utrilla},
  {Valentini}, {van Elteren}, {Van Hemelryck}, {van Leeuwen}, {Varadi},
  {Vecchiato}, {Veljanoski}, {Via}, {Vicente}, {Vogt}, {Voss}, {Votruba},
  {Voutsinas}, {Walmsley}, {Weiler}, {Weingrill}, {Werner}, {Wevers},
  {Whitehead}, {Wyrzykowski}, {Yoldas}, {{\v{Z}}erjal}, {Zucker}, {Zurbach},
  {Zwitter}, {Alecu}, {Allen}, {Allende Prieto}, {Amorim},
  {Anglada-Escud{\'e}}, {Arsenijevic}, {Azaz}, {Balm}, {Beck}, {Bernstein},
  {Bigot}, {Bijaoui}, {Blasco}, {Bonfigli}, {Bono}, {Boudreault}, {Bressan},
  {Brown}, {Brunet}, {Bunclark}, {Buonanno}, {Butkevich}, {Carret}, {Carrion},
  {Chemin}, {Ch{\'e}reau}, {Corcione}, {Darmigny}, {de Boer}, {de Teodoro}, {de
  Zeeuw}, {Delle Luche}, {Domingues}, {Dubath}, {Fodor}, {Fr{\'e}zouls},
  {Fries}, {Fustes}, {Fyfe}, {Gallardo}, {Gallegos}, {Gardiol}, {Gebran},
  {Gomboc}, {G{\'o}mez}, {Grux}, {Gueguen}, {Heyrovsky}, {Hoar}, {Iannicola},
  {Isasi Parache}, {Janotto}, {Joliet}, {Jonckheere}, {Keil}, {Kim},
  {Klagyivik}, {Klar}, {Knude}, {Kochukhov}, {Kolka}, {Kos}, {Kutka}, {Lainey},
  {LeBouquin}, {Liu}, {Loreggia}, {Makarov}, {Marseille}, {Martayan},
  {Martinez-Rubi}, {Massart}, {Meynadier}, {Mignot}, {Munari}, {Nguyen},
  {Nordlander}, {Ocvirk}, {O'Flaherty}, {Olias Sanz}, {Ortiz}, {Osorio},
  {Oszkiewicz}, {Ouzounis}, {Palmer}, {Park}, {Pasquato}, {Peltzer}, {Peralta},
  {P{\'e}turaud}, {Pieniluoma}, {Pigozzi}, {Poels}, {Prat}, {Prod'homme},
  {Raison}, {Rebordao}, {Risquez}, {Rocca-Volmerange}, {Rosen}, {Ruiz-Fuertes},
  {Russo}, {Sembay}, {Serraller Vizcaino}, {Short}, {Siebert}, {Silva},
  {Sinachopoulos}, {Slezak}, {Soffel}, {Sosnowska}, {Strai{\v{z}}ys}, {ter
  Linden}, {Terrell}, {Theil}, {Tiede}, {Troisi}, {Tsalmantza}, {Tur},
  {Vaccari}, {Vachier}, {Valles}, {Van Hamme}, {Veltz}, {Virtanen}, {Wallut},
  {Wichmann}, {Wilkinson}, {Ziaeepour}, \& {Zschocke}}]{2016A&A...595A...1G}
{Gaia Collaboration}, {Prusti}, T., {de Bruijne}, J.~H.~J., {et~al.} 2016,
  \aap, 595, A1, \dodoi{10.1051/0004-6361/201629272}

\bibitem[{{Gaia Collaboration} {et~al.}(2022{\natexlab{a}}){Gaia
  Collaboration}, {Arenou}, {Babusiaux}, {Barstow}, {Faigler}, {Jorissen},
  {Kervella}, {Mazeh}, {Mowlavi}, {Panuzzo}, {Sahlmann}, {Shahaf}, {Sozzetti},
  {Bauchet}, {Damerdji}, {Gavras}, {Giacobbe}, {Gosset}, {Halbwachs}, {Holl},
  {Lattanzi}, {Leclerc}, {Morel}, {Pourbaix}, {Re Fiorentin}, {Sadowski},
  {S{\'e}gransan}, {Siopis}, {Teyssier}, {Zwitter}, {Planquart}, {Brown},
  {Vallenari}, {Prusti}, {de Bruijne}, {Biermann}, {Creevey}, {Ducourant},
  {Evans}, {Eyer}, {Guerra}, {Hutton}, {Jordi}, {Klioner}, {Lammers},
  {Lindegren}, {Luri}, {Mignard}, {Panem}, {Randich}, {Sartoretti}, {Soubiran},
  {Tanga}, {Walton}, {Bailer-Jones}, {Bastian}, {Drimmel}, {Jansen}, {Katz},
  {van Leeuwen}, {Bakker}, {Cacciari}, {Casta{\~n}eda}, {De Angeli},
  {Fabricius}, {Fouesneau}, {Fr{\'e}mat}, {Galluccio}, {Guerrier}, {Heiter},
  {Masana}, {Messineo}, {Nicolas}, {Nienartowicz}, {Pailler}, {Riclet}, {Roux},
  {Seabroke}, {Sordo}, {Th{\'e}venin}, {Gracia-Abril}, {Portell}, {Altmann},
  {Andrae}, {Audard}, {Bellas-Velidis}, {Benson}, {Berthier}, {Blomme},
  {Burgess}, {Busonero}, {Busso}, {C{\'a}novas}, {Carry}, {Cellino}, {Cheek},
  {Clementini}, {Davidson}, {de Teodoro}, {Nu{\~n}ez Campos}, {Delchambre},
  {Dell'Oro}, {Esquej}, {Fern{\'a}ndez-Hern{\'a}ndez}, {Fraile}, {Garabato},
  {Garc{\'\i}a-Lario}, {Haigron}, {Hambly}, {Harrison}, {Hern{\'a}ndez},
  {Hestroffer}, {Hodgkin}, {Jan{\ss}en}, {Jevardat de Fombelle}, {Jordan},
  {Krone-Martins}, {Lanzafame}, {L{\"o}ffler}, {Marchal}, {Marrese},
  {Moitinho}, {Muinonen}, {Osborne}, {Pancino}, {Pauwels}, {Recio-Blanco},
  {Reyl{\'e}}, {Riello}, {Rimoldini}, {Roegiers}, {Rybizki}, {Sarro}, {Smith},
  {Utrilla}, {van Leeuwen}, {Abbas}, {{\'A}brah{\'a}m}, {Abreu Aramburu},
  {Aerts}, {Aguado}, {Ajaj}, {Aldea-Montero}, {Altavilla}, {{\'A}lvarez},
  {Alves}, {Anders}, {Anderson}, {Anglada Varela}, {Antoja}, {Baines}, {Baker},
  {Balaguer-N{\'u}{\~n}ez}, {Balbinot}, {Balog}, {Barache}, {Barbato},
  {Barros}, {Bartolom{\'e}}, {Bassilana}, {Becciani}, {Bellazzini},
  {Berihuete}, {Bernet}, {Bertone}, {Bianchi}, {Binnenfeld}, {Blanco-Cuaresma},
  {Blazere}, {Boch}, {Bombrun}, {Bossini}, {Bouquillon}, {Bragaglia},
  {Bramante}, {Breedt}, {Bressan}, {Brouillet}, {Brugaletta}, {Bucciarelli},
  {Burlacu}, {Butkevich}, {Buzzi}, {Caffau}, {Cancelliere}, {Cantat-Gaudin},
  {Carballo}, {Carlucci}, {Carnerero}, {Carrasco}, {Casamiquela}, {Castellani},
  {Castro-Ginard}, {Chaoul}, {Charlot}, {Chemin}, {Chiaramida}, {Chiavassa},
  {Chornay}, {Comoretto}, {Contursi}, {Cooper}, {Cornez}, {Cowell}, {Crifo},
  {Cropper}, {Crosta}, {Crowley}, {Dafonte}, {Dapergolas}, {David}, {de
  Laverny}, {De Luise}, {De March}, {De Ridder}, {de Souza}, {de Torres}, {del
  Peloso}, {del Pozo}, {Delbo}, {Delgado}, {Delisle}, {Demouchy},
  {Dharmawardena}, {Diakite}, {Diener}, {Distefano}, {Dolding}, {Enke},
  {Fabre}, {Fabrizio}, {Fedorets}, {Fernique}, {Figueras}, {Fournier},
  {Fouron}, {Fragkoudi}, {Gai}, {Garcia-Gutierrez}, {Garcia-Reinaldos},
  {Garc{\'\i}a-Torres}, {Garofalo}, {Gavel}, {Gerlach}, {Geyer}, {Gilmore},
  {Girona}, {Giuffrida}, {Gomel}, {Gomez}, {Gonz{\'a}lez-N{\'u}{\~n}ez},
  {Gonz{\'a}lez-Santamar{\'\i}a}, {Gonz{\'a}lez-Vidal}, {Granvik}, {Guillout},
  {Guiraud}, {Guti{\'e}rrez-S{\'a}nchez}, {Guy}, {Hatzidimitriou}, {Hauser},
  {Haywood}, {Helmer}, {Helmi}, {Sarmiento}, {Hidalgo}, {H{\l}adczuk}, {Hobbs},
  {Holland}, {Huckle}, {Jardine}, {Jasniewicz}, {Jean-Antoine Piccolo},
  {Jim{\'e}nez-Arranz}, {Juaristi Campillo}, {Julbe}, {Karbevska}, {Khanna},
  {Kordopatis}, {Korn}, {K{\'o}sp{\'a}l}, {Kostrzewa-Rutkowska},
  {Kruszy{\'n}ska}, {Kun}, {Laizeau}, {Lambert}, {Lanza}, {Lasne}, {Le
  Campion}, {Lebreton}, {Lebzelter}, {Leccia}, {Lecoeur-Taibi}, {Liao},
  {Licata}, {Lindstr{\o}m}, {Lister}, {Livanou}, {Lobel}, {Lorca}, {Loup},
  {Madrero Pardo}, {Magdaleno Romeo}, {Managau}, {Mann}, {Manteiga},
  {Marchant}, {Marconi}, {Marcos}, {Marcos Santos}, {Mar{\'\i}n Pina},
  {Marinoni}, {Marocco}, {Marshall}, {Polo}, {Mart{\'\i}n-Fleitas}, {Marton},
  {Mary}, {Masip}, {Massari}, {Mastrobuono-Battisti}, {McMillan}, {Messina},
  {Michalik}, {Millar}, {Mints}, {Molina}, {Molinaro}, {Moln{\'a}r}, {Monari},
  {Mongui{\'o}}, {Montegriffo}, {Montero}, {Mor}, {Mora}, {Morbidelli},
  {Morris}, {Muraveva}, {Murphy}, {Musella}, {Nagy}, {Noval}, {Oca{\~n}a},
  {Ogden}, {Ordenovic}, {Osinde}, {Pagani}, {Pagano}, {Palaversa}, {Palicio},
  {Pallas-Quintela}, {Panahi}, {Payne-Wardenaar}, {Pe{\~n}alosa Esteller},
  {Penttil{\"a}}, {Pichon}, {Piersimoni}, {Pineau}, {Plachy}, {Plum}, {Poggio},
  {Pr{\v{s}}a}, {Pulone}, {Racero}, {Ragaini}, {Rainer}, {Raiteri}, {Ramos},
  {Ramos-Lerate}, {Regibo}, {Richards}, {Rios Diaz}, {Ripepi}, {Riva}, {Rix},
  {Rixon}, {Robichon}, {Robin}, {Robin}, {Roelens}, {Rogues}, {Rohrbasser},
  {Romero-G{\'o}mez}, {Rowell}, {Royer}, {Ruz Mieres}, {Rybicki}, {S{\'a}ez
  N{\'u}{\~n}ez}, {Sagrist{\`a} Sell{\'e}s}, {Salguero}, {Samaras}, {Sanchez
  Gimenez}, {Sanna}, {Santove{\~n}a}, {Sarasso}, {Schultheis}, {Sciacca},
  {Segol}, {Segovia}, {Semeux}, {Siddiqui}, {Siebert}, {Siltala}, {Silvelo},
  {Slezak}, {Slezak}, {Smart}, {Snaith}, {Solano}, {Solitro}, {Souami},
  {Souchay}, {Spagna}, {Spina}, {Spoto}, {Steele}, {Steidelm{\"u}ller},
  {Stephenson}, {S{\"u}veges}, {Surdej}, {Szabados}, {Szegedi-Elek}, {Taris},
  {Taylor}, {Teixeira}, {Tolomei}, {Tonello}, {Torra}, {Torra}, {Torralba
  Elipe}, {Trabucchi}, {Tsounis}, {Turon}, {Ulla}, {Unger}, {Vaillant}, {van
  Dillen}, {van Reeven}, {Vanel}, {Vecchiato}, {Viala}, {Vicente}, {Voutsinas},
  {Weiler}, {Wevers}, {Wyrzykowski}, {Yoldas}, {Yvard}, {Zhao}, {Zorec}, \&
  {Zucker}}]{GaiaDR3}
{Gaia Collaboration}, {Arenou}, F., {Babusiaux}, C., {et~al.}
  2022{\natexlab{a}}, arXiv e-prints, arXiv:2206.05595.
\newblock \doarXiv{2206.05595}

\bibitem[{{Gaia Collaboration} {et~al.}(2022{\natexlab{b}}){Gaia
  Collaboration}, {Vallenari}, {Brown}, {Prusti}, {de Bruijne}, {Arenou},
  {Babusiaux}, {Biermann}, {Creevey}, {Ducourant}, {Evans}, {Eyer}, {Guerra},
  {Hutton}, {Jordi}, {Klioner}, {Lammers}, {Lindegren}, {Luri}, {Mignard},
  {Panem}, {Pourbaix}, {Randich}, {Sartoretti}, {Soubiran}, {Tanga}, {Walton},
  {Bailer-Jones}, {Bastian}, {Drimmel}, {Jansen}, {Katz}, {Lattanzi}, {van
  Leeuwen}, {Bakker}, {Cacciari}, {Casta{\~n}eda}, {De Angeli}, {Fabricius},
  {Fouesneau}, {Fr{\'e}mat}, {Galluccio}, {Guerrier}, {Heiter}, {Masana},
  {Messineo}, {Mowlavi}, {Nicolas}, {Nienartowicz}, {Pailler}, {Panuzzo},
  {Riclet}, {Roux}, {Seabroke}, {Sordo{\o}rcit}, {Th{\'e}venin},
  {Gracia-Abril}, {Portell}, {Teyssier}, {Altmann}, {Andrae}, {Audard},
  {Bellas-Velidis}, {Benson}, {Berthier}, {Blomme}, {Burgess}, {Busonero},
  {Busso}, {C{\'a}novas}, {Carry}, {Cellino}, {Cheek}, {Clementini},
  {Damerdji}, {Davidson}, {de Teodoro}, {Nu{\~n}ez Campos}, {Delchambre},
  {Dell'Oro}, {Esquej}, {Fern{\'a}ndez-Hern{\'a}ndez}, {Fraile}, {Garabato},
  {Garc{\'\i}a-Lario}, {Gosset}, {Haigron}, {Halbwachs}, {Hambly}, {Harrison},
  {Hern{\'a}ndez}, {Hestroffer}, {Hodgkin}, {Holl}, {Jan{\ss}en}, {Jevardat de
  Fombelle}, {Jordan}, {Krone-Martins}, {Lanzafame}, {L{\"o}ffler}, {Marchal},
  {Marrese}, {Moitinho}, {Muinonen}, {Osborne}, {Pancino}, {Pauwels},
  {Recio-Blanco}, {Reyl{\'e}}, {Riello}, {Rimoldini}, {Roegiers}, {Rybizki},
  {Sarro}, {Siopis}, {Smith}, {Sozzetti}, {Utrilla}, {van Leeuwen}, {Abbas},
  {{\'A}brah{\'a}m}, {Abreu Aramburu}, {Aerts}, {Aguado}, {Ajaj},
  {Aldea-Montero}, {Altavilla}, {{\'A}lvarez}, {Alves}, {Anders}, {Anderson},
  {Anglada Varela}, {Antoja}, {Baines}, {Baker}, {Balaguer-N{\'u}{\~n}ez},
  {Balbinot}, {Balog}, {Barache}, {Barbato}, {Barros}, {Barstow},
  {Bartolom{\'e}}, {Bassilana}, {Bauchet}, {Becciani}, {Bellazzini},
  {Berihuete}, {Bernet}, {Bertone}, {Bianchi}, {Binnenfeld}, {Blanco-Cuaresma},
  {Blazere}, {Boch}, {Bombrun}, {Bossini}, {Bouquillon}, {Bragaglia},
  {Bramante}, {Breedt}, {Bressan}, {Brouillet}, {Brugaletta}, {Bucciarelli},
  {Burlacu}, {Butkevich}, {Buzzi}, {Caffau}, {Cancelliere}, {Cantat-Gaudin},
  {Carballo}, {Carlucci}, {Carnerero}, {Carrasco}, {Casamiquela}, {Castellani},
  {Castro-Ginard}, {Chaoul}, {Charlot}, {Chemin}, {Chiaramida}, {Chiavassa},
  {Chornay}, {Comoretto}, {Contursi}, {Cooper}, {Cornez}, {Cowell}, {Crifo},
  {Cropper}, {Crosta}, {Crowley}, {Dafonte}, {Dapergolas}, {David}, {David},
  {de Laverny}, {De Luise}, {De March}, {De Ridder}, {de Souza}, {de Torres},
  {del Peloso}, {del Pozo}, {Delbo}, {Delgado}, {Delisle}, {Demouchy},
  {Dharmawardena}, {Di Matteo}, {Diakite}, {Diener}, {Distefano}, {Dolding},
  {Edvardsson}, {Enke}, {Fabre}, {Fabrizio}, {Faigler}, {Fedorets}, {Fernique},
  {Fienga}, {Figueras}, {Fournier}, {Fouron}, {Fragkoudi}, {Gai},
  {Garcia-Gutierrez}, {Garcia-Reinaldos}, {Garc{\'\i}a-Torres}, {Garofalo},
  {Gavel}, {Gavras}, {Gerlach}, {Geyer}, {Giacobbe}, {Gilmore}, {Girona},
  {Giuffrida}, {Gomel}, {Gomez}, {Gonz{\'a}lez-N{\'u}{\~n}ez},
  {Gonz{\'a}lez-Santamar{\'\i}a}, {Gonz{\'a}lez-Vidal}, {Granvik}, {Guillout},
  {Guiraud}, {Guti{\'e}rrez-S{\'a}nchez}, {Guy}, {Hatzidimitriou}, {Hauser},
  {Haywood}, {Helmer}, {Helmi}, {Sarmiento}, {Hidalgo}, {Hilger},
  {H{\l}adczuk}, {Hobbs}, {Holland}, {Huckle}, {Jardine}, {Jasniewicz},
  {Jean-Antoine Piccolo}, {Jim{\'e}nez-Arranz}, {Jorissen}, {Juaristi
  Campillo}, {Julbe}, {Karbevska}, {Kervella}, {Khanna}, {Kontizas},
  {Kordopatis}, {Korn}, {K{\'o}sp{\'a}l}, {Kostrzewa-Rutkowska},
  {Kruszy{\'n}ska}, {Kun}, {Laizeau}, {Lambert}, {Lanza}, {Lasne}, {Le
  Campion}, {Lebreton}, {Lebzelter}, {Leccia}, {Leclerc}, {Lecoeur-Taibi},
  {Liao}, {Licata}, {Lindstr{\o}m}, {Lister}, {Livanou}, {Lobel}, {Lorca},
  {Loup}, {Madrero Pardo}, {Magdaleno Romeo}, {Managau}, {Mann}, {Manteiga},
  {Marchant}, {Marconi}, {Marcos}, {Marcos Santos}, {Mar{\'\i}n Pina},
  {Marinoni}, {Marocco}, {Marshall}, {Polo}, {Mart{\'\i}n-Fleitas}, {Marton},
  {Mary}, {Masip}, {Massari}, {Mastrobuono-Battisti}, {Mazeh}, {McMillan},
  {Messina}, {Michalik}, {Millar}, {Mints}, {Molina}, {Molinaro}, {Moln{\'a}r},
  {Monari}, {Mongui{\'o}}, {Montegriffo}, {Montero}, {Mor}, {Mora},
  {Morbidelli}, {Morel}, {Morris}, {Muraveva}, {Murphy}, {Musella}, {Nagy},
  {Noval}, {Oca{\~n}a}, {Ogden}, {Ordenovic}, {Osinde}, {Pagani}, {Pagano},
  {Palaversa}, {Palicio}, {Pallas-Quintela}, {Panahi}, {Payne-Wardenaar},
  {Pe{\~n}alosa Esteller}, {Penttil{\"a}}, {Pichon}, {Piersimoni}, {Pineau},
  {Plachy}, {Plum}, {Poggio}, {Pr{\v{s}}a}, {Pulone}, {Racero}, {Ragaini},
  {Rainer}, {Raiteri}, {Rambaux}, {Ramos}, {Ramos-Lerate}, {Re Fiorentin},
  {Regibo}, {Richards}, {Rios Diaz}, {Ripepi}, {Riva}, {Rix}, {Rixon},
  {Robichon}, {Robin}, {Robin}, {Roelens}, {Rogues}, {Rohrbasser},
  {Romero-G{\'o}mez}, {Rowell}, {Royer}, {Ruz Mieres}, {Rybicki}, {Sadowski},
  {S{\'a}ez N{\'u}{\~n}ez}, {Sagrist{\`a} Sell{\'e}s}, {Sahlmann}, {Salguero},
  {Samaras}, {Sanchez Gimenez}, {Sanna}, {Santove{\~n}a}, {Sarasso},
  {Schultheis}, {Sciacca}, {Segol}, {Segovia}, {S{\'e}gransan}, {Semeux},
  {Shahaf}, {Siddiqui}, {Siebert}, {Siltala}, {Silvelo}, {Slezak}, {Slezak},
  {Smart}, {Snaith}, {Solano}, {Solitro}, {Souami}, {Souchay}, {Spagna},
  {Spina}, {Spoto}, {Steele}, {Steidelm{\"u}ller}, {Stephenson}, {S{\"u}veges},
  {Surdej}, {Szabados}, {Szegedi-Elek}, {Taris}, {Taylo}, {Teixeira},
  {Tolomei}, {Tonello}, {Torra}, {Torra}, {Torralba Elipe}, {Trabucchi},
  {Tsounis}, {Turon}, {Ulla}, {Unger}, {Vaillant}, {van Dillen}, {van Reeven},
  {Vanel}, {Vecchiato}, {Viala}, {Vicente}, {Voutsinas}, {Weiler}, {Wevers},
  {Wyrzykowski}, {Yoldas}, {Yvard}, {Zhao}, {Zorec}, {Zucker}, \&
  {Zwitter}}]{Vallenari2022}
{Gaia Collaboration}, {Vallenari}, A., {Brown}, A.~G.~A., {et~al.}
  2022{\natexlab{b}}, arXiv e-prints, arXiv:2208.00211.
\newblock \doarXiv{2208.00211}

\bibitem[{{Gao} {et~al.}(2014){Gao}, {Liu}, {Zhang}, {Justham}, {Deng}, \&
  {Yang}}]{Gao2014}
{Gao}, S., {Liu}, C., {Zhang}, X., {et~al.} 2014, \apjl, 788, L37,
  \dodoi{10.1088/2041-8205/788/2/L37}

\bibitem[{{Giesers} {et~al.}(2018){Giesers}, {Dreizler}, {Husser}, {Kamann},
  {Anglada Escud{\'e}}, {Brinchmann}, {Carollo}, {Roth}, {Weilbacher}, \&
  {Wisotzki}}]{Giesers2018}
{Giesers}, B., {Dreizler}, S., {Husser}, T.-O., {et~al.} 2018, \mnras, 475,
  L15, \dodoi{10.1093/mnrasl/slx203}

\bibitem[{{Giesers} {et~al.}(2019){Giesers}, {Kamann}, {Dreizler}, {Husser},
  {Askar}, {G{\"o}ttgens}, {Brinchmann}, {Latour}, {Weilbacher}, {Wendt}, \&
  {Roth}}]{Giesers2019}
{Giesers}, B., {Kamann}, S., {Dreizler}, S., {et~al.} 2019, \aap, 632, A3,
  \dodoi{10.1051/0004-6361/201936203}

\bibitem[{{Gilkis} {et~al.}(2021){Gilkis}, {Shenar}, {Ramachandran}, {Jermyn},
  {Mahy}, {Oskinova}, {Arcavi}, \& {Sana}}]{Gilkis2021}
{Gilkis}, A., {Shenar}, T., {Ramachandran}, V., {et~al.} 2021, \mnras, 503,
  1884, \dodoi{10.1093/mnras/stab383}

\bibitem[{{Gonz{\'a}lez Hern{\'a}ndez} {et~al.}(2008){Gonz{\'a}lez
  Hern{\'a}ndez}, {Rebolo}, {Israelian}, {Filippenko}, {Chornock}, {Tominaga},
  {Umeda}, \& {Nomoto}}]{Gonzalez-Hernandez2004}
{Gonz{\'a}lez Hern{\'a}ndez}, J.~I., {Rebolo}, R., {Israelian}, G., {et~al.}
  2008, \apj, 679, 732, \dodoi{10.1086/586888}

\bibitem[{{GRAVITY Collaboration} {et~al.}(2018){GRAVITY Collaboration},
  {Karl}, {Pfuhl}, {Eisenhauer}, {Genzel}, {Grellmann}, {Habibi}, {Abuter},
  {Accardo}, {Amorim}, {Anugu}, {{\'A}vila}, {Benisty}, {Berger}, {Blind},
  {Bonnet}, {Bourget}, {Brandner}, {Brast}, {Buron}, {Caratti O Garatti},
  {Chapron}, {Cl{\'e}net}, {Collin}, {Coud{\'e} Du Foresto}, {de Wit}, {de
  Zeeuw}, {Deen}, {Delplancke-Str{\"o}bele}, {Dembet}, {Derie}, {Dexter},
  {Duvert}, {Ebert}, {Eckart}, {Esselborn}, {F{\'e}dou}, {Finger}, {Garcia},
  {Garcia Dabo}, {Garcia Lopez}, {Gao}, {Gendron}, {Gillessen}, {Gont{\'e}},
  {Gordo}, {Gr{\"o}zinger}, {Guajardo}, {Guieu}, {Haguenauer}, {Hans},
  {Haubois}, {Haug}, {Hau{\ss}mann}, {Henning}, {Hippler}, {Horrobin}, {Huber},
  {Hubert}, {Hubin}, {Jakob}, {Jochum}, {Jocou}, {Kaufer}, {Kellner},
  {Kendrew}, {Kern}, {Kervella}, {Kiekebusch}, {Klein}, {K{\"o}hler}, {Kolb},
  {Kulas}, {Lacour}, {Lapeyr{\`e}re}, {Lazareff}, {Le Bouquin}, {L{\'e}na},
  {Lenzen}, {L{\'e}v{\^e}que}, {Lin}, {Lippa}, {Magnard}, {Mehrgan},
  {M{\'e}rand}, {Moulin}, {M{\"u}ller}, {M{\"u}ller}, {Neumann}, {Oberti},
  {Ott}, {Pallanca}, {Panduro}, {Pasquini}, {Paumard}, {Percheron}, {Perraut},
  {Perrin}, {Pfl{\"u}ger}, {Duc}, {Plewa}, {Popovic}, {Rabien}, {Ram{\'\i}rez},
  {Ramos}, {Rau}, {Riquelme}, {Rodr{\'\i}guez-Coira}, {Rohloff}, {Rosales},
  {Rousset}, {Sanchez-Bermudez}, {Scheithauer}, {Sch{\"o}ller}, {Schuhler},
  {Spyromilio}, {Straub}, {Straubmeier}, {Sturm}, {Suarez}, {Tristram},
  {Ventura}, {Vincent}, {Waisberg}, {Wank}, {Widmann}, {Wieprecht}, {Wiest},
  {Wiezorrek}, {Wittkowski}, {Woillez}, {Wolff}, {Yazici}, {Ziegler}, \&
  {Zins}}]{Karl18}
{GRAVITY Collaboration}, {Karl}, M., {Pfuhl}, O., {et~al.} 2018, \aap, 620,
  A116, \dodoi{10.1051/0004-6361/201833575}

\bibitem[{{Gray} {et~al.}(2003){Gray}, {Corbally}, {Garrison}, {McFadden}, \&
  {Robinson}}]{Gray03}
{Gray}, R.~O., {Corbally}, C.~J., {Garrison}, R.~F., {McFadden}, M.~T., \&
  {Robinson}, P.~E. 2003, \aj, 126, 2048, \dodoi{10.1086/378365}

\bibitem[{{Gray} \& {Corbally}(2009)}]{Gray09}
{Gray}, R.~O., \& {Corbally}, Christopher, J. 2009, {Stellar Spectral
  Classification}

\bibitem[{{Halbwachs} {et~al.}(2022){Halbwachs}, {Pourbaix}, {Arenou},
  {Galluccio}, {Guillout}, {Bauchet}, {Marchal}, {Sadowski}, \&
  {Teyssier}}]{Halbwachs2022}
{Halbwachs}, J.-L., {Pourbaix}, D., {Arenou}, F., {et~al.} 2022, arXiv
  e-prints, arXiv:2206.05726.
\newblock \doarXiv{2206.05726}

\bibitem[{{Henden}(2019)}]{Henden19}
{Henden}, A.~A. 2019, \jaavso, 47, 130

\bibitem[{{Henden} {et~al.}(2016){Henden}, {Templeton}, {Terrell}, {Smith},
  {Levine}, \& {Welch}}]{Henden2016}
{Henden}, A.~A., {Templeton}, M., {Terrell}, D., {et~al.} 2016, VizieR Online
  Data Catalog, II/336

\bibitem[{{Holl} {et~al.}(2022){Holl}, {Sozzetti}, {Sahlmann}, {Giacobbe},
  {S{\'e}gransan}, {Unger}, {Delisle}, {Barbato}, {Lattanzi}, {Morbidelli}, \&
  {Sosnowska}}]{Holl2022}
{Holl}, B., {Sozzetti}, A., {Sahlmann}, J., {et~al.} 2022, arXiv e-prints,
  arXiv:2206.05439.
\newblock \doarXiv{2206.05439}

\bibitem[{{Howard} {et~al.}(2010){Howard}, {Johnson}, {Marcy}, {Fischer},
  {Wright}, {Bernat}, {Henry}, {Peek}, {Isaacson}, {Apps}, {Endl}, {Cochran},
  {Valenti}, {Anderson}, \& {Piskunov}}]{Howard10}
{Howard}, A.~W., {Johnson}, J.~A., {Marcy}, G.~W., {et~al.} 2010, \apj, 721,
  1467, \dodoi{10.1088/0004-637X/721/2/1467}

\bibitem[{{Husser} {et~al.}(2013){Husser}, {Wende-von Berg}, {Dreizler},
  {Homeier}, {Reiners}, {Barman}, \& {Hauschildt}}]{phoenix2013}
{Husser}, T.~O., {Wende-von Berg}, S., {Dreizler}, S., {et~al.} 2013, \aap,
  553, A6, \dodoi{10.1051/0004-6361/201219058}

\bibitem[{{Janssens} {et~al.}(2022){Janssens}, {Shenar}, {Sana}, {Faigler},
  {Langer}, {Marchant}, {Mazeh}, {Sch{\"u}rmann}, \& {Shahaf}}]{Janssens2022}
{Janssens}, S., {Shenar}, T., {Sana}, H., {et~al.} 2022, \aap, 658, A129,
  \dodoi{10.1051/0004-6361/202141866}

\bibitem[{{Jayasinghe} {et~al.}(2021){Jayasinghe}, {Stanek}, {Thompson},
  {Kochanek}, {Rowan}, {Vallely}, {Strassmeier}, {Weber}, {Hinkle}, {Hambsch},
  {Martin}, {Prieto}, {Pessi}, {Huber}, {Auchettl}, {Lopez}, {Ilyin},
  {Badenes}, {Howard}, {Isaacson}, \& {Murphy}}]{Jayasinghe2021}
{Jayasinghe}, T., {Stanek}, K.~Z., {Thompson}, T.~A., {et~al.} 2021, \mnras,
  504, 2577, \dodoi{10.1093/mnras/stab907}

\bibitem[{{Jayasinghe} {et~al.}(2022){Jayasinghe}, {Thompson}, {Kochanek},
  {Stanek}, {Rowan}, {Martin}, {El-Badry}, {Vallely}, {Hinkle}, {Huber},
  {Isaacson}, {Tayar}, {Auchettl}, {Ilyin}, {Howard}, \&
  {Badenes}}]{Jayasinghe2022}
{Jayasinghe}, T., {Thompson}, T.~A., {Kochanek}, C.~S., {et~al.} 2022, \mnras,
  \dodoi{10.1093/mnras/stac2187}

\bibitem[{{Ji} {et~al.}(2020){Ji}, {Li}, {Simon}, {Marshall}, {Vivas}, {Pace},
  {Bechtol}, {Drlica-Wagner}, {Koposov}, {Hansen}, {Allam}, {Gruendl},
  {Johnson}, {McNanna}, {No{\"e}l}, {Tucker}, \& {Walker}}]{Ji2020}
{Ji}, A.~P., {Li}, T.~S., {Simon}, J.~D., {et~al.} 2020, \apj, 889, 27,
  \dodoi{10.3847/1538-4357/ab6213}

\bibitem[{{Kelson}(2003)}]{Kelson2003}
{Kelson}, D.~D. 2003, \pasp, 115, 688, \dodoi{10.1086/375502}

\bibitem[{{Kennicutt} {et~al.}(2003){Kennicutt}, {Bresolin}, \&
  {Garnett}}]{Kennicutt2003}
{Kennicutt}, Robert~C., J., {Bresolin}, F., \& {Garnett}, D.~R. 2003, \apj,
  591, 801, \dodoi{10.1086/375398}

\bibitem[{{Kirby} {et~al.}(2015){Kirby}, {Simon}, \& {Cohen}}]{Kirby15}
{Kirby}, E.~N., {Simon}, J.~D., \& {Cohen}, J.~G. 2015, \apj, 810, 56,
  \dodoi{10.1088/0004-637X/810/1/56}

\bibitem[{{Kobayashi} {et~al.}(2020){Kobayashi}, {Karakas}, \&
  {Lugaro}}]{kobayashi2020}
{Kobayashi}, C., {Karakas}, A.~I., \& {Lugaro}, M. 2020, \apj, 900, 179,
  \dodoi{10.3847/1538-4357/abae65}

\bibitem[{{Lallement} {et~al.}(2014){Lallement}, {Vergely}, {Valette},
  {Puspitarini}, {Eyer}, \& {Casagrande}}]{lallement2014}
{Lallement}, R., {Vergely}, J.~L., {Valette}, B., {et~al.} 2014, \aap, 561,
  A91, \dodoi{10.1051/0004-6361/201322032}

\bibitem[{{Lallement} {et~al.}(2018){Lallement}, {Capitanio}, {Ruiz-Dern},
  {Danielski}, {Babusiaux}, {Vergely}, {Elyajouri}, {Arenou}, \&
  {Leclerc}}]{lallement2018}
{Lallement}, R., {Capitanio}, L., {Ruiz-Dern}, L., {et~al.} 2018, \aap, 616,
  A132, \dodoi{10.1051/0004-6361/201832832}

\bibitem[{{Lam} {et~al.}(2022){Lam}, {Lu}, {Udalski}, {Bond}, {Bennett},
  {Skowron}, {Mr{\'o}z}, {Poleski}, {Sumi}, {Szyma{\'n}ski}, {Koz{\l}owski},
  {Pietrukowicz}, {Soszy{\'n}ski}, {Ulaczyk}, {Wyrzykowski}, {Miyazaki},
  {Suzuki}, {Koshimoto}, {Rattenbury}, {Hosek}, {Abe}, {Barry}, {Bhattacharya},
  {Fukui}, {Fujii}, {Hirao}, {Itow}, {Kirikawa}, {Kondo}, {Matsubara},
  {Matsumoto}, {Muraki}, {Olmschenk}, {Ranc}, {Okamura}, {Satoh}, {Silva},
  {Toda}, {Tristram}, {Vandorou}, {Yama}, {Abrams}, {Agarwal}, {Rose}, \&
  {Terry}}]{Lam2022}
{Lam}, C.~Y., {Lu}, J.~R., {Udalski}, A., {et~al.} 2022, \apjl, 933, L23,
  \dodoi{10.3847/2041-8213/ac7442}

\bibitem[{{Lamberts} {et~al.}(2016){Lamberts}, {Garrison-Kimmel}, {Clausen}, \&
  {Hopkins}}]{Lamberts2016}
{Lamberts}, A., {Garrison-Kimmel}, S., {Clausen}, D.~R., \& {Hopkins}, P.~F.
  2016, \mnras, 463, L31, \dodoi{10.1093/mnrasl/slw152}

\bibitem[{{Lehmer} {et~al.}(2021){Lehmer}, {Eufrasio}, {Basu-Zych}, {Doore},
  {Fragos}, {Garofali}, {Kovlakas}, {Williams}, {Zezas}, \&
  {Santana-Silva}}]{Lehmer2021}
{Lehmer}, B.~D., {Eufrasio}, R.~T., {Basu-Zych}, A., {et~al.} 2021, \apj, 907,
  17, \dodoi{10.3847/1538-4357/abcec1}

\bibitem[{{Lejeune} {et~al.}(1997){Lejeune}, {Cuisinier}, \&
  {Buser}}]{Lejeune1997}
{Lejeune}, T., {Cuisinier}, F., \& {Buser}, R. 1997, \aaps, 125, 229,
  \dodoi{10.1051/aas:1997373}

\bibitem[{{Lejeune} {et~al.}(1998){Lejeune}, {Cuisinier}, \&
  {Buser}}]{Lejeune1998}
---. 1998, \aaps, 130, 65, \dodoi{10.1051/aas:1998405}

\bibitem[{{Lennon} {et~al.}(2021){Lennon}, {Dufton}, {Villase{\~n}or}, {Evans},
  {Langer}, {Saxton}, {Monageng}, \& {Toonen}}]{Lennon2021}
{Lennon}, D.~J., {Dufton}, P.~L., {Villase{\~n}or}, J.~I., {et~al.} 2021, arXiv
  e-prints, arXiv:2111.12173.
\newblock \doarXiv{2111.12173}

\bibitem[{{Lennon} {et~al.}(2022){Lennon}, {Dufton}, {Villase{\~n}or}, {Evans},
  {Langer}, {Saxton}, {Monageng}, \& {Toonen}}]{Lennon2022}
---. 2022, \aap, 665, A180, \dodoi{10.1051/0004-6361/202142413}

\bibitem[{{Lind} {et~al.}(2011){Lind}, {Asplund}, {Barklem}, \&
  {Belyaev}}]{Lind2011}
{Lind}, K., {Asplund}, M., {Barklem}, P.~S., \& {Belyaev}, A.~K. 2011, \aap,
  528, A103, \dodoi{10.1051/0004-6361/201016095}

\bibitem[{{Lindegren} {et~al.}(2021{\natexlab{a}}){Lindegren}, {Klioner},
  {Hern{\'a}ndez}, {Bombrun}, {Ramos-Lerate}, {Steidelm{\"u}ller}, {Bastian},
  {Biermann}, {de Torres}, {Gerlach}, {Geyer}, {Hilger}, {Hobbs}, {Lammers},
  {McMillan}, {Stephenson}, {Casta{\~n}eda}, {Davidson}, {Fabricius},
  {Gracia-Abril}, {Portell}, {Rowell}, {Teyssier}, {Torra}, {Bartolom{\'e}},
  {Clotet}, {Garralda}, {Gonz{\'a}lez-Vidal}, {Torra}, {Abbas}, {Altmann},
  {Anglada Varela}, {Balaguer-N{\'u}{\~n}ez}, {Balog}, {Barache}, {Becciani},
  {Bernet}, {Bertone}, {Bianchi}, {Bouquillon}, {Brown}, {Bucciarelli},
  {Busonero}, {Butkevich}, {Buzzi}, {Cancelliere}, {Carlucci}, {Charlot},
  {Cioni}, {Crosta}, {Crowley}, {del Peloso}, {del Pozo}, {Drimmel}, {Esquej},
  {Fienga}, {Fraile}, {Gai}, {Garcia-Reinaldos}, {Guerra}, {Hambly}, {Hauser},
  {Jan{\ss}en}, {Jordan}, {Kostrzewa-Rutkowska}, {Lattanzi}, {Liao}, {Licata},
  {Lister}, {L{\"o}ffler}, {Marchant}, {Masip}, {Mignard}, {Mints}, {Molina},
  {Mora}, {Morbidelli}, {Murphy}, {Pagani}, {Panuzzo}, {Pe{\~n}alosa Esteller},
  {Poggio}, {Re Fiorentin}, {Riva}, {Sagrist{\`a} Sell{\'e}s}, {Sanchez
  Gimenez}, {Sarasso}, {Sciacca}, {Siddiqui}, {Smart}, {Souami}, {Spagna},
  {Steele}, {Taris}, {Utrilla}, {van Reeven}, \& {Vecchiato}}]{Lindegren21}
{Lindegren}, L., {Klioner}, S.~A., {Hern{\'a}ndez}, J., {et~al.}
  2021{\natexlab{a}}, \aap, 649, A2, \dodoi{10.1051/0004-6361/202039709}

\bibitem[{{Lindegren} {et~al.}(2021{\natexlab{b}}){Lindegren}, {Bastian},
  {Biermann}, {Bombrun}, {de Torres}, {Gerlach}, {Geyer}, {Hern{\'a}ndez},
  {Hilger}, {Hobbs}, {Klioner}, {Lammers}, {McMillan}, {Ramos-Lerate},
  {Steidelm{\"u}ller}, {Stephenson}, \& {van Leeuwen}}]{Lindegren2021b}
{Lindegren}, L., {Bastian}, U., {Biermann}, M., {et~al.} 2021{\natexlab{b}},
  \aap, 649, A4, \dodoi{10.1051/0004-6361/202039653}

\bibitem[{{Lipartito} {et~al.}(2021){Lipartito}, {Bailey}, {Brandt}, {Mazin},
  {Mateo}, {Spencer}, \& {Roederer}}]{Lipartito+Bailey+Brandt+etal_2021}
{Lipartito}, I., {Bailey}, John~I., I., {Brandt}, T.~D., {et~al.} 2021, \aj,
  162, 285, \dodoi{10.3847/1538-3881/ac2ccd}

\bibitem[{{Liu} {et~al.}(2019){Liu}, {Zhang}, {Howard}, {Bai}, {Lu}, {Soria},
  {Justham}, {Li}, {Zheng}, {Wang}, {Belczynski}, {Casares}, {Zhang}, {Yuan},
  {Dong}, {Lei}, {Isaacson}, {Wang}, {Bai}, {Shao}, {Gao}, {Wang}, {Niu},
  {Cui}, {Zheng}, {Mu}, {Zhang}, {Wang}, {Heger}, {Qi}, {Liao}, {Lattanzi},
  {Gu}, {Wang}, {Wu}, {Shao}, {Shen}, {Wang}, {Bregman}, {Di Stefano}, {Liu},
  {Han}, {Zhang}, {Wang}, {Ren}, {Zhang}, {Zhang}, {Wang}, {Cabrera-Lavers},
  {Corradi}, {Rebolo}, {Zhao}, {Zhao}, {Chu}, \& {Cui}}]{Liu2019}
{Liu}, J., {Zhang}, H., {Howard}, A.~W., {et~al.} 2019, \nat, 575, 618,
  \dodoi{10.1038/s41586-019-1766-2}

\bibitem[{{Luo} {et~al.}(2022){Luo}, {Zhao}, {Zhao}, \& {et al.}}]{Luo22}
{Luo}, A.~L., {Zhao}, Y.~H., {Zhao}, G., \& {et al.} 2022, VizieR Online Data
  Catalog, V/156

\bibitem[{{Mahy} {et~al.}(2022){Mahy}, {Sana}, {Shenar}, {Sen}, {Langer},
  {Marchant}, {Abdul-Masih}, {Banyard}, {Bodensteiner}, {Bowman}, {Dsilva},
  {Fabry}, {Hawcroft}, {Janssens}, {Van Reeth}, \& {Eldridge}}]{Mahy2022}
{Mahy}, L., {Sana}, H., {Shenar}, T., {et~al.} 2022, \aap, 664, A159,
  \dodoi{10.1051/0004-6361/202243147}

\bibitem[{{Ma{\'\i}z Apell{\'a}niz} {et~al.}(2016){Ma{\'\i}z Apell{\'a}niz},
  {Sota}, {Arias}, {Barb{\'a}}, {Walborn}, {Sim{\'o}n-D{\'\i}az}, {Negueruela},
  {Marco}, {Le{\~a}o}, {Herrero}, {Gamen}, \& {Alfaro}}]{Maiz2016}
{Ma{\'\i}z Apell{\'a}niz}, J., {Sota}, A., {Arias}, J.~I., {et~al.} 2016,
  \apjs, 224, 4, \dodoi{10.3847/0067-0049/224/1/4}

\bibitem[{{Marchant} {et~al.}(2021){Marchant}, {Pappas}, {Gallegos-Garcia},
  {Berry}, {Taam}, {Kalogera}, \& {Podsiadlowski}}]{Marchant2021}
{Marchant}, P., {Pappas}, K. M.~W., {Gallegos-Garcia}, M., {et~al.} 2021, \aap,
  650, A107, \dodoi{10.1051/0004-6361/202039992}

\bibitem[{{Martin} {et~al.}(2005){Martin}, {Fanson}, {Schiminovich},
  {Morrissey}, {Friedman}, {Barlow}, {Conrow}, {Grange}, {Jelinsky},
  {Milliard}, {Siegmund}, {Bianchi}, {Byun}, {Donas}, {Forster}, {Heckman},
  {Lee}, {Madore}, {Malina}, {Neff}, {Rich}, {Small}, {Surber}, {Szalay},
  {Welsh}, \& {Wyder}}]{Martin2005}
{Martin}, D.~C., {Fanson}, J., {Schiminovich}, D., {et~al.} 2005, \apjl, 619,
  L1, \dodoi{10.1086/426387}

\bibitem[{{Martins} \& {Palacios}(2013)}]{MartinsPalacios2013}
{Martins}, F., \& {Palacios}, A. 2013, \aap, 560, A16,
  \dodoi{10.1051/0004-6361/201322480}

\bibitem[{{Mashian} \& {Loeb}(2017)}]{Mashian2017}
{Mashian}, N., \& {Loeb}, A. 2017, \mnras, 470, 2611,
  \dodoi{10.1093/mnras/stx1410}

\bibitem[{{McWilliam}(1998)}]{mcwilliam1998}
{McWilliam}, A. 1998, \aj, 115, 1640, \dodoi{10.1086/300289}

\bibitem[{{Mel{\'e}ndez} {et~al.}(2014){Mel{\'e}ndez}, {Schirbel}, {Monroe},
  {Yong}, {Ram{\'\i}rez}, \& {Asplund}}]{melendez2014}
{Mel{\'e}ndez}, J., {Schirbel}, L., {Monroe}, T.~R., {et~al.} 2014, \aap, 567,
  L3, \dodoi{10.1051/0004-6361/201424172}

\bibitem[{{Mirabel} \& {Rodrigues}(2003)}]{Mirabel2003}
{Mirabel}, I.~F., \& {Rodrigues}, I. 2003, Science, 300, 1119,
  \dodoi{10.1126/science.1083451}

\bibitem[{{Moe} {et~al.}(2019){Moe}, {Kratter}, \& {Badenes}}]{Moe2019}
{Moe}, M., {Kratter}, K.~M., \& {Badenes}, C. 2019, \apj, 875, 61,
  \dodoi{10.3847/1538-4357/ab0d88}

\bibitem[{{Morrissey} {et~al.}(2007){Morrissey}, {Conrow}, {Barlow}, {Small},
  {Seibert}, {Wyder}, {Budav{\'a}ri}, {Arnouts}, {Friedman}, {Forster},
  {Martin}, {Neff}, {Schiminovich}, {Bianchi}, {Donas}, {Heckman}, {Lee},
  {Madore}, {Milliard}, {Rich}, {Szalay}, {Welsh}, \& {Yi}}]{Morrissey07}
{Morrissey}, P., {Conrow}, T., {Barlow}, T.~A., {et~al.} 2007, \apjs, 173, 682,
  \dodoi{10.1086/520512}

\bibitem[{{Morton}(2015)}]{morton2015}
{Morton}, T.~D. 2015, {isochrones: Stellar model grid package}.
\newblock \doeprint{1503.010}

\bibitem[{{Naoz} {et~al.}(2013){Naoz}, {Farr}, {Lithwick}, {Rasio}, \&
  {Teyssandier}}]{Naoz2013}
{Naoz}, S., {Farr}, W.~M., {Lithwick}, Y., {Rasio}, F.~A., \& {Teyssandier}, J.
  2013, \mnras, 431, 2155, \dodoi{10.1093/mnras/stt302}

\bibitem[{{Neijssel} {et~al.}(2019){Neijssel}, {Vigna-G{\'o}mez}, {Stevenson},
  {Barrett}, {Gaebel}, {Broekgaarden}, {de Mink}, {Sz{\'e}csi}, {Vinciguerra},
  \& {Mandel}}]{Neijssel2019}
{Neijssel}, C.~J., {Vigna-G{\'o}mez}, A., {Stevenson}, S., {et~al.} 2019,
  \mnras, 490, 3740, \dodoi{10.1093/mnras/stz2840}

\bibitem[{{Newman} {et~al.}(2013){Newman}, {Cooper}, {Davis}, {Faber}, {Coil},
  {Guhathakurta}, {Koo}, {Phillips}, {Conroy}, {Dutton}, {Finkbeiner}, {Gerke},
  {Rosario}, {Weiner}, {Willmer}, {Yan}, {Harker}, {Kassin}, {Konidaris},
  {Lai}, {Madgwick}, {Noeske}, {Wirth}, {Connolly}, {Kaiser}, {Kirby},
  {Lemaux}, {Lin}, {Lotz}, {Luppino}, {Marinoni}, {Matthews}, {Metevier}, \&
  {Schiavon}}]{Newman13}
{Newman}, J.~A., {Cooper}, M.~C., {Davis}, M., {et~al.} 2013, \apjs, 208, 5,
  \dodoi{10.1088/0067-0049/208/1/5}

\bibitem[{{Olejak} {et~al.}(2020){Olejak}, {Belczynski}, {Bulik}, \&
  {Sobolewska}}]{Olejak2020}
{Olejak}, A., {Belczynski}, K., {Bulik}, T., \& {Sobolewska}, M. 2020, \aap,
  638, A94, \dodoi{10.1051/0004-6361/201936557}

\bibitem[{{Onken} {et~al.}(2019){Onken}, {Wolf}, {Bessell}, {Chang}, {Da
  Costa}, {Luvaul}, {Mackey}, {Schmidt}, \& {Shao}}]{Onken19}
{Onken}, C.~A., {Wolf}, C., {Bessell}, M.~S., {et~al.} 2019, \pasa, 36, e033,
  \dodoi{10.1017/pasa.2019.27}

\bibitem[{{Orosz} {et~al.}(2001){Orosz}, {Kuulkers}, {van der Klis},
  {McClintock}, {Garcia}, {Callanan}, {Bailyn}, {Jain}, \&
  {Remillard}}]{Orosz2001}
{Orosz}, J.~A., {Kuulkers}, E., {van der Klis}, M., {et~al.} 2001, \apj, 555,
  489, \dodoi{10.1086/321442}

\bibitem[{{{\"O}zel} {et~al.}(2010){{\"O}zel}, {Psaltis}, {Narayan}, \&
  {McClintock}}]{Ozel2010}
{{\"O}zel}, F., {Psaltis}, D., {Narayan}, R., \& {McClintock}, J.~E. 2010,
  \apj, 725, 1918, \dodoi{10.1088/0004-637X/725/2/1918}

\bibitem[{{Panizo-Espinar} {et~al.}(2022){Panizo-Espinar}, {Armas Padilla},
  {Mu{\~n}oz-Darias}, {Koljonen}, {C{\'u}neo}, {S{\'a}nchez-Sierras}, {Mata
  S{\'a}nchez}, {Casares}, {Corral-Santana}, {Fender}, {Jim{\'e}nez-Ibarra},
  {Ponti}, {Steeghs}, \& {Torres}}]{Pan22}
{Panizo-Espinar}, G., {Armas Padilla}, M., {Mu{\~n}oz-Darias}, T., {et~al.}
  2022, \aap, 664, A100, \dodoi{10.1051/0004-6361/202243426}

\bibitem[{{Paxton} {et~al.}(2011){Paxton}, {Bildsten}, {Dotter}, {Herwig},
  {Lesaffre}, \& {Timmes}}]{paxton2011}
{Paxton}, B., {Bildsten}, L., {Dotter}, A., {et~al.} 2011, \apjs, 192, 3,
  \dodoi{10.1088/0067-0049/192/1/3}

\bibitem[{{Paxton} {et~al.}(2019){Paxton}, {Smolec}, {Schwab}, {Gautschy},
  {Bildsten}, {Cantiello}, {Dotter}, {Farmer}, {Goldberg}, {Jermyn}, {Kanbur},
  {Marchant}, {Thoul}, {Townsend}, {Wolf}, {Zhang}, \& {Timmes}}]{paxton2019}
{Paxton}, B., {Smolec}, R., {Schwab}, J., {et~al.} 2019, \apjs, 243, 10,
  \dodoi{10.3847/1538-4365/ab2241}

\bibitem[{{Pinsonneault} {et~al.}(2018){Pinsonneault}, {Elsworth}, {Tayar},
  {Serenelli}, {Stello}, {Zinn}, {Mathur}, {Garc{\'\i}a}, {Johnson}, {Hekker},
  {Huber}, {Kallinger}, {M{\'e}sz{\'a}ros}, {Mosser}, {Stassun}, {Girardi},
  {Rodrigues}, {Silva Aguirre}, {An}, {Basu}, {Chaplin}, {Corsaro}, {Cunha},
  {Garc{\'\i}a-Hern{\'a}ndez}, {Holtzman}, {J{\"o}nsson}, {Shetrone}, {Smith},
  {Sobeck}, {Stringfellow}, {Zamora}, {Beers}, {Fern{\'a}ndez-Trincado},
  {Frinchaboy}, {Hearty}, \& {Nitschelm}}]{Pin2018}
{Pinsonneault}, M.~H., {Elsworth}, Y.~P., {Tayar}, J., {et~al.} 2018, \apjs,
  239, 32, \dodoi{10.3847/1538-4365/aaebfd}

\bibitem[{{Price-Whelan} {et~al.}(2017){Price-Whelan}, {Hogg},
  {Foreman-Mackey}, \& {Rix}}]{Price-Whelan17}
{Price-Whelan}, A.~M., {Hogg}, D.~W., {Foreman-Mackey}, D., \& {Rix}, H.-W.
  2017, \apj, 837, 20, \dodoi{10.3847/1538-4357/aa5e50}

\bibitem[{{Prochaska} \& {McWilliam}(2000)}]{prochaska2000a}
{Prochaska}, J.~X., \& {McWilliam}, A. 2000, \apjl, 537, L57,
  \dodoi{10.1086/312749}

\bibitem[{{Prochaska} {et~al.}(2000){Prochaska}, {Naumov}, {Carney},
  {McWilliam}, \& {Wolfe}}]{prochaska2000b}
{Prochaska}, J.~X., {Naumov}, S.~O., {Carney}, B.~W., {McWilliam}, A., \&
  {Wolfe}, A.~M. 2000, \aj, 120, 2513, \dodoi{10.1086/316818}

\bibitem[{{Radovan} {et~al.}(2010){Radovan}, {Cabak}, {Laiterman}, {Lockwood},
  \& {Vogt}}]{Radovan2010}
{Radovan}, M.~V., {Cabak}, G.~F., {Laiterman}, L.~H., {Lockwood}, C.~T., \&
  {Vogt}, S.~S. 2010, in Society of Photo-Optical Instrumentation Engineers
  (SPIE) Conference Series, Vol. 7735, Ground-based and Airborne
  Instrumentation for Astronomy III, ed. I.~S. {McLean}, S.~K. {Ramsay}, \&
  H.~{Takami}, 77354K, \dodoi{10.1117/12.857726}

\bibitem[{{Raithel} {et~al.}(2018){Raithel}, {Sukhbold}, \&
  {{\"O}zel}}]{Raithel2018}
{Raithel}, C.~A., {Sukhbold}, T., \& {{\"O}zel}, F. 2018, \apj, 856, 35,
  \dodoi{10.3847/1538-4357/aab09b}

\bibitem[{{Ramachandran} {et~al.}(2019){Ramachandran}, {Hamann}, {Oskinova},
  {Gallagher}, {Hainich}, {Shenar}, {Sander}, {Todt}, \&
  {Fulmer}}]{Ramachandran2019}
{Ramachandran}, V., {Hamann}, W.~R., {Oskinova}, L.~M., {et~al.} 2019, \aap,
  625, A104, \dodoi{10.1051/0004-6361/201935365}

\bibitem[{{Reggiani} {et~al.}(2022{\natexlab{a}}){Reggiani}, {Ji},
  {Schlaufman}, {Frebel}, {Necib}, {Nelson}, {Hawkins}, \&
  {Galarza}}]{reggiani2022b}
{Reggiani}, H., {Ji}, A.~P., {Schlaufman}, K.~C., {et~al.} 2022{\natexlab{a}},
  \aj, 163, 252, \dodoi{10.3847/1538-3881/ac62d9}

\bibitem[{{Reggiani} \& {Mel{\'e}ndez}(2018)}]{reggiani2018}
{Reggiani}, H., \& {Mel{\'e}ndez}, J. 2018, \mnras, 475, 3502,
  \dodoi{10.1093/mnras/sty104}

\bibitem[{{Reggiani} {et~al.}(2022{\natexlab{b}}){Reggiani}, {Schlaufman},
  {Healy}, {Lothringer}, \& {Sing}}]{reggiani2022a}
{Reggiani}, H., {Schlaufman}, K.~C., {Healy}, B.~F., {Lothringer}, J.~D., \&
  {Sing}, D.~K. 2022{\natexlab{b}}, \aj, 163, 159,
  \dodoi{10.3847/1538-3881/ac4d9f}

\bibitem[{{Reggiani} {et~al.}(2019){Reggiani}, {Amarsi}, {Lind}, {Barklem},
  {Zatsarinny}, {Bartschat}, {Fursa}, {Bray}, {Spina}, \&
  {Mel{\'e}ndez}}]{reggiani2019}
{Reggiani}, H., {Amarsi}, A.~M., {Lind}, K., {et~al.} 2019, \aap, 627, A177,
  \dodoi{10.1051/0004-6361/201935156}

\bibitem[{{Reggiani} {et~al.}(2022{\natexlab{c}}){Reggiani}, {Rainot}, {Sana},
  {Almeida}, {Caballero-Nieves}, {Kratter}, {Lacour}, {Le Bouquin}, \&
  {Zinnecker}}]{MReggiani2022}
{Reggiani}, M., {Rainot}, A., {Sana}, H., {et~al.} 2022{\natexlab{c}}, \aap,
  660, A122, \dodoi{10.1051/0004-6361/202142418}

\bibitem[{{Riello} {et~al.}(2021){Riello}, {De Angeli}, {Evans}, {Montegriffo},
  {Carrasco}, {Busso}, {Palaversa}, {Burgess}, {Diener}, {Davidson}, {Rowell},
  {Fabricius}, {Jordi}, {Bellazzini}, {Pancino}, {Harrison}, {Cacciari}, {van
  Leeuwen}, {Hambly}, {Hodgkin}, {Osborne}, {Altavilla}, {Barstow}, {Brown},
  {Castellani}, {Cowell}, {De Luise}, {Gilmore}, {Giuffrida}, {Hidalgo},
  {Holland}, {Marinoni}, {Pagani}, {Piersimoni}, {Pulone}, {Ragaini}, {Rainer},
  {Richards}, {Sanna}, {Walton}, {Weiler}, \& {Yoldas}}]{Riello2021}
{Riello}, M., {De Angeli}, F., {Evans}, D.~W., {et~al.} 2021, \aap, 649, A3,
  \dodoi{10.1051/0004-6361/202039587}

\bibitem[{{Rivinius} {et~al.}(2020){Rivinius}, {Baade}, {Hadrava}, {Heida}, \&
  {Klement}}]{Rivinius2020}
{Rivinius}, T., {Baade}, D., {Hadrava}, P., {Heida}, M., \& {Klement}, R. 2020,
  \aap, 637, L3, \dodoi{10.1051/0004-6361/202038020}

\bibitem[{{Russell} {et~al.}(2022){Russell}, {Del Santo}, {Marino}, {Segreto},
  {Motta}, {Bahramian}, {Corbel}, {D'A{\`\i}}, {Salvo}, {Miller-Jones},
  {Pinto}, {Pintore}, \& {Tzioumis}}]{Russell2022}
{Russell}, T.~D., {Del Santo}, M., {Marino}, A., {et~al.} 2022, \mnras, 513,
  6196, \dodoi{10.1093/mnras/stac1332}

\bibitem[{{Sahu} {et~al.}(2022){Sahu}, {Anderson}, {Casertano}, {Bond},
  {Udalski}, {Dominik}, {Calamida}, {Bellini}, {Brown}, {Rejkuba}, {Bajaj},
  {Kains}, {Ferguson}, {Fryer}, {Yock}, {Mr{\'o}z}, {Koz{\l}owski},
  {Pietrukowicz}, {Poleski}, {Skowron}, {Soszy{\'n}ski}, {Szyma{\'n}ski},
  {Ulaczyk}, {Wyrzykowski}, {Barry}, {Bennett}, {Bond}, {Hirao}, {Silva},
  {Kondo}, {Koshimoto}, {Ranc}, {Rattenbury}, {Sumi}, {Suzuki}, {Tristram},
  {Vandorou}, {Beaulieu}, {Marquette}, {Cole}, {Fouqu{\'e}}, {Hill}, {Dieters},
  {Coutures}, {Dominis-Prester}, {Bennett}, {Bachelet}, {Menzies}, {Albrow},
  {Pollard}, {Gould}, {Yee}, {Allen}, {Almeida}, {Christie}, {Drummond},
  {Gal-Yam}, {Gorbikov}, {Jablonski}, {Lee}, {Maoz}, {Manulis}, {McCormick},
  {Natusch}, {Pogge}, {Shvartzvald}, {J{\o}rgensen}, {Alsubai}, {Andersen},
  {Bozza}, {Novati}, {Burgdorf}, {Hinse}, {Hundertmark}, {Husser}, {Kerins},
  {Longa-Pe{\~n}a}, {Mancini}, {Penny}, {Rahvar}, {Ricci}, {Sajadian},
  {Skottfelt}, {Snodgrass}, {Southworth}, {Tregloan-Reed}, {Wambsganss},
  {Wertz}, {Tsapras}, {Street}, {Bramich}, {Horne}, {Steele}, \& {RoboNet
  Collaboration}}]{Sahu2022}
{Sahu}, K.~C., {Anderson}, J., {Casertano}, S., {et~al.} 2022, \apj, 933, 83,
  \dodoi{10.3847/1538-4357/ac739e}

\bibitem[{{Sana} {et~al.}(2009){Sana}, {Gosset}, \& {Evans}}]{Sana2009}
{Sana}, H., {Gosset}, E., \& {Evans}, C.~J. 2009, \mnras, 400, 1479,
  \dodoi{10.1111/j.1365-2966.2009.15545.x}

\bibitem[{{Sana} {et~al.}(2012){Sana}, {de Mink}, {de Koter}, {Langer},
  {Evans}, {Gieles}, {Gosset}, {Izzard}, {Le Bouquin}, \&
  {Schneider}}]{Sana2012}
{Sana}, H., {de Mink}, S.~E., {de Koter}, A., {et~al.} 2012, Science, 337, 444,
  \dodoi{10.1126/science.1223344}

\bibitem[{{Saracino} {et~al.}(2022){Saracino}, {Kamann}, {Guarcello}, {Usher},
  {Bastian}, {Cabrera-Ziri}, {Gieles}, {Dreizler}, {Da Costa}, {Husser}, \&
  {H{\'e}nault-Brunet}}]{Saracino2022}
{Saracino}, S., {Kamann}, S., {Guarcello}, M.~G., {et~al.} 2022, \mnras, 511,
  2914, \dodoi{10.1093/mnras/stab3159}

\bibitem[{{Shahaf} {et~al.}(2022){Shahaf}, {Bashi}, {Mazeh}, {Faigler},
  {Arenou}, {El-Badry}, \& {Rix}}]{Shahaf2022}
{Shahaf}, S., {Bashi}, D., {Mazeh}, T., {et~al.} 2022, arXiv e-prints,
  arXiv:2209.00828.
\newblock \doarXiv{2209.00828}

\bibitem[{{Shao} \& {Li}(2019)}]{Shao2019}
{Shao}, Y., \& {Li}, X.-D. 2019, \apj, 885, 151,
  \dodoi{10.3847/1538-4357/ab4816}

\bibitem[{{Shenar} {et~al.}(2020){Shenar}, {Bodensteiner}, {Abdul-Masih},
  {Fabry}, {Mahy}, {Marchant}, {Banyard}, {Bowman}, {Dsilva}, {Hawcroft},
  {Reggiani}, \& {Sana}}]{Shenar2020}
{Shenar}, T., {Bodensteiner}, J., {Abdul-Masih}, M., {et~al.} 2020, \aap, 639,
  L6, \dodoi{10.1051/0004-6361/202038275}

\bibitem[{{Shenar} {et~al.}(2022){Shenar}, {Sana}, {Mahy}, {El-Badry},
  {Marchant}, {Langer}, {Hawcroft}, {Fabry}, {Sen}, {Almeida}, {Abdul-Masih},
  {Bodensteiner}, {Crowther}, {Gieles}, {Gromadzki}, {H{\'e}nault-Brunet},
  {Herrero}, {de Koter}, {Iwanek}, {Koz{\l}owski}, {Lennon}, {Ma{\'\i}z
  Apell{\'a}niz}, {Mr{\'o}z}, {Moffat}, {Picco}, {Pietrukowicz}, {Poleski},
  {Rybicki}, {Schneider}, {Skowron}, {Skowron}, {Soszy{\'n}ski},
  {Szyma{\'n}ski}, {Toonen}, {Udalski}, {Ulaczyk}, {Vink}, \&
  {Wrona}}]{Shenar2022}
{Shenar}, T., {Sana}, H., {Mahy}, L., {et~al.} 2022, Nature Astronomy, 6, 1085,
  \dodoi{10.1038/s41550-022-01730-y}

\bibitem[{{Shikauchi} {et~al.}(2022){Shikauchi}, {Tanikawa}, \&
  {Kawanaka}}]{Shikauchi2022}
{Shikauchi}, M., {Tanikawa}, A., \& {Kawanaka}, N. 2022, \apj, 928, 13,
  \dodoi{10.3847/1538-4357/ac5329}

\bibitem[{{Simon} \& {Geha}(2007)}]{SG07}
{Simon}, J.~D., \& {Geha}, M. 2007, \apj, 670, 313, \dodoi{10.1086/521816}

\bibitem[{{Simon} {et~al.}(2017){Simon}, {Li}, {Drlica-Wagner}, {Bechtol},
  {Marshall}, {James}, {Wang}, {Strigari}, {Balbinot}, {Kuehn}, {Walker},
  {Abbott}, {Allam}, {Annis}, {Benoit-L{\'e}vy}, {Brooks}, {Buckley-Geer},
  {Burke}, {Carnero Rosell}, {Carrasco Kind}, {Carretero}, {Cunha}, {D'Andrea},
  {da Costa}, {DePoy}, {Desai}, {Doel}, {Fernandez}, {Flaugher}, {Frieman},
  {Garc{\'\i}a-Bellido}, {Gaztanaga}, {Goldstein}, {Gruen}, {Gutierrez},
  {Kuropatkin}, {Maia}, {Martini}, {Menanteau}, {Miller}, {Miquel}, {Neilsen},
  {Nord}, {Ogando}, {Plazas}, {Romer}, {Rykoff}, {Sanchez}, {Santiago},
  {Scarpine}, {Schubnell}, {Sevilla-Noarbe}, {Smith}, {Sobreira}, {Suchyta},
  {Swanson}, {Tarle}, {Whiteway}, {Yanny}, \& {DES Collaboration}}]{Simon2017}
{Simon}, J.~D., {Li}, T.~S., {Drlica-Wagner}, A., {et~al.} 2017, \apj, 838, 11,
  \dodoi{10.3847/1538-4357/aa5be7}

\bibitem[{{Skrutskie} {et~al.}(2006){Skrutskie}, {Cutri}, {Stiening},
  {Weinberg}, {Schneider}, {Carpenter}, {Beichman}, {Capps}, {Chester},
  {Elias}, {Huchra}, {Liebert}, {Lonsdale}, {Monet}, {Price}, {Seitzer},
  {Jarrett}, {Kirkpatrick}, {Gizis}, {Howard}, {Evans}, {Fowler}, {Fullmer},
  {Hurt}, {Light}, {Kopan}, {Marsh}, {McCallon}, {Tam}, {Van Dyk}, \&
  {Wheelock}}]{Skrutskie2006}
{Skrutskie}, M.~F., {Cutri}, R.~M., {Stiening}, R., {et~al.} 2006, \aj, 131,
  1163, \dodoi{10.1086/498708}

\bibitem[{{Sneden}(1973)}]{sneden1973}
{Sneden}, C.~A. 1973, PhD thesis, THE UNIVERSITY OF TEXAS AT AUSTIN.

\bibitem[{{Sohn} {et~al.}(2007){Sohn}, {Majewski}, {Mu{\~n}oz}, {Kunkel},
  {Johnston}, {Ostheimer}, {Guhathakurta}, {Patterson}, {Siegel}, \&
  {Cooper}}]{Sohn07}
{Sohn}, S.~T., {Majewski}, S.~R., {Mu{\~n}oz}, R.~R., {et~al.} 2007, \apj, 663,
  960, \dodoi{10.1086/518302}

\bibitem[{{Soubiran} {et~al.}(2018){Soubiran}, {Jasniewicz}, {Chemin},
  {Zurbach}, {Brouillet}, {Panuzzo}, {Sartoretti}, {Katz}, {Le Campion},
  {Marchal}, {Hestroffer}, {Th{\'e}venin}, {Crifo}, {Udry}, {Cropper},
  {Seabroke}, {Viala}, {Benson}, {Blomme}, {Jean-Antoine}, {Huckle}, {Smith},
  {Baker}, {Damerdji}, {Dolding}, {Fr{\'e}mat}, {Gosset}, {Guerrier}, {Guy},
  {Haigron}, {Jan{\ss}en}, {Plum}, {Fabre}, {Lasne}, {Pailler}, {Panem},
  {Riclet}, {Royer}, {Tauran}, {Zwitter}, {Gueguen}, \& {Turon}}]{Soubiran2018}
{Soubiran}, C., {Jasniewicz}, G., {Chemin}, L., {et~al.} 2018, \aap, 616, A7,
  \dodoi{10.1051/0004-6361/201832795}

\bibitem[{{Stefanik} {et~al.}(1999){Stefanik}, {Latham}, \&
  {Torres}}]{Stefanik1999}
{Stefanik}, R.~P., {Latham}, D.~W., \& {Torres}, G. 1999, in Astronomical
  Society of the Pacific Conference Series, Vol. 185, IAU Colloq. 170: Precise
  Stellar Radial Velocities, ed. J.~B. {Hearnshaw} \& C.~D. {Scarfe}, 354

\bibitem[{{Su{\'a}rez-Andr{\'e}s} {et~al.}(2015){Su{\'a}rez-Andr{\'e}s},
  {Gonz{\'a}lez Hern{\'a}ndez}, {Israelian}, {Casares}, \&
  {Rebolo}}]{Suarez-Andres2015}
{Su{\'a}rez-Andr{\'e}s}, L., {Gonz{\'a}lez Hern{\'a}ndez}, J.~I., {Israelian},
  G., {Casares}, J., \& {Rebolo}, R. 2015, \mnras, 447, 2261,
  \dodoi{10.1093/mnras/stu2612}

\bibitem[{{Sukhbold} {et~al.}(2016){Sukhbold}, {Ertl}, {Woosley}, {Brown}, \&
  {Janka}}]{Sukhbold2016}
{Sukhbold}, T., {Ertl}, T., {Woosley}, S.~E., {Brown}, J.~M., \& {Janka}, H.~T.
  2016, \apj, 821, 38, \dodoi{10.3847/0004-637X/821/1/38}

\bibitem[{{Suzuki} {et~al.}(2019){Suzuki}, {Gupta}, {Okawa}, \&
  {Maeda}}]{Suzuki2019}
{Suzuki}, H., {Gupta}, P., {Okawa}, H., \& {Maeda}, K.-i. 2019, \mnras, 486,
  L52, \dodoi{10.1093/mnrasl/slz058}

\bibitem[{{Tanikawa} {et~al.}(2022){Tanikawa}, {Hattori}, {Kawanaka},
  {Kinugawa}, {Shikauchi}, \& {Tsuna}}]{Tanikawa2022}
{Tanikawa}, A., {Hattori}, K., {Kawanaka}, N., {et~al.} 2022, arXiv e-prints,
  arXiv:2209.05632, \dodoi{10.48550/arXiv.2209.05632}

\bibitem[{{Thompson} {et~al.}(2019){Thompson}, {Kochanek}, {Stanek}, {Badenes},
  {Post}, {Jayasinghe}, {Latham}, {Bieryla}, {Esquerdo}, {Berlind}, {Calkins},
  {Tayar}, {Lindegren}, {Johnson}, {Holoien}, {Auchettl}, \&
  {Covey}}]{Thompson2019}
{Thompson}, T.~A., {Kochanek}, C.~S., {Stanek}, K.~Z., {et~al.} 2019, Science,
  366, 637, \dodoi{10.1126/science.aau4005}

\bibitem[{{Tokovinin} \& {Moe}(2020)}]{Tokov2020}
{Tokovinin}, A., \& {Moe}, M. 2020, \mnras, 491, 5158,
  \dodoi{10.1093/mnras/stz3299}

\bibitem[{{Tokovinin} {et~al.}(2006){Tokovinin}, {Thomas}, {Sterzik}, \&
  {Udry}}]{Tokov2006}
{Tokovinin}, A., {Thomas}, S., {Sterzik}, M., \& {Udry}, S. 2006, \aap, 450,
  681, \dodoi{10.1051/0004-6361:20054427}

\bibitem[{{Tonry} {et~al.}(2012){Tonry}, {Stubbs}, {Lykke}, {Doherty},
  {Shivvers}, {Burgett}, {Chambers}, {Hodapp}, {Kaiser}, {Kudritzki},
  {Magnier}, {Morgan}, {Price}, \& {Wainscoat}}]{Tonry2012}
{Tonry}, J.~L., {Stubbs}, C.~W., {Lykke}, K.~R., {et~al.} 2012, \apj, 750, 99,
  \dodoi{10.1088/0004-637X/750/2/99}

\bibitem[{{van den Heuvel} {et~al.}(2017){van den Heuvel}, {Portegies Zwart},
  \& {de Mink}}]{VandenHeuvel2017}
{van den Heuvel}, E.~P.~J., {Portegies Zwart}, S.~F., \& {de Mink}, S.~E. 2017,
  \mnras, 471, 4256, \dodoi{10.1093/mnras/stx1430}

\bibitem[{{van den Heuvel} \& {Tauris}(2020{\natexlab{a}})}]{vandenHeuvel2020}
{van den Heuvel}, E. P.~J., \& {Tauris}, T.~M. 2020{\natexlab{a}}, Science,
  368, eaba3282, \dodoi{10.1126/science.aba3282}

\bibitem[{{van den Heuvel} \&
  {Tauris}(2020{\natexlab{b}})}]{VandenHeuvel2020Sci}
---. 2020{\natexlab{b}}, Science, 368, eaba3282,
  \dodoi{10.1126/science.aba3282}

\bibitem[{{Vigna-G{\'o}mez} {et~al.}(2022){Vigna-G{\'o}mez}, {Liu},
  {Aguilera-Dena}, {Grishin}, {Ramirez-Ruiz}, \&
  {Soares-Furtado}}]{VignaGomez2022}
{Vigna-G{\'o}mez}, A., {Liu}, B., {Aguilera-Dena}, D.~R., {et~al.} 2022,
  \mnras, 515, L50, \dodoi{10.1093/mnrasl/slac067}

\bibitem[{{Vines} \& {Jenkins}(2022)}]{Vines_Jenkins2022}
{Vines}, J.~I., \& {Jenkins}, J.~S. 2022, \mnras, 513, 2719,
  \dodoi{10.1093/mnras/stac956}

\bibitem[{{Vogt} {et~al.}(2014){Vogt}, {Radovan}, {Kibrick}, {Butler},
  {Alcott}, {Allen}, {Arriagada}, {Bolte}, {Burt}, {Cabak}, {Chloros},
  {Cowley}, {Deich}, {Dupraw}, {Earthman}, {Epps}, {Faber}, {Fischer}, {Gates},
  {Hilyard}, {Holden}, {Johnston}, {Keiser}, {Kanto}, {Katsuki}, {Laiterman},
  {Lanclos}, {Laughlin}, {Lewis}, {Lockwood}, {Lynam}, {Marcy}, {McLean},
  {Miller}, {Misch}, {Peck}, {Pfister}, {Phillips}, {Rivera}, {Sandford},
  {Saylor}, {Stover}, {Thompson}, {Walp}, {Ward}, {Wareham}, {Wei}, \&
  {Wright}}]{Vogt2014}
{Vogt}, S.~S., {Radovan}, M., {Kibrick}, R., {et~al.} 2014, \pasp, 126, 359,
  \dodoi{10.1086/676120}

\bibitem[{{Wang} {et~al.}(2022){Wang}, {Liao}, {Giacobbo}, {Olejak}, {Gao}, \&
  {Liu}}]{Wang2022}
{Wang}, Y., {Liao}, S., {Giacobbo}, N., {et~al.} 2022, \aap, 665, A111,
  \dodoi{10.1051/0004-6361/202243684}

\bibitem[{{Wiktorowicz} {et~al.}(2020){Wiktorowicz}, {Lu}, {Wyrzykowski},
  {Zhang}, {Liu}, {Justham}, \& {Belczynski}}]{Wiktorowicz2020}
{Wiktorowicz}, G., {Lu}, Y., {Wyrzykowski}, {\L}., {et~al.} 2020, \apj, 905,
  134, \dodoi{10.3847/1538-4357/abc699}

\bibitem[{{Wiktorowicz} {et~al.}(2019){Wiktorowicz}, {Wyrzykowski},
  {Chruslinska}, {Klencki}, {Rybicki}, \& {Belczynski}}]{Wiktorowicz2019}
{Wiktorowicz}, G., {Wyrzykowski}, {\L}., {Chruslinska}, M., {et~al.} 2019,
  \apj, 885, 1, \dodoi{10.3847/1538-4357/ab45e6}

\bibitem[{{Wolf} {et~al.}(2018){Wolf}, {Onken}, {Luvaul}, {Schmidt}, {Bessell},
  {Chang}, {Da Costa}, {Mackey}, {Martin-Jones}, {Murphy}, {Preston}, {Scalzo},
  {Shao}, {Smillie}, {Tisserand}, {White}, \& {Yuan}}]{Wolf2018}
{Wolf}, C., {Onken}, C.~A., {Luvaul}, L.~C., {et~al.} 2018, \pasa, 35, e010,
  \dodoi{10.1017/pasa.2018.5}

\bibitem[{{Wright} {et~al.}(2010){Wright}, {Eisenhardt}, {Mainzer}, {Ressler},
  {Cutri}, {Jarrett}, {Kirkpatrick}, {Padgett}, {McMillan}, {Skrutskie},
  {Stanford}, {Cohen}, {Walker}, {Mather}, {Leisawitz}, {Gautier}, {McLean},
  {Benford}, {Lonsdale}, {Blain}, {Mendez}, {Irace}, {Duval}, {Liu}, {Royer},
  {Heinrichsen}, {Howard}, {Shannon}, {Kendall}, {Walsh}, {Larsen}, {Cardon},
  {Schick}, {Schwalm}, {Abid}, {Fabinsky}, {Naes}, \& {Tsai}}]{Wright2010}
{Wright}, E.~L., {Eisenhardt}, P. R.~M., {Mainzer}, A.~K., {et~al.} 2010, \aj,
  140, 1868, \dodoi{10.1088/0004-6256/140/6/1868}

\bibitem[{{Wright}(2005)}]{Wright2005}
{Wright}, J.~T. 2005, \pasp, 117, 657, \dodoi{10.1086/430369}

\bibitem[{{Yalinewich} {et~al.}(2018){Yalinewich}, {Beniamini}, {Hotokezaka},
  \& {Zhu}}]{Yalinewich2018}
{Yalinewich}, A., {Beniamini}, P., {Hotokezaka}, K., \& {Zhu}, W. 2018, \mnras,
  481, 930, \dodoi{10.1093/mnras/sty2327}

\bibitem[{{Yamaguchi} {et~al.}(2018){Yamaguchi}, {Kawanaka}, {Bulik}, \&
  {Piran}}]{Yamaguchi2018}
{Yamaguchi}, M.~S., {Kawanaka}, N., {Bulik}, T., \& {Piran}, T. 2018, \apj,
  861, 21, \dodoi{10.3847/1538-4357/aac5ec}

\bibitem[{{Yee} {et~al.}(2017){Yee}, {Petigura}, \& {von Braun}}]{Yee17}
{Yee}, S.~W., {Petigura}, E.~A., \& {von Braun}, K. 2017, \apj, 836, 77,
  \dodoi{10.3847/1538-4357/836/1/77}

\end{thebibliography}
